%% file: cps-resilience-vision.tex
\documentclass[manuscript,screen]{acmart}

\input{macros}

\usepackage{pifont}
\usepackage{multirow}
\usepackage{float}
\usepackage{subfig}
\usepackage{soul}
\usepackage{temporal}
\usepackage{wrapfig}

\makeatletter
\renewcommand\@authorfont{\small}
\renewcommand\@affiliationfont{\footnotesize}
\makeatother

\AtBeginDocument{%
  \providecommand\BibTeX{{%
    \normalfont B\kern-0.5em{\scshape i\kern-0.25em b}\kern-0.8em\TeX}}}

\setcopyright{acmcopyright}
\copyrightyear{2025}
\acmYear{2025}
\acmDOI{XXXXXXX.XXXXXXX}

\acmJournal{CSUR}
\acmVolume{37}
\acmNumber{4}
\acmArticle{111}
\acmMonth{8}

\acmPrice{15.00}
\acmISBN{978-1-4503-XXXX-X/18/06}

\begin{document}


\title{Digital Guardians: The Past and The Future of Cyber-Physical Resilience}



\author{Saurabh Bagchi*}
\affiliation{%
  \institution{Purdue University}
  \city{West Lafayette, IN}
  \country{United States of America}
  }
  \thanks{*Other than the first two authors, all other authors are listed alphabetically. Corresponding author: Saurabh Bagchi.}
\email{sbagchi@purdue.edu}

\author{Hyunseung Kim*}
\affiliation{%
  \institution{Purdue University}
  \city{West Lafayette, IN}
  \country{United States of America}
  }
\email{kim4061@purdue.edu}

\author{Tarek Abdelzaher}
\affiliation{%
  \institution{University of Illinois Urbana-Champaign}
  \city{Champaign, IL}
  \country{United States of America}
  }
\email{zaher@illinois.edu}

\author{Homa Alemzadeh}
\affiliation{%
  \institution{The University of Virginia}
  \city{Charlottesville, VA}
  \country{United States of America}
  }
\email{alemzadeh@virginia.edu}

\author{Somali Chaterji}
\affiliation{%
  \institution{Purdue University}
  \city{West Lafayette, IN}
  \country{United States of America}
  }
\email{schaterj@purdue.edu}

\author{Glen Chou}
\affiliation{%
  \institution{Georgia Institute of Technology}
  \city{Atlanta, GA}
  \country{United States of America}
  }
\email{chou@gatech.edu}

\author{Yuying Duan}
\affiliation{%
  \institution{University of Notre Dame}
  \city{Notre Dame, IN}
  \country{United States of America}
  }
\email{yduan2@nd.edu}

\author{Fanxin Kong}
\affiliation{%
  \institution{University of Notre Dame}
  \city{Notre Dame, IN}
  \country{United States of America}
  }
\email{fkong@nd.edu}

\author{Michael Lemmon}
\affiliation{%
  \institution{University of Notre Dame}
  \city{Notre Dame, IN}
  \country{United States of America}
  }
\email{lemmon@nd.edu}

\author{Yin Li}
\affiliation{%
  \institution{University of Wisconsin–Madison}
  \city{Madison, WI}
  \country{United States of America}
  }
\email{yin.li@wisc.edu}

\author{Mengyu Liu}
\affiliation{%
  \institution{University of Notre Dame}
  \city{Notre Dame, IN}
  \country{United States of America}
  }
\email{mliu9@nd.edu}

\author{Wenhao Luo}
\affiliation{%
  \institution{University of Illinois Chicago}
  \city{Chicago, IL}
  \country{United States of America}
  }
\email{wenhao@uic.edu}

\author{Meiyi Ma}
\affiliation{%
  \institution{Vanderbilt University}
  \city{Nashville, TN}
  \country{United States of America}
  }
\email{meiyi.ma@vanderbilt.edu}

\author{Sibin Mohan}
\affiliation{%
  \institution{George Washington University}
  \city{Washington, DC}
  \country{United States of America}
  }
\email{sibin.mohan@gwu.edu}

\author{Ayan Mukhopadhyay}
\affiliation{%
  \institution{William \& Mary}
  \city{Williamsburg, VA}
  \country{United States of America}
  }
\email{amukhopadhyay@wm.edu}

\author{Melkior Ornik}
\affiliation{%
  \institution{University of Illinois Urbana-Champaign}
  \city{Champaign, IL}
  \country{United States of America}
  }
\email{mornik@illinois.edu}

\author{Dimitra Panagou}
\affiliation{%
  \institution{University of Michigan}
  \city{Ann Arbor, MI}
  \country{United States of America}
  }
\email{dpanagou@umich.edu}

\author{Kristin Yvonne Rozier}
\affiliation{%
  \institution{Iowa State University}
  \city{Ames, IA}
  \country{United States of America}
  }
\email{kyrozier@iastate.edu}

\author{Ivan Ruchkin}
\affiliation{%
  \institution{University of Florida}
  \city{Gainesville, FL}
  \country{United States of America}
  }
\email{iruchkin@ece.ufl.edu}

\author{Huajie Shao}
\affiliation{%
  \institution{William \& Mary}
  \city{Williamsburg, VA}
  \country{United States of America}
  }
\email{hshao@wm.edu}

\author{Sze Zheng Yong}
\affiliation{%
  \institution{Northeastern University}
  \city{Boston, MA}
  \country{United States of America}
  }
\email{s.yong@northeastern.edu}

\author{Majid Zamani}
\affiliation{%
  \institution{University of Colorado Boulder}
  \city{Boulder, CO}
  \country{United States of America}
  }
\email{majid.zamani@colorado.edu}

\author{Xugui Zhou}
\affiliation{%
  \institution{Louisiana State University}
  \city{Baton Rouge, LA}
  \country{United States of America}
  }
\email{xuguizhou@lsu.edu}

\renewcommand{\shortauthors}{Bagchi, et al.}

\begin{abstract}

Resilience in cyber-physical systems (CPS) is the fundamental ability to maintain safety and critical functionality despite adverse "perturbations," which includes security attacks, environmental disruptions, and hardware or software failures. This survey provides a comprehensive review of CPS resilience, framing the field through five interconnected themes that are required in an integrated whole to achieve real-world resilience.

The article first posits that resilience is a system-wide property emerging from interactions between hardware, software, and human users. Second, it addresses the challenges of learning-enabled CPS, which often operate in data-scarce environments characterized by imbalanced or noisy data, requiring innovative solutions like synthetic data generation and foundation model adaptation. Third, the survey examines proactive measures for resilience, which include distinctive aspects of verification, testing, and redundancy. Fourth, it explores recovery mechanisms, moving beyond traditional fault models to design "just good enough" recovery strategies that prioritize safety-critical functions during perturbations. Finally, it highlights the central role of the human, focusing on the different levels of human intervention, the necessity of trust calibration, and the requirement for explainable AI to support human-CPS teaming.

These themes are illustrated through representative application domains, primarily Connected and Autonomous Transportation Systems (CATS) and Medical CPS (MCPS). By integrating the five interconnected themes, this survey provides a systematic roadmap for achieving the resilient CPS in increasingly complex and adversarial environments.
\end{abstract}




\received{XXX}
\received[revised]{XXX}
\received[accepted]{XXX}


\maketitle

\saurabh{For CSUR, references count toward the 35 page budget, unlike in any sane conference that we submit to. So, to pare the number of references down, each of you go through your writeups and if you are using > 2 citations for the same point, please reduce that down to 1 or 2.}

\input{Sections/introduction}

\input{Sections/applications}

\input{Sections/related}

\input{Sections/theme1-system}

\input{Sections/theme2-learning}

\input{Sections/theme3-proactive}

\input{Sections/theme4-recovery}
\input{Sections/theme5-human}

\input{Sections/look-ahead}


\input{Sections/acknowledgments}

\bibliographystyle{ACM-Reference-Format}

\bibliography{refs/ref-sample-base, refs/ref-chaterji, refs/ref-saurabh, refs/ref-Huajie, refs/ref-tarek, refs/ref-hyunseung, refs/ref-xugui, refs/ref-glen, refs/ref-lemmon, refs/ref-ivan, refs/ref-ayan,refs/ref-meiyi.bib,refs/ref-yong, refs/ref-yin, refs/ref-fanxinmengyu, refs/ref-panagou,refs/refs2,refs/KYR,refs/mybibfile, refs/ref-wenhao, refs/ref-sibin, refs/ref-homa}


\end{document}

%% file: macros.tex
\newif\ifdraft
\draftfalse

\usepackage{tikz}
\usetikzlibrary{arrows.meta, positioning, fit}
\usepackage{adjustbox}
\usepackage{booktabs}  
\usepackage{adjustbox} 
\usepackage{xspace}
\usepackage{comment}
\usepackage{multirow}
\usepackage{amsmath}
\usepackage{amsfonts}
\usepackage{subfig}
\usepackage[font={bf,it,small},labelsep=colon,justification=centering]{caption}
\usepackage[super]{nth}
\usepackage{xurl}
\usepackage{setspace}
\usepackage{titlesec}
\usepackage[normalem]{ulem}
\usepackage{algorithm}
\usepackage[hang,flushmargin]{footmisc} 
\usepackage{xcolor}
\usepackage{todonotes}
\usepackage{lettrine}
\usetikzlibrary{positioning,fit,calc}

\newcommand{\saurabh}[1]{\ifdraft{\textcolor{blue}{[Saurabh: #1]}}\fi}

\definecolor{darkgreen}{rgb}{0.0, 0.2, 0.13}

\newcommand{\lemmon}[1]{\ifdraft{\textcolor{blue}{[Lemmon: #1]}}\fi}

\newenvironment{sibin}{\par\color{blue}[Sibin:]}{\par}

\newcommand{\chop}[1]{}
\newcommand{\eg}{{\em e.g.},\xspace}

\newcommand{\etal}{{\em et al.},\xspace}

%% file: Sections/introduction.tex
\section{Introduction}

Resilience in cyber-physical systems (CPS) is the ability to maintain critical functions and safety in the face of adverse events like security attacks, environmental disruptions, unintended human-system interactions, and natural failures in the hardware, the software, or their interfaces. Resilience goes beyond the traditional notion of detection and recovery; it focuses not just on preventing failures but also on ensuring the system can withstand, recover from, and adapt to adverse events. We use the term {\em ``perturbations''} to include all the types of adverse events mentioned above.

In this article, we shed new light on the five key themes required to achieve CPS resilience, under practical real-world constraints. These constraints come from the fact that these systems often include many legacy elements, which cannot be ``upgraded'' for reliability or security. Further, CPS will have interactions with human users at different cognitive and application-specific skill levels; as such, it is impossible to instantiate practical systems for all types of interactions. Finally, there might exist policy and administrative restrictions on what is allowable for making CPS resilient. For example, different stakeholders may own different parts of a CPS --- think of any of a variety of large-scale CPS that surround us today, like energy distribution, transportation, and industrial supply chain. The different stakeholders may have only partially aligned incentives, and there are regulatory and competitive factors that prevent cooperation.

Under this set of real-world constraints, we posit that there are \textbf{five key themes} for CPS resilience. Individually, elements within each theme have been discussed in prior literature; what this article brings is a systematic way of tying each theme to the system goal of CPS resilience and the interplay among these aspects.

\begin{enumerate}

\item {\em Resilience is a system-wide property}. This implies that resilience is {\em not} an individual-component property; rather, resilience needs to be ensured through interactions among hardware-software components, as well as between such components and the human users of the system. On the flip side, a reduction in system resilience can happen due to vulnerable interactions. This shines the spotlight on the robustness of decision and component boundaries for ensuring system resilience.

\item {\em Data for learning in CPS is a challenge}. Data in any learning-enabled system is a key ingredient for its resilience. Distinctively, data in CPS is expensive to collect and expensive to label. Thus, for the foreseeable future, we expect to be in a regime of low availability of data: it will be hugely disbalanced, limited for failures, and barely available if at all for catastrophic failures. This clues us into solution directions, e.g., creating synthetic data and learning from near-failures. Also, data will have various modalities and will need to be integrated for integrated situational awareness, leading to resilience. 
     
\item {\em Verification, testing, and redundancy for CPS resilience}. This theme is well understood in the context of the resilience of any system. For CPS, this theme brings out some distinctive aspects. For one, timeliness must be part of the equation, as functionality must be provided within the required time bounds. Achieving high coverage in testing CPS is a challenge when there is close interaction between the cyber and the physical elements. A simple but useful imperative is to achieve high coverage for the combinations of the two kinds of elements.

\item {\em Role of recovery}. This is a complement to the above theme and focuses on mitigation and recovery. Resilience incorporates the ability to handle perturbations beyond the pre-specified, and therefore designed for, fault models. The key question here is how do we recover from such situations, rarely to a fully functional state. Rather, we need to understand and design for “just good enough” recovery, one that keeps the security- or safety-critical functionality of the system available.

\item {\em Role of the human}. Frequently mentioned in design documents and papers on CPS resilience, a gap often exists between how the human role is considered in the design and implemented in the deployment. A deep consideration of this factor opens up areas of inquiry. For example, the fact that human users will behave only in a partially rational manner must factor into the fault model. As another example, explainability and trustworthiness of the algorithms become crucial, with explainability being a mix of ante-hoc and post-hoc reasoning. 
  
\end{enumerate}

\begin{figure}
\centering
\includegraphics[width=0.80\textwidth,clip]{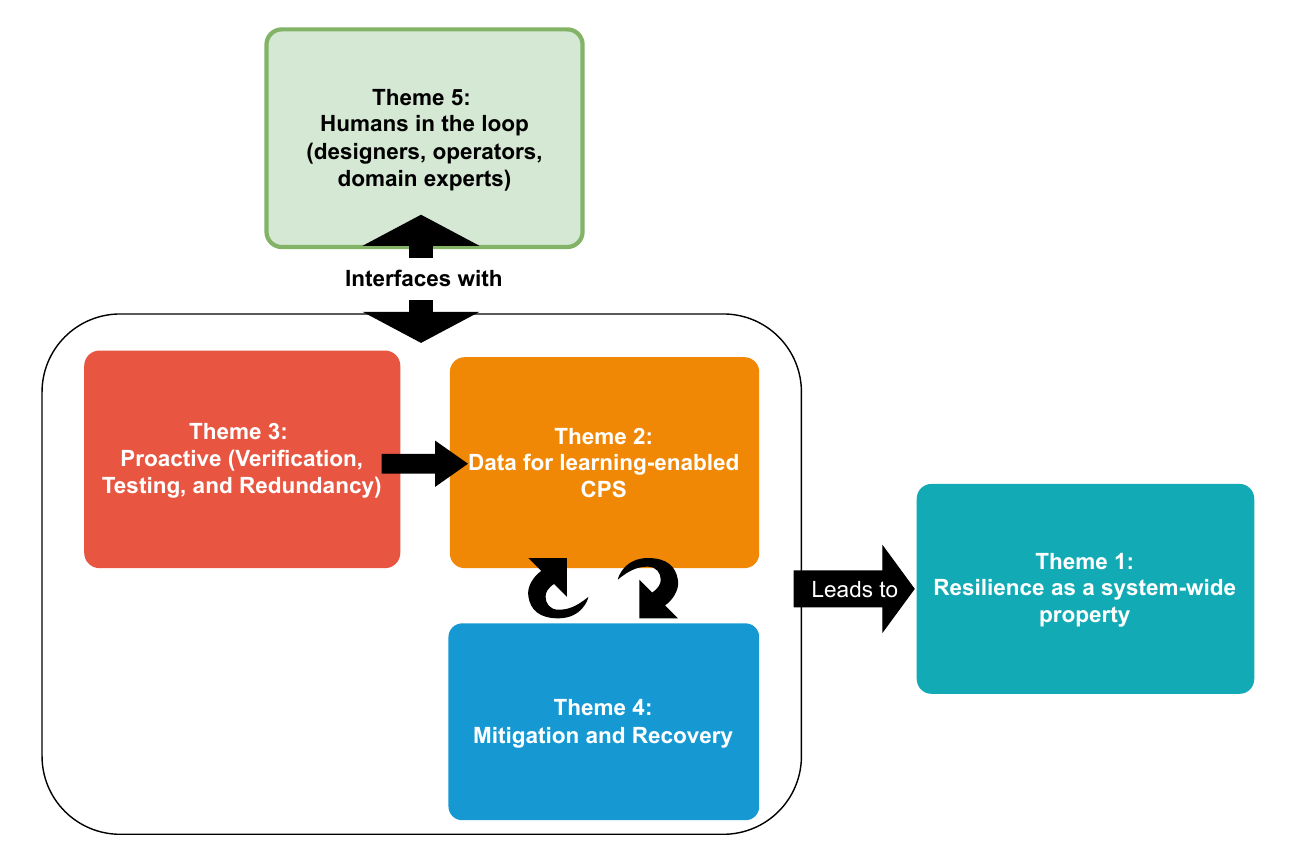}
\caption{High-level view of the interplay between the various themes developed in this article leading to resilient CPS 
}
\label{fig:overview}
\vspace{-6mm}
\end{figure}

A high-level overview of how the different themes interact among themselves is shown in Figure~\ref{fig:overview}. This shows that the proactive, learning-enabled CPS, and mitigation and recovery themes work in an integrated manner to lead to CPS resilience being a system-wide property, rather than a property for individual or sets of components. Humans interface with the overall system at different spatial and temporal granularities, through design, deployment, operation, and getting the results and providing feedback. 

\noindent {\bf Exemplars of CPS Resilience in the World}.

The positive examples of resilience in CPS often show up through the Sherlockian syndrome of ``the dog that did not bark.'' For example, the internet infrastructure in Japan proved to be largely resilient through the Great East Japan Earthquake of 2011. This was attributed to proactive provisioning of redundancy, such as through backup network lines, as well as to the rapid restoration effort by Japanese telecom companies like NTT. For a more recent case, and one with mixed examples, consider Ukraine’s power grid in the ongoing Russo-Ukrainian war. Initially, the power grid withstood the destruction surprisingly well, preventing large-scale blackouts. The nation had made dramatic improvements to the resilience of its power grid after a malware attack had severely disrupted the power grid in 2015. However, under more intense attacks in 2024 and 2025, which have been informed by previous defenses, the Ukrainian power grid has now been brought to its knees, operating at only about a third of its pre-invasion generation capacity.
This is caused by coordinated attacks against multiple parts of its infrastructure --- physical (natural gas, hydroelectric, and thermal power) and cyber (large-scale cyberattacks, coordinated with the physical attacks). 

Sometimes, a successful case of resilience is harder to spot due to the far more newsworthy headlines of the devastating impact of failures. A case in point is the ALERTCalifornia wildfire detection system, a sophisticated network of over 1,100 high-definition, pan-tilt-zoom cameras equipped with AI software. In the first two months of its deployment in 2023, the system had correctly identified 77 fires before any 911 calls came in. A notable success was the River Fire in San Diego County in August 2024, where the AI software classified the smoke plume as dangerous and alerted authorities nine minutes before the first public report. This early warning allowed for a rapid aerial and ground response that kept the fire to just 54 acres. This is a textbook example of the three constituent elements coming together to achieve CPS resilience --- cyber (the software, the networked aspect), physical (the cameras mounted on high places), and human (the personnel actually performing the response). However, the LA wildfire in January 2025 caused devastating loss of property (\$250B) and lives (31 confirmed deaths), and such failures of CPS resilience and infrastructure gathered far more public attention.

%% file: Sections/applications.tex
\section{Representative Application Domains}
\label{sec:applications}


Cyber-physical systems (CPS) are increasingly being used for "smart" infrastructure systems providing critical services to human society.  Examples are readily found in smart sewer systems \cite{montestruque2015globally}, medical CPS (MCPS) ~\cite{10.1145/1837274.1837463}, smart power grids \cite{yu2016smart} and smart highway systems \cite{7535338}.  For infrastructure CPS, resilience moves beyond concerns over component reliability and security to concerns over the system's capacity to serve the target parts of the society, which is almost always at a large scale. Next, we are going to discuss the CPS resilience challenges in two representative application domains --- Connected and Autonomous Transportation System (CATS) and Medical CPS. In the end, after the description of the five themes, we will bring back how these themes collectively largely address these challenges.  

\subsection{Connected and Autonomous Transportation Systems (CATS)}
\label{sec:application-cats}

The safety and reliability of autonomous vehicles can be significantly enhanced with \textit{connected autonomy}, in which driving decisions are made with the assistance of, or in cooperation with, other vehicles or infrastructure elements (road-side sensors, edge computing servers). This can reduce cost and improve safety by overcoming limitations of in-vehicle autonomy~\cite{rana2023connected}.
An autonomous vehicle has three  components: Perception, Planning, and Control. \textit{Perception} processes the output of depth perception sensors like LiDAR or stereo cameras to build a time-varying model of the physical environment around the vehicle, updated in real-time. 
\textit{Planning} uses this model to devise a strategy for reaching the  destination at three different timescales --- duration of the trip, seconds for nearby environment, and sub-second for immediate environment.
\textit{Control} takes this planned trajectory and translates it into actions for the vehicle's actuators, such as steering, acceleration, and braking (at timescales of a few milliseconds).

As of today, \textit{\textbf{all three components (perception, planning, and control) operate onboard the autonomous vehicle itself}}.
A connected autonomous transportation system (CATS) is a class of CPS in which some of these components reside in other vehicles, or in the infrastructure. This system can enhance the functionality and safety of autonomous vehicles in several ways.
For example, in \textit{\textbf{V2V Cooperative Perception}} each vehicle shares the processed data from its own sensors with other vehicles~\cite{zhang2021emp,qiu2022autocast}, increasing safety by overcoming sensor line-of-sight limitations. In  \textbf{\textit{Infrastructure-aided Perception}}, an array of roadside sensors~\cite{he2021vi, shi2022vips} can track objects, use an edge computing platform to compute a time-varying model of the environment, and deliver it wirelessly to each vehicle. Finally, in \textit{\textbf{Infrastructure-aided Planning}}, the edge computing platform can plan paths and trajectories on behalf of the vehicle, and deliver these wirelessly to each vehicle, where the on-board control component implements them.
Even in a highly connected CATS setting, each vehicle still requires a dependable local perception stack because wireless connectivity can be intermittent, delayed, or unavailable. Recent adaptive onboard perception systems, including 2D approaches such as LiteReconfig~\cite{litereconfig} and 3D approaches such as Agile3D~\cite{agile3d}, illustrate this need by dynamically reconfiguring perception pipelines under changing scene complexity, latency constraints, and resource contention. Thus, adaptive local perception complements V2V and infrastructure-assisted autonomy by providing robust fallback and graceful degradation when external support is unreliable.
However, the Control component \textit{cannot be offloaded to the infrastructure}, since control tasks, \eg steering, acceleration, and braking, operate   require millisecond-level response times that can only be achieved with an onboard control system.

This vision of CATS raises several challenges for reliability and security of CPS, some quite distinctive to this application domain. This vision relies on wireless networks, short range or long range, to be provisioned and operated dependably. It relies on trust at multiple end points, both static endpoints (such as roadside units or RSUs) and mobile endpoints (obviously, other vehicles). With respect to any vehicle, the connectivity graph is also dynamic highlighting another challenge for reliability and security. 

\subsection{Medical Cyber-Physical Systems (MCPS)}
\label{sec:application-mcps}
Advances in sensing, computing, networking, and machine learning technologies have resulted in the ubiquitous deployment of medical cyber-physical systems in various clinical and personalized settings, ranging from implantable and wearable devices such as pacemakers, physiological monitors, insulin pumps, and artificial pancreas systems (APS), to smart hospital devices such as infusion pumps, intensive care units (ICUs) monitors, and surgical robots, to image-guided surgery and radiation therapy systems. 

Within an emergency room or ICU, MCPS are essentially networked embedded systems that must reliably monitor patient vitals in a highly dynamic, uncertain, and unsecured environment.  However, with the aging US population, MCPS are increasingly being used to connect an outpatient's embedded medical devices to a centralized medical server.  The sum of all these interconnected outpatient devices and servers forms an enterprise-scale MCPS whose main function is to improve the overall health of the served populations.  Although component reliability and security are important issues, the overall effectiveness of the system now depends on the extent to which the MCPS improves general public health.  This shifts the mission from ensuring favorable individual health outcomes to ensuring the average outcome for all individuals in the community.  The success of that system-wide MCPS mission depends on the quality of the data used to train diagnostic models running on these embedded devices; data that are subject to abrupt distribution shifts triggered by internal and exogenous forces.  These distribution shifts in MCPS training data present enormous challenges to enterprise-scale MCPS resilience that can only be met if the system can detect and adaptively react to such shifts. 


Medical Cyber-Physical Systems (MCPS) \cite{lee2011challenges} used in remote patient monitoring (RPM) provide a compelling example of CPS's requiring system-wide resilience.  
RPM or Telemonitoring \cite{pare2007systematic} uses wearable digital devices to record a patient's
physiological function outside of a traditional hospital setting.  The data generated by these devices is transmitted to hospital servers.  Over the past decade \cite{varma2023remote}, this data has been used by data analytic models to notify the user and care provider regarding the need for follow-up care or critical medical interventions.
RPM is used for diabetes management \cite{salehi2020assessment}, post-operative cardiac care \cite{atilgan2021remote} , and end-of-life palliative care  \cite{padros2023smart}.  
In these settings, MCPS function is evaluated through metrics measuring its impact
on overall health outcomes and overall healthcare costs.  
These metrics are not a sole function of device reliability and accuracy, they also
hinge on the community-wide impact such systems have on healthcare outcomes and costs.   
In viewing MCPS as providing a community-wide service,  the resilience of such systems must embrace a range of  issues including device reliability, data integrity and security, and the robustness of data analytic models.

The growing use of personal devices (smartwatches, smart rings) for health monitoring opens a new aspect of MCPS resilience.  With the data collected from these personal devices, MCPS take on the additional role of monitoring and maintaining
community public health in a way that is resilient to inevitable shifts in community data.  
MCPS resilience to community shifts is challenging due to the lack of large labeled datasets.  One promising approach for addressing the lack of large labeled datasets is through the use
of optimal transport algorithms for transductive label augmentation.
Realizing this promise faces important technical issues regarding privacy, data imbalance, and distributed datasets.  Recent research on the role of OT in federated learning may provide an avenue for addressing these technical issues.

%% file: Sections/related.tex
\section{Relation to Prior Surveys}
\label{sec:related}

Resilience in cyber-physical systems (CPSs) has emerged as a critical area of research, driven by the increasing reliance on CPS infrastructures in safety-critical domains such as energy, transportation, and healthcare. In response, several surveys have reviewed techniques and frameworks for achieving resilience in CPS. However, many of these works adopt narrowly scoped perspectives. As a result, they often fail to capture the broader, system-level dimensions of resilience that are essential for real-world deployments. Key aspects such as human-system interaction, data scarcity, and operational adaptability remain underexplored in much of the existing literature.
Segovia-Ferreira \etal~\cite{segovia2024survey} conduct a focused study on control-theoretic methods and industrial control systems. Their work provides a structured taxonomy of resilient control techniques and cyber-resilience techniques. It also discusses various evaluation metrics. However, it pays limited attention to resilience challenges that arise from human-system interactions and data generation challenges.
Ratasich \etal~\cite{ratasich2019roadmap} examine resilience in IoT-based CPSs with a particular emphasis on fault detection, localization, and recovery in resource-constrained and real-time environments. Their survey addresses key challenges such as limited computation, strict latency requirements, and constrained communication, and illustrates their discussion with application case studies like resilient smart mobility. While this perspective is valuable within the IoT domain, it remains narrow in scope with respect to the broader CPS settings. The work does not extend to broader CPS settings involving heterogeneous components, nor does it consider system-wide concerns such as formal verification, adaptive recovery beyond predefined fault models, and learning-based approaches under uncertainty.
Cassottana \etal~\cite{cassottana2023resilience} take a quantitative approach by reviewing mathematical models and metrics for assessing CPS resilience. Their work offers a systematic framework to evaluate the resilience of a CPS, before and after a disruption. However, the survey lacks discussion on data-driven learning and the integration of human factors and cyber-physical system components that together shape resilience.
Kim \etal~\cite{kim2022survey}provide a comprehensive survey of network-layer security threats in CPSs. While they discuss resilience in the context of detection and mitigation techniques, the scope of resilience is limited to maintaining secure communication and preventing adversarial disruptions. Similarly, Yu \etal~\cite{yu2023survey} provide a structured overview of attack and defense methods in CPSs, yet their focus remains on security threats and detection techniques. Broader system-level concerns are not within the scope of their discussion.


To address these gaps, our survey provides a comprehensive view of CPS resilience by organizing the discussion around five interconnected themes: (1) system-wide resilience across components, (2) CPS learning challenged by scarce, imbalanced, and multimodal data, (3) CPS resilience via specification, coverage, and redundancy, (4) recovery for CPS resilience, and (5) human in CPS. To the best of our knowledge, no prior work offers a comprehensive and integrative view of resilience that addresses the unique constraints and real-world complexities of CPSs.

%% file: Sections/theme1-system.tex
\section{Theme 1: Resilience as System-wide Property}
\label{sec:theme1-system}

Resilience is fundamentally a system-wide property that reflects the ability of CPS to anticipate, adapt to, and recover from adverse conditions or disruptions such as deliberate attacks, accidental faults, human errors, or naturally occurring threats~\cite{NIST_SP_1500_202_2017,chen2022stl,hollnagel2006resilience}. 

\subsection{Systems-Theoretic Hazard Modeling and Analysis}
\begin{figure}
\centering
\includegraphics[width=0.70\textwidth,clip]{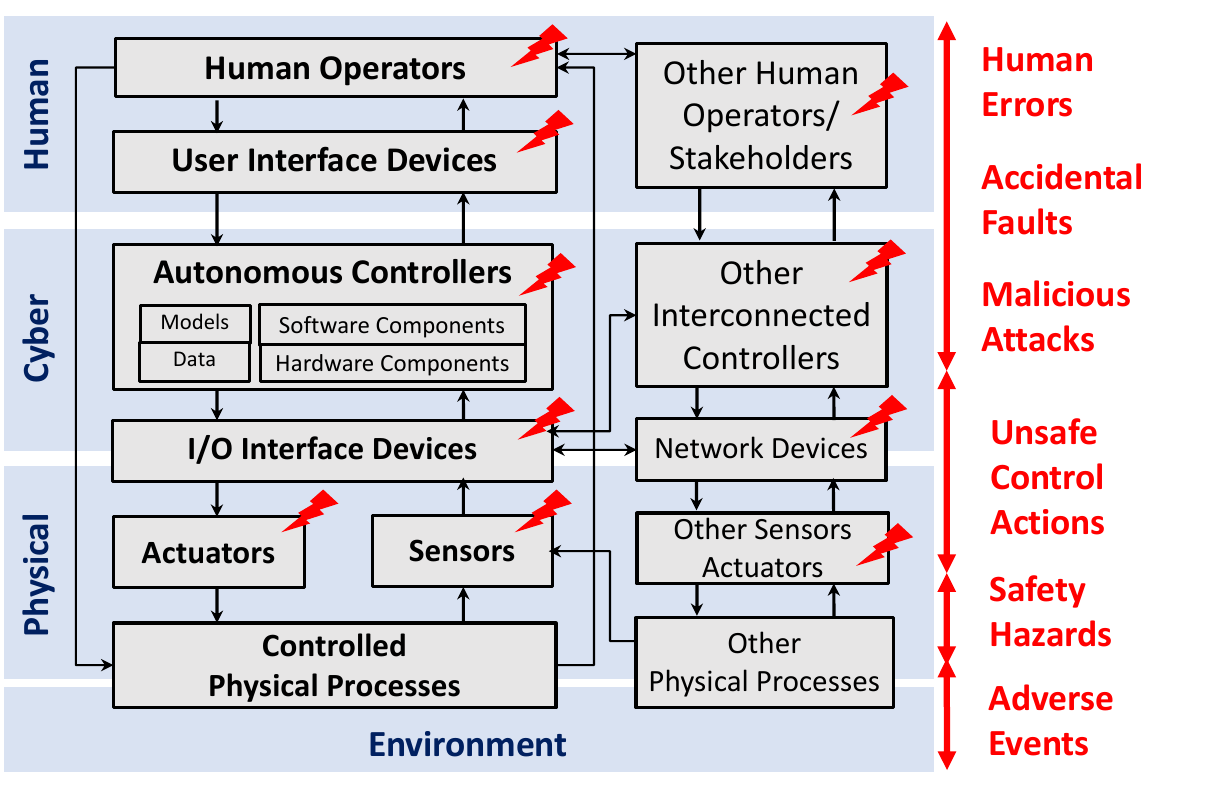}
\vspace{-2mm}
\caption{Hierarchical System Control Structures in Interconnected CPS: \\Accidental and Malicious Threats to System Resilience across the Human, Cyber, Physical and Environment layers}
\label{fig:CPS-System}
\vspace{-4mm}
\end{figure}

Systems and control-theoretic approaches to CPS model accidents and safety hazards as the result of context-dependent constraint violations across multiple levels of hierarchical CPS control structure~\cite{leveson2011engineering,leveson2004new} (see Figure \ref{fig:CPS-System}). These levels include humans and autonomous controller actions, cyber and physical system components, and the interactions among them. Accidental faults, malicious attacks, and unintended human errors targeting CPS can originate from different system components, including software, hardware, datasets and models for ML, network and interface devices, sensors, actuators, or other interconnected controllers and devices and get propagated through the cyber layer, corrupt cyber states, and cause erroneous controller actions. Then erroneous control actions when occurred under specific system contexts, defined by the combination of operational, environmental, and physical conditions, could lead to safety and security violations and result in accidents. 

\subsection{Assume-guarantee Contracts for System Resilience} Since resilience is a system-wide property, which needs to hold between levels of abstraction, between system components, and across temporal modes, a common structure underlying design-for-resilience takes the form of an assume-guarantee contract (AGC)~\cite{DBR21,Dab21,DRB22,DBR23}. 
An AGC consists of two parts. Firstly, there is an \emph{assumption} (which may be empty) that captures the input to the exchange under description, such as the output from a triggering component or action, the environmental conditions that must hold, or other system state parameters. Secondly, there is a \emph{guarantee} that describes the outputs of exchange under description, which are guaranteed to hold due to our compositional verification efforts. For example, a software subroutine might have an associated AGC stating the enabling parameters for it to be executed and the guaranteed result of executing the subroutine. Similarly, an AGC might describe the interaction between hardware and software components, such that a software component can trigger a hardware action, like moving a vehicle, or that a hardware component can trigger a software action, like a sensor value resulting in a change of plan. AGCs are organized hierarchically and represent a key building block for resilient CPS: at the highest level, we have an assumption that if the CPS becomes operational, that it will behave resiliently in that it upholds certain guarantees for operation until mission termination. Such a design is scalable and represents best-practices for resilient CPS design. For an example case study, in the realm of connected and autonomous transportation systems, the NASA Lunar Gateway Vehicle System Manager uses AGCs to frame system requirements (in English as well as in mathematical logic), conduct design-time formal verification and simulation, and create runtime monitors to support resilient behavior during system operation.

\subsection{System-Wide Recovery}
One way to ensure resilience in CPS is to ensure that the system remains \textit{safe} in the presence of failures and attacks. 
Hence, the system takes \textit{proactive measures} --- whether or not a fault or attack occurs.
This, coupled with analyses that identify \textit{when} to carry out the proactive actions, can significantly increase the safety of the system.
To further improve the resilience of the system, we should ensure that the \textit{(a)} analyses methods and the \textit{(b)} proactive measures are \textit{tamper-proof} by using hardware and software roots of trust.

One of the best proactive measures is the use of \textit{system-wide resets}~\cite{resecure,resecure-iot,groundhog} that are triggered independent of the existence/detection of faults or attacks.
Post reset, the system can be restored to a known safe/secure state --- by loading the system with trusted software, perhaps from a read-only memory (ROM) or a hardware root of trust.
This ensures that the system is reset to a known safe state and the attackers/faults have been cleared out of the system. 
Within this broad approach, there are important questions that need to answered: 
\begin{enumerate}
\item \textbf{When to reset?} As mentioned earlier, the system should reset independent of the existence or detection of faults or attacks. 
If we wait for the attacker or fault to be detected, it may be too late. Thus, if we reset the system too infrequently, it may lead to a significant increase in the risk of attacks and faults.
On the other hand, if we reset the system too often, it may lead to unnecessary downtime and a significant decrease in the quality of service.
\item \textbf{how to reset?} The reset mechanism should be tamper-proof so that attackers or faults cannot interfere with the process or even prevent the reset from happening.
\end{enumerate}

For the first issue, we can use the fact that CPS are often deterministic in nature and are dealing with physical systems that are subject to physical laws.
Hence, we can use/develop analyses techniques such as real-time reachability analysis~\cite{resecure} that can predict when a CPS is likely to get close to violating is safety constraints~\footnote{Note that we assume the main aim of an adversary is to push the system beyond its safety limits. Hence, the prevention of safety violations will also prevent many classes of attacks from succeeding.}.
Once we know how long it takes for the system to reach the safety boundaries, we can compute the time when the system should reset --- essentially \textit{before} it breaches the safety boundaries and with enough time left to reset the system, load the trusted software and push the system back into a safer state. 
These specifics of these processes will vary depending on the type of CPS but we have shown it to work for a wide variety of systems --- from avionices and building automation systems~\cite{resecure} to IoT systems~\cite{resecure-iot} and even distributed cryptography systems~\cite{groundhog}.

To decide on how to reset, again there are multiple options, depending on what has been established as a root of trust in the system.
For instance, \textit{hardware mechanisms} such as external timers and watchdogs can be used to trigger resets in a tamper-proof manner~\cite{resecure}.
Alternatively, specialized hardware features such as ARM Trustzone~\cite{arm-trustzone} can also be used to initiate resets~\cite{resecure-iot}.
If the operating system (OS) is considered to be trusted then it can also carry out this functionality.
In a ditributed setting, one of more of the nodes can be designated as a trusted node and it can initiate the reset process~\cite{groundhog}.

\subsection{Resilience through Obfuscation}
Another proactive way to improve the resilience of CPS as a whole is by \textit{obfuscating} the \textit{behavior} of the system --- for instance, by adding \textit{noise} to the execution patterns and timing of the system.
Obviously, indiscriminate addition of noise can be quite detrimental since that can severely degrade the performance and timing guarantees of the system.
Hence, we need to do this in a deliberate manner so that the system appears to be \textit{normal} to an external observer/attacker but, in reality, it is not.
%
This is where the notion of ``\textit{schedule indistinguishability}''~\cite{indistinguishability} comes into play.
We adapt ideas from the differential privacy world and add noise, in an algorithmically determined manner, to the execution schedules of tasks in the system so that an attacker/observer cannot distinguish between the real execution schedule and the obfuscated one.
This method has been shown to be effective against timing-based attacks in autonomous and other CPS, including streaming across the internet.

%% file: Sections/theme2-learning.tex
\section{Theme 2: Data for Learning-Enabled CPS}
\label{sec:theme2-learning}
\begin{wrapfigure}{r}{0.25\textwidth}
\vspace{-0.2in}
  \begin{center}
    \includegraphics[width=0.25\textwidth]{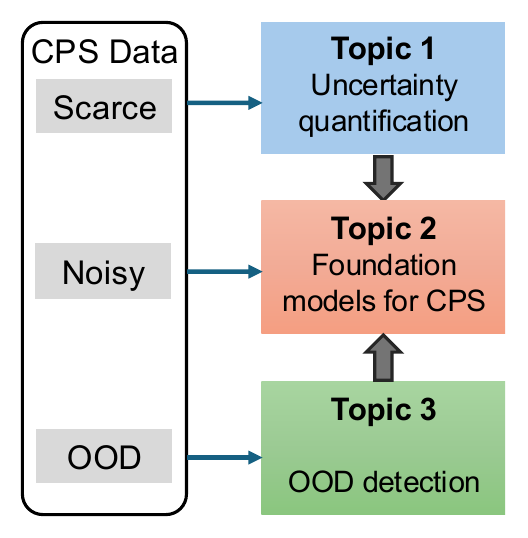}
  \end{center}
  \vspace{-0.2in}
  \caption{The overview of learning-enabled CPS.}\label{fig:theme2}
  \vspace{-0.2in}
\end{wrapfigure}

\textbf{Overview.} When CPS operate in dynamic and unseen environments, their sensory data may be noisy, scarce, or distributionally shifted. In addition, the collected data sometimes is irregularly sampled due to variations in sensor types. Under such scenarios, it is hard to ensure the safety, reliability, and generalizability of learning-enabled CPS. Resilience of learning-enabled CPS refers to systems that can handle uncertainty, identify anomalies, and adapt to unseen circumstances, even where data is noisy, scarce, or out-of-distribution (see Figure \ref{fig:theme2}). The goal of this theme is to present the significant challenges and existing solutions for learning-enabled CPS operating in such environments.

\subsection{Learning in Data-Scarce Environments}

Across a range of critical domains, including the two exemplar domains of medical CPS and connected and autonomous transportation, a crucial need is to develop autonomy in changing, possibly hostile conditions where there no capability to learn, train, or model the environment prior to mission commencement. This challenge meets three questions:
\begin{enumerate}
\item[(i)] \textit{What to learn?} In order to quantify the usefulness of agent's actions --- including motion, sensing, and communication --- at reducing the uncertainty about its operating environment, we need to introduce a metric of \textit{mission-oriented value of information}~\cite{Foretal22, Shaetal23, Putetal25}. This metric must naturally depend on the variance in mission success across all possible ground truths consistent with available data, as well as the expected decrease in this variance after a single additional observation.

\item[(ii)] \textit{How to learn?} Given a set of observations, the \textit{learning} element of a standard learn-and-control loop requires an agent to produce -- implicitly or explicitly -- an environment model that best fits the given observations. Many current learning approaches rely on function approximation methods~\cite{Fujetal18,Ornetal19,Adcetal21}, with samples provided at each time step during agent's mission, possibly through multiple episodes. However, in a setting of sparse data, the only available knowledge might be outdated or otherwise deficient~\cite{Yuetal19,Dixetal20,OrnTop21,Putetal24d}. 
Recent semi-supervised learning methods also show that, even under sparse supervision, learning can be improved by leveraging unlabeled data to regularize intermediate feature representations rather than relying only on output-level pseudo-label consistency. For example, SWSEG optimizes both alignment and uniformity of backbone features via a Sliced-Wasserstein objective, yielding stronger segmentation performance in low-label regimes~\cite{SWSEG}.
Thus, research needs to establish meaningful approaches for \textit{varying importance} of available samples, ensuring that available data is maximally used, but not afforded undue trust.

\item[(iii)] \textit{How to act?} The agent's decisions are a consequence of two goals. In the short term, the agent seeks to collect data that enable it to understand the ever-changing environment. In the long term, the agent uses this understanding to move towards completing its mission. Such a mission may be complex and not described in terms of time-invariant rewards, but possibly expressible through the formal framework of temporal logic, including both objectives to reach and safety constraints. A natural approach to designing an efficient policy that completes the agent's mission is one of \textit{exploration vs. exploitation}~\cite{Ornetal18,ThaOrn22}. Namely, after converting the agent's task into a time-varying reward function --- equivalently, a time-invariant reward function on a product space of the agent's state space and an automaton that describes the task --- it makes sense to provide \textit{bonus rewards} for actions which reduce the agent's uncertainty about the system, using the value of information described above. When the uncertainty is large, the agent will thus choose actions that reduce it; on the other hand, when the agent has a good understanding of the environment, it will chose actions that drive it towards mission completion. Unlike standard methods of planning in learned environments which make no use of uncertainty quantification and rather often assume perfect learning from a perfect model, it is critical to \textit{certifiably bound} the unavoidable suboptimality of the agent's plans~\cite{Aueetal08,ShaOrn22,PadOrn25}. Consequently, combining the uncertainty-based performance estimation with the active learning and planning may produce \textit{explicit overall bounds on the agent's performance in a data-sparse environment}.
\end{enumerate}

Current state-of-the-art learning methods for control synthesis are often meaningless in data-sparse settings. There is no opportunity to train prior to the mission execution, and each sample collected during the mission might be collected from a different underlying distribution~\cite{Woletal12,Thaetal22,Zhaetal24,Putetal25b}. Understanding that the planner may not be able to exactly learn the correct environment from the small amount of available data, it makes sense to instead identify and exploit invariant features of the environment across space and time to propose notions of (i) mission-relevant uncertainty, quantifying the variance in mission performance across all possible environments supported by obtained data and (ii) sensitivity to data sparsity, describing the trade-off between mission performance and increased need for data. In turn, these metrics are crucial in developing uncertainty-aware plans which enable agents to proceed with their objective while continually learning about the operating environment.

\subsection{Resilience to Uncertainties}
Quantifying prediction uncertainty is fundamental to building resilient learning-enabled CPS. Two primary types of uncertainty are commonly considered: aleatory uncertainty and epistemic uncertainty. Aleatory uncertainty, or statistical uncertainty, refers to the inherent randomness or variability in the system. It is generally considered irreducible and arises from intrinsic stochasticity in the environment or sensor noise. For example, random fluctuations in vehicle sensor readings caused by electronic interference. In contrast, epistemic uncertainty, also known as model or knowledge uncertainty, stems from incomplete observations, limited training data, or inadequate models. Unlike aleatory uncertainty, epistemic uncertainty is reducible through additional data collection, model refinement, or improved domain understanding. Examples include uncertainty about a patient's physiological state due to missing sensor inputs, or about potential emergent pathologies stemming from limited knowledge of disease progression.

When data is sparse, one approach is to modeling epistemic uncertainty by determining \textit{what possible environments are supported by the collected data and prior knowledge}. 
One option is thus to extend the commonly used notion of \textit{maximum likelihood}, to inference across unknown stochastic processes~\cite{Joh88,Cheetal08,Benetal20, Putetal24}. However, while classical work on maximum likelihood only identifies the model that generates the highest probability of observed data, recognizing that determining a single exact environment model is unlikely, it makes sense to generate a distribution of likelihoods over a set of all possible environmental models. Having defined a likelihood-based probability distribution, one possible path of integrating mission-awareness is through the notion of \textit{regret}~\cite{Ahmetal13,FosSim20}; namely, considering the sensitivity of performance of a given control law to model mismatch across likely models.
Equipped with the mission-aware uncertainty metric, a natural strategy~\cite{Pazetal16,ThaOrn20,NawOrn20} to combine learning and planning is to mix exploration --- enticing the agent to take actions which result in high value of mission-aware information --- with exploitation --- driving the agent to take actions which seem to progress towards mission completion given the current environment model. 

\subsection{Adapting Pre-trained Foundation Models for Learning-enabled CPS}

Foundation models, such as multimodal large language models (MLLMs)~\cite{wu2023multimodal} and vision-language-action (VLA) models~\cite{zitkovich2023rt2}, are capable of integrating signals across multiple modalities and controlling physical systems through those multimodal inputs. A key advantage of foundation models lies in their ability to generalize to downstream CPS tasks with limited data, often via zero-shot or few-shot learning~\cite{brown2020language,wei2022finetuned}. Further, recent advances have introduced lightweight variants deployable on edge devices~\cite{yao2024minicpm}, as well as models capable of incorporating domain-specific knowledge~\cite{kimura2024case,kimura2024vibrofm}. As such, these foundation models hold great promise for enabling a wide range of learning-enabled CPS applications. 

\noindent{\em Adapting pre-trained foundation models for CPS applications}. A promising direction is to adapt existing, pre-trained foundation models to CPS applications. However, beyond the need for computational efficiency, this adaptation, presents two major challenges. \textit{First}, CPS operating in dynamic environments often encounters noisy, sparse, and out-of-distribution (OOD) data. Such discrepancy between training and inference data can significantly degrade model performance and reliability. \textit{Second}, sensor characteristics during CPS deployment may differ from those used during pre-training in terms of sensing range, resolution, sampling rate, or even modality. This mismatch introduces a fundamental domain gap that can prevent foundation models from fully leveraging available data. Addressing these challenges requires both theoretical analysis to characterize the conditions under which such adaptation is feasible, and practical algorithms that enable effective and robust adaptation in real-world environments.

Learning with noisy, sparse, and out-of-distribution data presents a long-standing challenge in machine learning~\cite{wang2020generalizing}. Prior approaches for adapting foundation models with limited data include learning from examples in the context prompt (in-context learning)~\cite{brown2020language}, constructing simple classifiers based on the pretrained representation~\cite{caron2021emerging}, learning lightweight adapters~\cite{hu2021lora}, or prepending learnable input tokens (\eg prompts)~\cite{jia2022visual}. An emerging solution, related to meta learning~\cite{hospedales2021meta}, involves finetuning a pretrained model on multiple auxiliary tasks pertaining to the target task~\cite{sanh2022multitask,muennighoff2023crosslingual}.
Our prior work investigated the theoretical justification of this multitask finetuning~\cite{xu2024towards}. In this setting, a pre-trained model is further fine-tuned with a set of relevant tasks before adapting to a target task. Each of these auxiliary tasks might have a small number of labeled samples, and categories of these samples might not overlap with those on the target task. Our analysis builds on a key intuition is that a sufficiently diverse set of relevant tasks can capture similar latent characteristics as the target task, thereby producing meaningful representation and reducing errors in the target task. Our theorem, further confirmed by empirical results, reveals that with limited labeled data from diverse tasks, finetuning can improve the prediction performance on a downstream task.
Despite these advances, adapting pre-trained foundation models with limited data remains an open challenge. 


Adapting pre-trained foundation models to new tasks involving different sensor data types (e.g., varying resolution) or incorporating additional sensing modalities presents an open problem. A promising direction is adapter-based methods, such as low-rank adaptation (LoRA)~\cite{hu2021lora}, which introduce parameter-efficient modules into a pre-trained model without modifying its architecture or core weights. Each adapter can be viewed as a functional ``patch'' to the foundation model, conceptually analogous to a software update. Such adapters have been widely used to customize large pre-trained language and image generation models, enabling flexible control over content and style without retraining the entire model and only using a small amount of data~\cite{hu2021lora,kumari2023multi}.
Our recent work~\cite{liu2025pave} demonstrated an early version of this idea in the context of video foundation models, introducing lightweight adapters for downstream tasks with additional input modalities, including audio, 3D cues, multi-view inputs. By adding only a small number of parameters and operations to the base model, and leaving its original architecture and weights untouched, the adapted model can support a range of complex tasks such as audio-visual question answering, 3D reasoning, and multi-view video recognition, with significant performance gains.
While adapter-based methods can bridge modest modality mismatches, more substantial sensing heterogeneity may require domain-specific architectural design. 


Effective adaptation and efficient deployment of pre-trained foundation models for CPS remains an open research challenge, pointing towards several promising avenues for future exploration. One compelling direction is to advance the theory and practice of data-efficient adaptation: can models be adapted to handle noisy and OOD data using as few samples as possible? Additionally, can synthetic data be leveraged to support this adaptation and reduce reliance on real-world data? Another important direction lies in compositionality: can multiple task-specific adapters be integrated into a unified model to enable improved generalization across a broad range of compositional tasks? Progress in these areas could significantly enhance the applicability and robustness of foundation models in learning-enabled CPS.

\subsection{Building customized foundation models for CPS applications}
While adapting existing pre-trained foundation models allows leveraging investments in prior model training, it is not always the best approach. Vanilla foundation models are not specialized. They embody large amounts of general knowledge that contributes to model bloat without necessarily being directly applicable to the domain at hand. An alternative solution is therefore to develop domain-specific foundation models from scratch that act as specialists in the given (CPS) application domain. Such models, being limited to a narrower domain, would be smaller and more efficient, sometimes called {\em micro foundation models\/}~\cite{kimura2024case,kimura2024vibrofm}. Developing foundation models for CPS applications introduces a slew of additional challenges~\cite{baris2025foundation}. Examples of specialized domains include climate and sustainability data analysis applications, energy optimization in data centers, network security analysis, and maintenance/diagnostics in complex systems, among others. Specialized foundation models can help recognize complex patterns in domain-specific time-series data to  advanced diagnostic, reasoning, and prediction capabilities. 

Models for CPS applications should be able to efficiently handle: (i) multimodal data of arbitrary modalities (such as data collected from domain specific sensors), (ii) time-series data with frequency domain signatures that encode complex recurrence patterns, (iii) spatial reasoning from multi-vantage data (such as data from multiple observation points), and (iv) large sensor and environment configuration space, as is common in CPS applications.
While some of these challenges also arise when adapting pre-trained models, developing models from scratch requires natively supporting these capabilities. This introduces new research questions in model design, training, and evaluation, especially under the typical constraints of limited labeled data and diverse deployment scenarios common in CPS applications. 


The state of the art must be advanced in several respects. First, much of today’s work on foundation models focuses on data modalities common to human communication and perception, such as vision, text, and audio. The diversity of envisioned intelligent CPS applications may employ, possibly proprietary, multimodal sensor data where sources are substantially more heterogeneous, calling for learning and inference mechanisms suitable for arbitrary multimodal time-series data. Also, many applications, such as those featuring environmental data, residential energy consumption data, data center temperature measurements, and similar sensor-based sources, will feature measurements of some external physical environment. Often the underlying physics have clear signatures in the frequency domain, making Fourier transforms (or spectrograms) a preferred input type. Pre-training must be cognizant of properties of such transforms in order to make learning more efficient. For example, the design of data augmentation pipelines (e.g., in such training components as contrastive learning) must be physically-informed to offer additional structure that can be leveraged in the augmentation design. Additionally, such applications will often use data sources at multiple locations. For example, weather data features deployments of large numbers of sensors (observing their physical environment from multiple vantage points). Thus, foundation models must be developed that can reason spatially about multi-vantage data sources to compute holistic representations of the observed physical phenomena from the multitude of individual multimodal and multi-vantage data streams. The FMs must support heterogeneity not only in sensing modalities but also in individual sensor properties such as calibration, measurement resolution, and sampling rates, and should neither require per-device re-training nor cause parameter explosion due to the large source (e.g., sensing device) configuration space. Moreover, the mix of sources used may differ substantially from installation to installation (e.g., from one data center to another in data center diagnostic applications) and may evolve over time. The foundation models should be easily customizable to their deployment environment. They should easily support upgrades and remain robustly operational in the presence of individual device/source failures, processing workflow changes, and disconnections.  

While normally one might rely on AI scaling laws to endow models with better capabilities, in the case of domain-specific CPS models, training data are usually a bottleneck. Thus, rather than relying on model size (and thus larger amounts of training data) one needs to rethink the training pipeline to learn more efficiently from limited amounts of data by injecting inductive biases inspired by the application domain.  
Recent work on domain-specific foundation models~\cite{kimura2024case,kimura2024vibrofm} has espoused this approach to help address the above challenges, improving the efficacy of multimodal self-supervised pre-training~\cite{kara2024phymask,kara2024freqmae,liu2024focal} and domain-specific data augmentation~\cite{wang2025dynagen,wang2024data,wang2023sudokusens,yao2018sensegan,liu2021contrastive,mahadik2016sarvavid} to increase both training performance and data size. 

\subsection{Resilience to Out-of-distribution (OOD) Detection}
OOD detection~\cite{mohammadi2020anomaly,ramakrishna2022efficient,kaur2023codit} is critical for ensuring the reliability and resilience of learning-enabled cyber-physical systems (CPS). In general, existing research on OOD detection in CPS can be categorized into two types: single-modal and multi-modal OOD detection. 

(i) \textit{Single-modal OOD detection.} This line of work primarily focuses on detecting OOD samples using data from a single sensor modality, such as time-series, image, or LiDAR data. For instance, Ramakrishna et al.~\cite{ramakrishna2022efficient} adopted a $\beta$-VAE to learn disentangled representations from images for detecting OOD frames. More recently, Kaur et al.~\cite{kaur2023codit} proposed a conformal anomaly detection framework for OOD detection in CPS time-series data, based on deviations from in-distribution temporal equivariance. Despite significant achievements~\cite{ramakrishna2022efficient, kaur2023codit}, single-modal approaches have been found to be susceptible to missed detections when that modality is affected by environmental factors such as fog or rain. 

(ii) \textit{Multi-modal OOD detection.}  Several studies~\cite{sun2020real, wang2021radar, duonggeneral2024} have leveraged multi-modal data to enhance OOD detection accuracy in CPS equipped with multiple sensors. For example, Wang \etal~\cite{wang2021radar} developed a multi-modal transformer network that fuses radar and LiDAR data to detect radar ghost targets in autonomous driving. Qiu \etal~\cite{qiu2022unsupervised} introduced an unsupervised contrastive learning method that employs generative adversarial networks to extract latent features from five different sensor modalities for accurate abnormal driving segment detection. While these approaches achieve strong performance in OOD detection, they generally lack interpretability, making it difficult to explain prediction results. To address this limitation, recent works~\cite{zhang2024holmes,zhang2024gpt} have incorporated LLMs to enhance the identification and reasoning of OOD samples. For instance, Sinha \etal~\cite{sinha2024real} proposed a hybrid framework that combines a fast VAE-based detector with a slow LLM-based reasoner for real-time anomaly detection and reasoning in CPS. Despite their significant strides, LLM-based OOD detection still faces key challenges. First, the system must determine when to trigger the slower reasoning module in place of the faster detector. Second, the latency introduced during the reasoning process remains a critical concern for real-time CPS applications. Third, it still remains a challenge to reduce the false positives or false negatives for OOD detection.

%% file: Sections/theme3-proactive.tex
\section{Theme 3: Verification, Testing, and Redundancy for CPS Resilience}
\label{sec:theme3-proactive}

Ensuring the resilience of safety-critical CPS over their lifecycle requires the ability to specify and analyze requirements that capture reliable, high-confidence, provable behaviors. We therefore use temporal logics to express requirements in a precise, unambiguous form that enables automated analysis, including formal verification and synthesis.
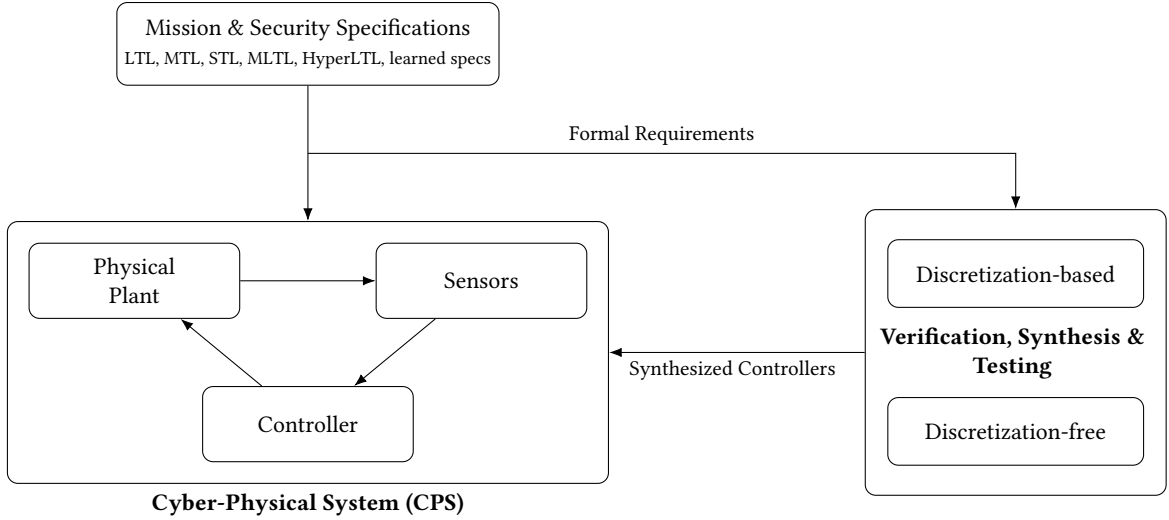
\begin{figure}[t]
    \centering
    \begin{tikzpicture}[
        >=Latex,
        block/.style={draw, rounded corners, align=center, minimum width=2.8cm, minimum height=1.0cm},
        bigblock/.style={draw, rounded corners, align=center, minimum width=4.0cm, minimum height=1.1cm},
        smallblock/.style={draw, rounded corners, align=center, minimum width=3.4cm, minimum height=0.9cm},
        lbl/.style={font=\small},
        node distance=1.2cm and 1.8cm
    ]

\node[block] (plant) {Physical\\Plant};
\node[block, right=1.8cm of plant] (sensors) {Sensors};
\node[block, below=1.4cm of $(plant)!0.5!(sensors)$] (controller) {Controller};

    \draw[->] (plant) -- node[lbl, above] {} (sensors);
    \draw[->] (sensors) -- node[lbl, right] {} (controller);
    \draw[->] (controller) -- node[lbl, left] {} (plant);

    \node[draw, rounded corners, fit=(plant) (sensors) (controller), inner sep=8pt] (cps) {};
    \node[below=3.5cm of cps.north, font=\bfseries] {Cyber-Physical System (CPS)};

    \node[bigblock, above=1.8cm of cps.north] (specs) {Mission \& Security Specifications\\
        \footnotesize LTL, MTL, STL, MLTL, HyperLTL, learned specs};

    \node[bigblock, right=3.4cm of cps.east, anchor=west, minimum height=3.8cm] (verifbox) {
        \begin{minipage}{3.5cm}\centering
        \textbf{Verification, Synthesis \& Testing}
        \end{minipage}
    };

    \node[smallblock, anchor=north, yshift=-0.4cm] (disc) at (verifbox.north) 
        {Discretization-based};

    \node[smallblock, anchor=south, yshift=0.4cm] (cont) at (verifbox.south) 
        {Discretization-free};

\draw[->] (specs.south) --
    node[lbl, left] {} 
    coordinate[pos=0.5] (midreq)
    (cps.north);

\draw[->] (midreq) --
    node[lbl, above, pos=0.5] {Formal Requirements}
    ($(verifbox|-midreq)$) 
    -- (verifbox.north);

    \draw[->] (verifbox.west) |- 
        node[lbl, below, xshift=-50pt] {Synthesized Controllers} 
        (cps.east);

    \end{tikzpicture}
    \caption{Theme 3 overview: formal specifications drive verification, synthesis, and testing, which in turn yield correct-by-construction controllers for the CPS.}
    \label{fig:theme3-overview}
\end{figure}

\subsection{Specifications of Resilience, Mission, and Security Properties}

\paragraph{Mission Requirements:}
Without clear mission and resilience requirements, it is difficult, if not impossible, to determine what resilience mechanisms must be designed. Temporal logics have proven effective for specifying behaviors precisely and unambiguously, and can be extended to address resilience challenges.

Linear Temporal Logic (LTL)~\cite{katoen08} is a convenient and expressive formalism for specifying properties over infinite executions (traces) of a system; a variant for finite traces has also been proposed~\cite{DeGia13}. 
The set of LTL formulas over a finite set of atomic propositions $\mathcal{AP}$ is defined by the grammar:
\[
\phi ::= a \in \mathcal{AP} \mid \neg \phi \mid \phi \lor \phi \mid \nextt \phi \mid \phi \until \phi.
\]
Here, $\neg$ and $\lor$ denote logical negation and disjunction, while $\nextt$ and $\until$ are temporal operators representing \texttt{next} (in the next discrete step) and \texttt{until} (the left property holds until the right becomes true), respectively. For convenience, additional logical and temporal operators are derived from this core syntax:
\[
\begin{aligned}
&\texttt{true} \triangleq a \lor \neg a, \quad \texttt{false} \triangleq \neg \texttt{true}, \quad
\varphi \wedge \psi \triangleq \neg (\neg \varphi \lor \neg \psi), \quad 
\varphi \rightarrow \psi \triangleq \neg \varphi \lor \psi, \quad \eventually \varphi \triangleq \texttt{false} \until \varphi, \quad 
\always \varphi \triangleq \neg \eventually \neg \varphi.
\end{aligned}
\]
Here, $\wedge$ and $\rightarrow$ denote conjunction and implication, while $\eventually$ and $\always$ represent the temporal operators \texttt{eventually} (at some point in the future) and \texttt{always} (at all times), respectively.

The semantics of LTL are defined inductively; see~\cite{katoen08} for details. LTL enables rigorous specification of system behaviors. For example, a safety property can be expressed as $\always \neg \phi$, asserting that a bad condition $\phi$ never occurs, while a reachability property $\eventually \phi$ requires that a desirable condition $\phi$ eventually holds. For an infinite trace $r \in (2^{\mathcal{AP}})^\omega$, we write $r \models \varphi$ if $r$ satisfies the LTL formula $\varphi$. The set of infinite traces satisfying an LTL formula can be accepted by either a non-deterministic B\"{u}chi automaton or a deterministic Rabin automaton~\cite{katoen08}.

Beyond LTL, several temporal logics refine expressiveness for CPS. Metric Temporal Logic (MTL)~\cite{AH90} augments temporal operators with time intervals and can be given continuous or pointwise semantics over finite or infinite runs. A variety of interval types is possible (open/closed, bounded/unbounded, integer/real endpoints); see \cite{OW08} for a survey. Signal Temporal Logic (STL) further extends MTL with real-valued predicates over real-time intervals~\cite{MN04,Ale13}, yielding a highly expressive language for detailed properties of real-time CPS.

However, greater expressiveness often comes at the cost of analyzability: satisfiability and model checking for general MTL are undecidable, and STL faces similar complexity challenges. In practice, one typically restricts to subsets of STL or MTL for real-time checks. Mission-time Linear Temporal Logic (MLTL), developed by NASA as a CPS specification logic~\cite{RRS14}, strikes a practical balance between expressiveness and computational complexity, enabling tractable formal verification of complex temporal behaviors. Aerospace operational concepts often describe requirements using integer-labeled timelines (e.g., clock ticks, radar sweeps, or Hertz) as in NASA's Automated Airspace Concept~\cite{EH10}, the U.S.~Navy's Aircraft Carrier Deck Scheduler~\cite{RCRBS11}, and the JAXA--NASA GPM Observatory~\cite{GPM}. MLTL efficiently captures such requirements, supports formal analysis, remains readable for certification authorities~\cite{FMRSWY21b}, and is efficiently monitorable in real time~\cite{LVR19}.

\noindent \textit{Security Requirements:}
LTL, MTL, STL, and MLTL formulas are typically used to specify safety and functional correctness requirements, and they evaluate whether a \emph{single} infinite trace satisfies a property. Security properties, by contrast, often depend on relationships among multiple traces of the system.

Alur \etal~\cite{ACZ06} showed that the modal $\mu$-calculus is insufficient to capture all opacity policies. To address such limitations, Clarkson and Schneider~\cite{Clark10} introduced \emph{hyperproperties}, specified using second-order logic over \emph{sets of execution traces}. Hyperproperties generalize linear-time properties~\cite{katoen08} by moving from sets of runs to \emph{sets of sets of runs}.

HyperLTL extends LTL with trace quantifiers, enabling simultaneous reasoning about multiple traces:
\begin{eqnarray*}
\psi ::= \exists \pi.\, \psi \mid \forall \pi.\, \psi \mid \phi, \quad \phi ::= a_\pi \mid \neg \phi \mid \phi \lor \phi \mid
\nextt \phi \mid \phi \until \phi.
\end{eqnarray*}
The quantifiers $\exists \pi$ and $\forall \pi$ denote ``there exists a trace $\pi$'' and ``for all traces $\pi$'', respectively. Formulas $\phi$ retain the usual LTL structure, except that atomic propositions are annotated with trace variables: for each $a \in \mathcal{AP}$ and trace variable $\pi$, $a_\pi$ refers to the valuation of $a$ along trace $\pi$. A HyperLTL formula with no free trace variables is called \emph{closed}.

HyperLTL can express a wide range of security properties, including opacity. For example, the following formula expresses language-based opacity, assuming that LTL formulas $\varsigma$ and $\varphi$ describe sets of traces that reveal and do not reveal secrets, respectively:
\[
\forall \pi\, \exists \pi'.\;  L(\pi) \models \varsigma \rightarrow (H(\pi) = H(\pi') \wedge L(\pi') \models \varphi).
\]
This states that for every trace $\pi$ satisfying the secret-revealing condition $\varsigma$, there exists a trace $\pi'$ that is observationally indistinguishable from $\pi$ (i.e., $H(\pi) = H(\pi')$) but satisfies the non-secret condition $\varphi$, thereby preserving opacity. These mission and security logics provide the formal backbone for the verification and testing techniques described in this theme.

\subsection{Learning Mission and Security Requirements}

Robustness and resilience in CPS must ultimately be defined in terms of concrete mathematical constraints, but specifying such constraints is challenging for typical end users. Although LTL formulas or state-space safe sets~\cite{DBLP:journals/ijrr/ChouBO21} can formally encode high- and low-level behavior requirements, they are difficult for untrained users to construct. This is problematic for CPS domains---from home robotics to urban air mobility---where non-experts interact directly with the system. Incorrect specifications can cause verification procedures to confirm the wrong properties, leading to violations of the intended behavior and potentially introducing adversarial vulnerabilities. This motivates rigorous specification checking and principled uncertainty quantification.

Recent work addresses this challenge by inferring specifications from natural language commands~\cite{DBLP:conf/icra/PanCB23} and from human demonstration trajectory data~\cite{DBLP:journals/ijrr/ChouBO21}, which are much easier for non-expert users to provide than formal constraints. These approaches use large language models to translate natural language into symbolic constraints in LTL, and exploit the Karush--Kuhn--Tucker optimality conditions of the demonstrator's control problem to identify specifications from limited human data.

Another direction is to learn \emph{neuro-symbolic} specifications that combine traditional symbolic variables and operators with neural functions and predicates. This allows one to express difficult-to-formalize concepts, such as the presence of a pedestrian in an image. For instance, the Neuro-Symbolic Assertion Language (NeSaL)~\cite{xie_neuro-symbolic_2022} specifies properties for image-based verification and approximation, supporting assurance of closed-loop CPS~\cite{geng_bridging_2024}. Neural operators can also learn real-world perturbations to images and enter into properties describing robustness to distribution shifts~\cite{wu_toward_2023}. 

Because specification inference is inherently ill-posed---multiple plausible specifications may explain the same data---it is crucial to quantify uncertainty in the learned specifications. Recent methods approximate the set of all constraints consistent with the observed data and then use this set to guide conservative downstream planning and control~\cite{DBLP:conf/corl/ChouBO20}, synthesizing controllers that satisfy all candidate safety constraints.

Accurate specification learning directly supports verification and testing by ensuring that resilience analyses are grounded in the \emph{intended} behavioral and safety requirements. Data-driven and neuro-symbolic specification learning thereby complements downstream verification methods, jointly advancing the provable robustness of CPS.

\subsection{Modeling Complex Sensing and Perception Modules}

Resilient CPS rely on models at varying levels of granularity to design components, test diverse scenarios, and monitor or predict uncertain environments online. Resilience testing, in particular, requires generating unexpected yet realistic inputs. At the same time, CPS resilience is often undermined by unreliable sensing and perception, so accurate, tractable models of sensing and perception errors are essential.

As sensor data become increasingly rich, it is difficult to construct realistic models against which to design, verify, test, or monitor CPS resilience. This is especially true for vision sensors and neural-network-based perception~\cite{mitra_formal_2025}, which operate in high-dimensional pixel spaces lacking realistic symbolic models of pixel evolution. Recent work introduces \emph{verifiable probabilistic abstractions} of sensing and perception, ranging from simple statistical intervals~\cite{pasareanu_closed-loop_2023}, to conservative inflations~\cite{cleaveland_conservative_2025}, and contraction methods for downstream resilient control~\cite{DBLP:conf/wafr/ChouOB22}. These are parametric perceptual models learned from data together with confidence intervals.

When combined with models of dynamics and control, abstractions of sensing and perception enable \emph{closed-loop verification} of resilience properties. For example, one can verify that a system accomplishes its task even when visual sensors produce degraded images (e.g., grainy or blurry), by carefully coupling statistical models of perception with deterministic reachability analysis for neural-network-based CPS~\cite{waite_state-dependent_2025}. A critical deployment step is to check model validity in the target environment; invalid models can invalidate previously obtained guarantees~\cite{peper_towards_2025}. Overall, such techniques provide a path to resilience against out-of-distribution inputs, as discussed in the previous theme.

A complementary approach uses generative learning. \emph{World models}, which predict future observations from past observations and actions, originated in reinforcement learning to efficiently sample realistic episodes~\cite{ha_world_2018}. Generative world models have been adopted in robotics and CPS for monitoring~\cite{acharya_competency_2022}, prediction~\cite{mao_how_2024}, and verification~\cite{katz_verification_2022}. A key challenge is to firmly ground these generative models in physical reality~\cite{peper_four_2025} to prevent hallucinations and support trustworthy resilience assessment. This would allow the CPS community to bring decades of experience in relating cyber and physical dynamics to the emerging challenge of physically unrealistic hallucinations in generative AI.

\subsection{Verification, Synthesis, and Redundancy with Known/Learned Specifications and Models}

\subsubsection{Discretization-based Techniques:}
A standard approach to designing correct-by-construction control software or performing formal verification is via \emph{symbolic abstraction}. Here, requirements are specified using temporal logic (e.g., LTL) or automata over infinite strings~\cite{katoen08}. A finite, discrete abstraction of the continuous dynamical system is constructed so that any controller synthesized on the abstraction can be systematically refined to a controller for the original system, and verification results transfer from the abstraction to the concrete system. These finite abstractions provide a unified modeling framework for both cyber and physical components. By composing their finite-state models, we obtain a complete finite-state representation of the CPS, enabling the use of techniques from discrete-event systems~\cite{DEDSBook,CassandrasBook} and automata games~\cite{Thomas95,ComputeGames,MalerPnueliSifakis95} for automatic controller synthesis and property verification. The synthesized discrete controller is then refined into a hybrid controller, or the guarantees are carried over to the concrete model.

Research on symbolic abstractions has evolved in three major directions. The first constructs non-deterministic finite abstractions that over-approximate the input-output behaviors of the original system, often motivated by sensing and actuation limitations~\cite{Raisch98,Lunze99}. Extensions based on Willems’ behavioral theory and $\ell$-complete systems~\cite{MoorRaisch99,willems} improve abstraction accuracy by increasing the history length $\ell$ of input/output symbols, at the cost of larger models. The second, inspired by bisimulation theory~\cite{milner,park}, builds quotient systems by partitioning the state space to obtain exact bisimulations between the original and abstract models~\cite{AHLP00,KloetzerBelta08,paulo}. Such partitions, however, may not exist or may not terminate even for simple linear systems~\cite{asarin}. The third, more recent direction introduces approximate equivalence relations, such as approximate bisimulation~\cite{girard}, which relax exact equivalence by allowing bounded output discrepancies between related states. This greatly expands the class of systems that admit finite abstractions and has led to a rich body of results~\cite{paulo,pola,girard2,majid,gunther2,majid8}.

Unfortunately, none of the finite abstractions in this literature is guaranteed to preserve security properties such as opacity. As shown in~\cite{Zhang2018OpacitySimilation}, standard (bi)simulation relations and their approximate variants, typically used in abstraction schemes, fail to preserve opacity. Recent results~\cite{Yinapproximate,LTYZ} adapt simulation relations to the opacity setting and develop, for the first time, abstraction-based techniques for opacity verification in CPS.

\subsubsection{Discretization-free Techniques:}
The above abstraction-based framework provides a systematic way to address mission and opacity properties in complex CPS, but it can face scalability challenges due to discretization of state and input spaces. This has motivated discretization-free approaches, particularly those based on barrier certificates.

Over the past two decades, \emph{barrier certificates}~\cite{prajna_safety_2004} have become a powerful tool for safety verification of dynamical systems, with automated search enabled by optimization techniques such as sum-of-squares programming~\cite{Parrilo_2003}. Consider a system $x(t+1)=f(x(t))$ with $x(t)\in X$ for all $t\in\mathbb{N}_0$. A function $B: X \to \mathbb{R}$ is called a \emph{barrier certificate (BC)} if
\begin{equation*}
    B\bigl(f(x)\bigr) \;\le\; B(x) \quad \text{for all } x \in X.
\end{equation*}
If there exists a BC such that $B(x)\le 0$ for all $x \in X_0$ (initial set) and $B(x) > 0$ for all $x \in X_1$ (unsafe set), then the system never reaches $X_1$ from any $x_0 \in X_0$. In automata-theoretic verification, a central challenge is to determine whether a set of states can be visited only finitely many times. This can be reduced to a safety-like condition by bounding the number of visits to a given region. Recent results~\cite{murali2023co} build on bounded synthesis and introduce an abstraction-free technique for automata-theoretic verification of discrete-time dynamical systems, using barrier-certificate-style constructions for $\omega$-regular properties.

Despite the power of verification, it is typically reactive—analyzing a system after it is designed. Correct-by-construction synthesis instead aims to \emph{generate} controllers or estimators that are guaranteed to satisfy safety or performance specifications by design, embedding correctness into the synthesis process. Safe controller synthesis constructs control strategies that prevent the system from entering unsafe states, often under uncertainty or adversarial disturbances~\cite{tabuada2009verification,belta2017formal}. This proactive perspective is central to resilient CPS.

\subsection{Provably Resilient Estimators in CPS}

When complete state information is unavailable, estimator synthesis is used to infer system states while respecting high-level requirements. This is particularly relevant for state estimation under sensor or actuator attacks and under bounded or stochastic disturbances~\cite{khajenejad2022resilient,pajic2016attack,khajenejad2023resilient,yong2018switching}.

Provably safe state estimation is crucial for CPS integrity in adversarial environments. Formal observers aim to characterize when agents, actuators, or sensors are compromised and to map such conditions into security or privacy policies. The field of secure state estimation has evolved from \emph{reactive} detection and mitigation of attacks~\cite{murguia2016cusum,mo2010false,jin2017adaptive} to \emph{pre-emptive} resilience in the estimation process, including characterizing upper bounds on the number of attacked sensors/actuators~\cite{dan2010stealth,yong2018switching,mishra2016secure}. These bounds, together with the notion of \emph{sensor redundancy} (multiple sensors measuring the same quantity), guide preventive attack mitigation: one can determine which sensors or actuators must be protected to guarantee resilient estimation prior to deployment~\cite{yang2019sensor,yang2020multi,yong2018switching,khajenejad2022resilient}. 

Hardware redundancy, such as triple modular redundancy (TMR) in flight and aerospace systems~\cite{yeh1996triple,berg2016verification}, will continue to play a key role in pre-emptive resilience. Determining which hardware elements should be duplicated or triplicated in CPS, while balancing cost and weight, remains an important open problem. Together with the verification, testing, and synthesis techniques above, resilient estimators and hardware redundancy form a critical layer of defense for CPS, enabling robust operation even under faults and attacks.

%% file: Sections/theme4-recovery.tex
\section{Theme 4: Recovery for CPS Resilience}
\label{sec:theme4-recovery}

\textbf{Overview:} Resilience requires the ability to recover from and adapt to changing conditions, failures, and attacks. The complexity and scale of uncertain, high-dimensional, networked CPS, such as mobile sensor and robot networks operating in unknown environments, pose significant challenges to the timely and safe response of CPS, so that they can efficiently tolerate disruptions and adapt to new conditions. For example, slowly time-varying environmental conditions may require replanning of the nominal objectives of the CPS, along with on-the-fly data collection for learning the new conditions, while unanticipated failures of CPS components require rapid detection and mitigation strategies. Furthermore, CPS can be vulnerable to adversarial attacks on their sensor and actuator systems, raising the need for resilience against such threats. Finally, networked CPS raise additional challenges in how the communication topology and the level of connectivity among agents offers resilience guarantees against the possibly compromised information flowing through the system. 


\noindent \textbf{Motivating Example:} To articulate the above aspects of CPS recovery in more detail, let us consider a nominal stochastic control process where an agent (or a collection of agents) must make sequential decisions (i.e., over time) under uncertainty. Typically, CPS, under stochastic control processes, operate under a \textit{sense-predict-plan-control} loop, where a perception component \textit{senses} the environment, a prediction component forecasts the evolution of the environment, a planner optimizes the expected performance of the agent under the predicted evolution,\footnote{The ``expectation'' is taken with respect to the stochastic plan, which subsumes the stochastic evolution of the environment.} and a controller optimizes the low-level control actions. For example, consider a drone that must navigate an uncertain environment and reach a target of interest. In this case, the drone's onboard sensors sense obstacles and terrain features, while a learned or model-based predictor forecasts how wind conditions or moving obstacles, e.g., other drones, may evolve. A high-level planner, typically pre-trained offline, generates a trajectory that balances risk and reward, and a low-level controller executes fine-grained motor commands to follow the trajectory safely and effectively.

\begin{figure}
\centering
\includegraphics[width=0.70\textwidth,clip]{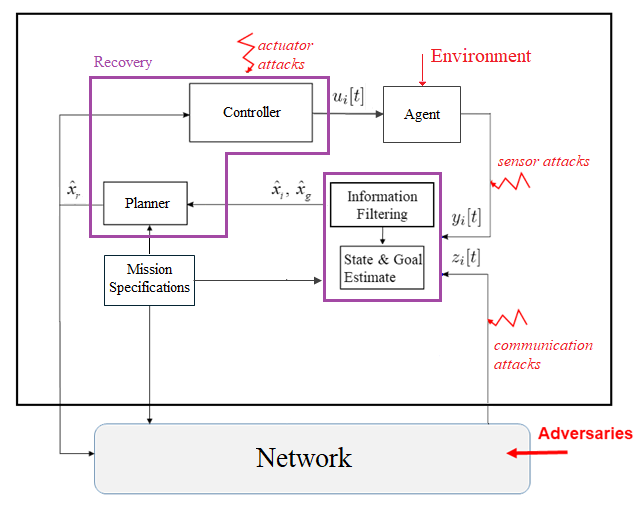}
\vspace{-2mm}
\caption{A recovery architecture against attacks and environmental uncertainty.}
\label{fig:recovery}
\vspace{-4mm}
\end{figure}

\subsection{Resilience as (Re)Planning in Stochastic Environments} 

We focus specifically on the planner for this discussion. The planner's decision-making process can be modeled as a Markov decision process (MDP), where the state captures the current configuration of the drone and relevant features of the environment, the action corresponds to a high-level navigational decision, and the transition dynamics reflect the stochastic evolution of the system due to environmental uncertainty. Given this formulation, the planner's goal is to compute a policy—a mapping from states to actions—that maximizes the expected cumulative reward, typically representing progress toward the target while avoiding risk. Such a policy can be computed using dynamic programming techniques when the model is known and tractable, or learned through reinforcement learning when the model is unknown or the environment is too complex to model explicitly. 

In many real-world scenarios, the environment in which a CPS operates is not stationary; its underlying dynamics and reward structures may evolve over time~\cite{ackerson1970state,campo1991state,luo_dynamic_2023,keplinger2025ns}. If such time-varying conditions are known beforehand—e.g., during training—and exploration across the temporal axis is feasible (such as through simulation), then the problem can be reduced to a stationary decision-making setting~\cite{lecarpentier2019non,keplinger2025ns}. In theory, the model designer can augment the state space with a time index, effectively treating time as a feature, and learn a time-dependent policy over a stationary MDP. However, in practice, environmental changes are often unexpected and unmodeled, making such augmentation ineffective. In these cases, the CPS must first detect that a change has occurred, then update its internal model of the world—potentially inferring how the environment's dynamics or reward structures have shifted—and subsequently replan or adapt its policy in accordance with the new conditions. 

These challenges are naturally captured by the framework of non-stationary Markov decision processes (NS-MDPs), where the transition dynamics or reward functions evolve over time in unknown ways~\cite{lecarpentier2019non,keplinger2025ns,chandak2020optimizing}. A central difficulty in such settings is that the agent must collect new data to update its understanding of the world, yet doing so requires acting in an environment whose behavior it does not fully understand. This creates an inherent tension between exploration and safety. A common approach to address this is risk-averse planning~\cite{lecarpentier2019non}, in which the agent initially follows a pessimistic policy that prioritizes safety and robustness while gathering data, and subsequently transitions to a more performant policy once sufficient knowledge has been acquired~\cite{luo_dynamic_2023,luo2024act}.
However, data collection and model updates do not guarantee improved performance. The agent's uncertainty during planning can be broadly categorized as epistemic or aleatoric~\cite{luo2024act}. Epistemic uncertainty arises from incomplete environmental knowledge and can, in principle, be reduced through active exploration. On the other hand, Aleatoric uncertainty is intrinsic to the environment—for instance, due to sensor noise, wind gusts, or moving obstacles—and cannot be eliminated through data alone~\cite{luo2024act}. In the drone scenario, epistemic uncertainty may stem from an incomplete map of wind patterns or terrain, which can be improved with data collection. Aleatoric uncertainty may arise from unpredictable gusts or transient obstructions, which persist despite data collection. Thus, while the drone can reduce epistemic uncertainty through targeted exploration, it must continue to operate under a risk-averse policy in regions of high aleatoric uncertainty to ensure safety and resilience.

In such non-stationary settings, the agent is often required to compute a plan online, adapting in real time to observed changes. Monte Carlo Tree Search (MCTS)~\cite{kocsis2006bandit} is a widely used approach in this context, allowing the agent to simulate possible futures and select actions based on sampled trajectories. Importantly, the agent need not discard its previously learned policy altogether. If the environment has only changed partially, the existing policy can still provide valuable priors~\cite{pettet2024decision}. This motivates using policy-augmented search techniques, where the prior policy guides the MCTS rollout distribution or initializes the search tree, balancing adaptation with prior knowledge~\cite{luo2024act}. Such hybrid strategies are particularly effective when changes are localized or gradual, enabling the agent to maintain performance while remaining responsive to environmental shifts.

\subsection{Guaranteed Transition between Nominal and Backup Systems} 

We know that the trading-off between exploration and exploitation in reinforcement learning is an open challenge, especially in non-stationary environments. Nevertheless, it offers a principled manner of how resilience can be viewed as the transition from the nominal system operation (under a nominal policy) to the backup system operation (under a backup policy). Continuing the example above, 
guaranteeing that the agent will always remain safe while navigating along a possibly unsafe nominal trajectory (which can be viewed as exploration) requires that the nominal trajectory can be always transitioned in a timely manner to a backup trajectory that is safe, albeit of possibly reduced functionality compared to the nominal one. In other words, for safe navigation among unknown obstacles that are detected on-the-fly, exploration could be viewed as finding an optimal trajectory (nominal) that maximizes the sensed area, albeit possibly risking safety, while exploitation could be viewed as finding a trajectory (backup) that remains in the sensed safe area, albeit possibly being suboptimal w.r.t. the nominal objective. Viewing this paradigm in resilience terms, the backup system is the one of reduced functionality that can keep the system ``running" if the nominal system fails. Hence, recovery entails that the nominal system (exploration) should be designed to always be able to transition to the backup system (exploitation) so that upon the disruption event, the system trajectories can be governed by the backup system.

The design of nominal-backup system pairs and policies for recovery raises significant challenges, including how to obtain and interconnect them in a computationally-efficient manner for real-time execution and adaptation. Considering the sensing-planning-control loop, the sensing component builds the sensed area, or currently known set, of the environment, which can have safe and unsafe parts, the planner generates a nominal path or trajectory for the robot to follow to a goal location, which can lie through the currently unknown, and hence possibly unsafe, part of the environment, and the controller tracks this nominal trajectory. Since the known set is built on-the-fly, the nominal trajectory needs to be verified online that it is safe to be tracked by the controller. Nominal system trajectories that are predicted to be unsafe should be mitigated by safe, backup system trajectories.

We have recently developed a computationally-efficient method for online safety verification \cite{DAgrawal_TRO24}, called Gatekeeper, which serves as a novel component between the planner and controller. Gatekeeper defines the committed trajectory that the system tracks as a composition (and hence trade-off) between the nominal trajectory and the backup trajectory. The construction of the committed trajectory and the verification of its safety is done recursively, and by construction assures safety under certain assumptions. 
By tracking the committed trajectory as generated by Gatekeeper at each iteration, the system is guaranteed \emph{to be able to, if needed,} converge to a safe backup set. 
The Gatekeeper principle has already been applied to recovery objectives such as energy renewal in mobile sensor networks \cite{naveed2024mesch} (enacting safe transitions to a charging station for a team of autonomous aerial robots), and landmark tracking in aerial surveillance \cite{cherenson2025autonomy} (for reducing navigation/localization errors under budget, state, and input constraints). 
While co-designing the nominal and backup systems remains an open challenge and problem, and depends highly on the mission at hand, generic methodologies for robust and adaptive safety control that can be used to define the backup set include Control Barrier Functions \cite{garg2024advances}, and their formulations that handle input constraints and disturbances  \cite{agrawal2021safe, breeden2021high}, online adaptation to structural parameters \cite{black2021fixed, kim2024learning} and risk-aware settings \cite{black2023safety, wang2025safe}, in tandem with classical planners such as MPC and RRT* for compatibility between nominal and backup trajectories \cite{agrawal2021constructive, kim2024visibility, cherenson2025autonomy}.

\subsection{Recovery from Adversarial Attacks} 

Recovery for CPS should be both safe and prompt to ensure system resilience. Recovery safety and recovery promptness are not independent of each other.
Recovery safety entails that the recovery process does not introduce new hazards or violate critical system constraints. Recovery promptness is essential to minimize performance degradation and prevent cascading failures. 

Adversarial attacks on CPS can be categorized by their purpose. The first type directly compromises safety, such as actuator signal injections that induce structural resonance, potentially causing severe damage. These are the main focus of recovery methods discussed in this subsection. The second type targets non-safety aspects, including system confidentiality, infrastructure availability, or privacy.

The attack surfaces through which adversarial actions are executed span both the cyber and physical domains. In the cyber domain, attackers may employ techniques such as false data injection, signal delay, or denial-of-service to disrupt communication and control logic. In the physical domain, attacks manipulate the surrounding environment to induce sensor misreadings. For example, through temperature changes, electromagnetic interference, or physical obstructions. Importantly, physical-domain attacks can be particularly challenging to detect because they often circumvent traditional software-based anomaly detection mechanisms by exploiting the physics of the system rather than its code.

Recovery strategies from adversarial attacks can be broadly classified into two categories: shallow recovery and deep recovery \cite{lu2024recovery}. Shallow recovery methods aim to restore system functionality by reconstructing or estimating the corrupted sensor or actuator data and allowing the original control logic to continue operating with minimal disruption \cite{abad2016reset}. These methods typically rely on statistical filtering, observer-based estimation, or state reconstruction using historical data. While shallow recovery is computationally lightweight, it assumes that the original controller remains effective and efficient after data restoration. This assumption may not hold in the presence of persistent attacks.

In contrast, deep recovery adopts a more comprehensive approach by engaging a dedicated recovery controller designed to steer the system back to a safe operational state \cite{liu2023learn, zhang2020real, zhang2021real, zhang2024fast}. These recovery controllers may be synthesized using model-based or data-driven techniques and are often equipped with formal safety or performance guarantees. Deep recovery mechanisms are especially valuable in scenarios where the original controller cannot ensure safe operation due to altered system dynamics, actuator saturation, or unrecoverable state divergence. While deep recovery generally involves higher computational cost and design complexity, it offers greater timeliness and adaptability in adversarial settings.


In addition, there is often a need for \textit{proactive measures} so that the system can recover \textit{before} it moves to an unsafe state --- either by an adversary or even a fault.
%
The proactive measures include, 
\textit{(a)} methods to restore the system to a "clean" state (perhaps by restarting the system and reloading from a read-only memory)~\cite{resecure, resecure-iot} and
\textit{(b)} addition of noise in a systematic manner~\cite{indistinguishability} to ensure that an adversary cannot predict the behavior of the system, and hence cannot exploit it.
These methods focus on ensuring the safety of the system --- in the presence of adversarial/faulty behavior --- and are complementary to the recovery methods discussed above.

\subsection{Recovery of Networks in Networked CPS} 
Networked CPS, such as mobile sensor networks and teams of networked autonomous robots, rely on the interconnection among components to enable collective decisions and system-level behaviors that emerge from local interactions. Communication topologies and levels of connectivity \cite{prorok2021beyond} that govern the evolving inter-component interactions in these networked systems are thus critical for determining the system resilience.
Therefore, leveraging system-level redundancy is essential for designing recovery strategies that enhance the resilience of networked CPS by initiating system-wide transformations to preserve or restore networking capabilities.

Designing effective recovery strategies for networked CPS is challenging because it involves balancing competing goals. While network redundancy enhances fault tolerance, maintaining additional connections can lead to conservative behaviors due to limited promixity-based communication capabilities that restrict the system's performance, especially when flexibility is essential. 
It is critical to jointly consider (i) how to define effective redundant network structures based on measurement of network resilience, (ii) how to control the behaviors of the networked system to satisfy the redundancy-prescribed specifications, and (iii) how to balance redundancy and performance during system reconfiguration for recovery that adapts to the rapidly unfolding situations. 
Graph-theoretic measures such as algebraic connectivity \cite{ong2023nonsmooth, luo2020behavior, yang2023minimally} have been widely used to design motion controllers that ensure the integrity of multi-robot communication networks as their topologies evolve during coordinated robot movement.
In the presence of a bounded number of random robot failures or malicious robots, resilience measures such as $ k-$connectivity \cite{luo2019minimum, luo2020minimally} and $r-$robustness \cite{LeBlanc2013} can be employed to design recovery strategies that re-configure multi-robot networks to attain networking capability with random node removals or mitigate the influence of informational adversaries \cite{cavorsi2023multirobot, usevitch2021resilient, lee2024maintaining}. The graph robustness notion in particular generalizes the notion of network connectivity, and captures how dense the communication structure should be for the network to be able to tolerate a known upper bound on the number of adversarial agents, i.e., so that the intact agents can achieve consensus on common values despite the effect of malicious information from the adversarial agents \cite{LeBlanc2013}. For a recent survey on network resilience the interested reader is referred to \cite{PIRANI2023111264}. Recent work \cite{yang2024decentralized, yang2024integrating} has proposed uncertainty-aware and data-driven approaches for maintaining multi-robot communication structures that adapt to positional noise and real-world communication performance, while being co-optimized with task-level objectives.

%% file: Sections/theme5-human.tex
\section{Theme 5: Role of the Human in CPS Resilience}
\label{sec:theme5-human}
\begin{figure}
    \centering
    \includegraphics[width=0.75\linewidth]{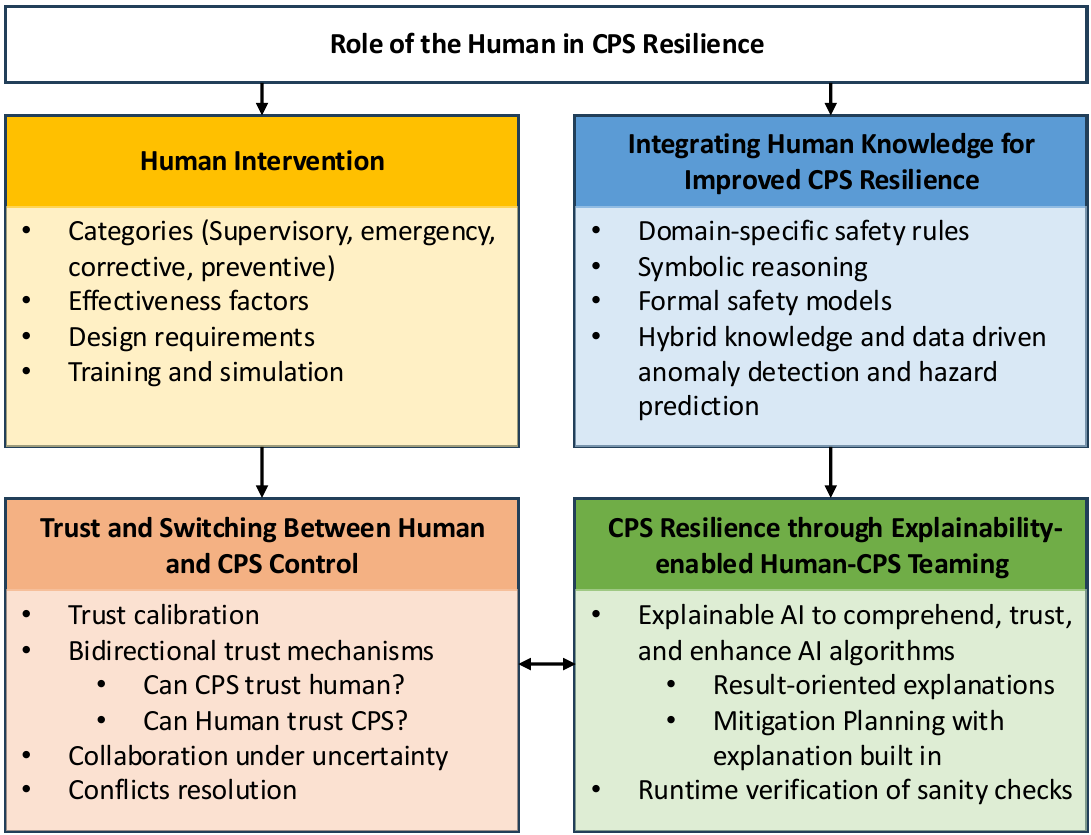}
    \caption{Overview of Theme 5: Role of the Human in CPS Resilience.}
    \label{fig:theme5}
\end{figure}

While much research has focused on the technical aspects of CPS resilience (e.g., fault tolerance, redundancy), the role of the human remains critical, especially in human-in-the-loop cyber-physical systems (HCPS), such as Advanced Driver Assistance Systems (ADAS) and a variety of medical CPS. This section explores how human interactions influence CPS resilience, categorizes types of human involvement, and identifies key challenges and opportunities for designing resilient CPS. Fig. \ref{fig:theme5} shows the overview of this section.
 
\subsection{Human Intervention}
Human intervention remains a cornerstone of resilience in CPS, particularly in contexts where system autonomy is insufficient to handle anomalies, unexpected events, or degraded conditions. While modern CPS increasingly integrate intelligent automation and fault-tolerant architectures, they cannot fully eliminate the need for timely and informed human interventions. The human operator often serves as a fallback mechanism, stepping in when automated systems reach the boundary of their capabilities, or when situational complexity exceeds preprogrammed responses \cite{leveson2011engineering, sheridan2014humans}.

Human interventions in CPS can be broadly categorized as {\em supervisory}, {\em emergency}, {\em corrective}, or {\em preventive}. In {\em supervisory} roles, humans oversee system behavior and intervene occasionally to correct deviations or refine performance. For example, a grid operator may fine-tune load balancing strategies in response to forecasted weather disturbances. {\em Emergency} interventions, on the other hand, are characterized by immediate human action during critical failures or hazards, such as when a pilot disables an aircraft’s autopilot system during turbulence or a driver takes control of a semi-autonomous vehicle to avoid a collision. {\em Corrective} interventions occur after faults are detected, often requiring human diagnosis and reconfiguration of the system, such as after a detected cyberattack on an industrial control system. {\em Preventive} interventions are proactive measures, like scheduling maintenance or isolating vulnerable subsystems prior to a high-risk event.

The effectiveness of human intervention is tightly coupled with several time-dependent factors, including detection latency, cognitive load, decision-making ability, and action execution time. Particularly in time-sensitive CPS, such as autonomous vehicles, smart manufacturing systems, or surgical robots, even brief delays in human reaction can lead to undesirable or dangerous outcomes.
In contrast, timely human interventions can significantly mitigate potential safety hazards and improve the system's resilience.
Cognitive psychology research shows that under high stress or low awareness, human performance deteriorates significantly, especially when transitioning from a passive monitoring role to an active control role \cite{parasuraman2000model}. This switch latency, i.e., the time taken for the human to re-engage with the system and execute an appropriate action, can be exacerbated by poor interface design, ambiguous system feedback, or insufficient training.

Designing CPS that support effective human intervention requires careful attention to system transparency, feedback clarity, and interface usability. The system must present its internal state, decision logic, and level of certainty in a manner that facilitates rapid human comprehension and trust calibration \cite{wischnewski2023measuring}. 
 Furthermore, systems should maintain historical context and provide situational summaries that allow human operators to quickly regain situational awareness during handovers. 
Training and simulation also play vital roles in improving the quality and timeliness of interventions. Repeated exposure to realistic failure scenarios enhances human readiness and helps develop robust mental models of system behavior. For instance, power grid operators and air traffic controllers regularly undergo simulator-based training to improve their ability to manage rare but high-impact incidents \cite{hollnagel2006resilience}. Such training should not only focus on routine tasks but also emphasize adaptive decision-making under uncertainty.

To this end, developing realistic, domain-specific human intervention simulators is essential for both operator training and system design validation. These simulators must support interactive, real-time scenarios that closely mirror CPS operations under various fault conditions, cyberattacks, or system degradations. For example, in the autonomous driving domain, our prior work \cite{zhou2022strategic,zhou2024runtime} introduced a rule-based driver simulator to assess its performance on improving the system resilience against perception attacks. 
A complementary study developed a human‑in‑the‑loop driver simulation framework to evaluate autonomous driving safety and vulnerability under varied driving conditions and behaviors \cite{DriveSim}.  
Other similar efforts include Shao \emph{et al.}'s CARLA-based closed-loop simulator for evaluating instruction-following and trustworthy AV behavior \cite{shao2024lmdrive}, Wei \emph{et al.}'s ChatSim for editable, photo-realistic, and human-centric scene simulation \cite{wei2024chatsim}, and Sim-on-Wheels, which overlays virtual hazards into live driving feeds to create hybrid virtual-real test environments \cite{shen2023simonwheels}. These tools allow for high-fidelity studies of interface design, human decision latency, and adaptive system response during emergencies.
Together, these simulators enhance operator preparedness and also guide the co-design of resilient CPS, providing insights into how human behavior, perception, and trust interact with autonomous decision-making. Future simulators should integrate richer multimodal sensing, real-time explainability, and adaptive scenario generation to evaluate how human-CPS teaming behaves under stress, deception, or degraded awareness.

In conclusion, as CPS continue to evolve toward higher autonomy, human intervention remains indispensable for maintaining system resilience. Ensuring that operators can effectively detect, interpret, and respond to anomalies in complex CPS requires an integrated approach that melds technical robustness, including clear system state and uncertainty reporting, with human-centered design, featuring transparent interfaces and intervention pathways. 

\subsection{Integrating Human Knowledge for Improved CPS Resilience}
Enhancing the resilience of CPS requires more than robust algorithms and fault-tolerant architectures; it demands the explicit integration of human and expert knowledge into both system design and runtime operation. Human intuition, domain-specific safety constraints, and prior operational experience often encode valuable insights that machine learning models or control algorithms alone may not capture. As a result, CPS research increasingly focuses on leveraging structured expert knowledge to guide ML algorithms, enhance interpretability, and improve responsiveness under uncertainty or attacks \cite{zhou2022robustness,rai2020driven}.

A growing body of work aims to formalize the integration of such knowledge into ML pipelines. Integrating logic-based constraints or domain-specific safety rules during training has become a key approach for aligning ML models with human expectations. These include soft-constraint formulations \cite{xu2018semantic}, logic circuits for structured generalization \cite{liang2019learning}, embedding logical rules in network architectures \cite{embedlogic2019chen}, and graph-based knowledge representations \cite{guo2016jointly}. Techniques like knowledge distillation have also been used to transfer structured or symbolic knowledge into compact, efficient networks while preserving behavior fidelity \cite{gou2021knowledge}.

In the context of CPS safety, domain knowledge is particularly valuable when formalized as Signal Temporal Logic (STL) properties. STL provides a rich framework for expressing temporal behaviors and constraints in continuous systems. Several recent studies have explored the integration of STL into learning-based monitoring and anomaly detection. For example, Bartocci \etal \cite{bartocci_2018} proposed methods to mine STL properties from data, while \cite{telexpaper} introduced approaches to learn STL formulas from positive examples. These works have demonstrated that logic-based specifications can guide anomaly detection in CPS applications such as industrial control and automotive systems \cite{JonesAnomaly}.

Our prior work \cite{dsn2021zhou} extended this paradigm by proposing a formal framework to synthesize STL-based safety monitors in artificial pancreas systems (APS), based on control-theoretic hazard analysis. Building on this foundation, we have developed a new approach that integrates STL safety requirements as soft constraints into sequential prediction models for real-time hazard detection and mitigation. This extension enables context-aware monitoring that not only detects but also predicts unsafe behaviors, and it is generalizable to a broader class of CPS, including ADAS \cite{zhou2023hybrid}.
Beyond STL, symbolic reasoning and customized learning constraints continue to play an important role. For example, Chen \etal \cite{embedlogic2019chen} proposed embedding logic rules directly into recurrent neural networks using feedback masking techniques. Similarly, reinforcement learning research has leveraged teacher policies and symbolic constraints to train more efficient and safer agents \cite{rusu2016policy}. These strategies highlight how structured expert knowledge can shape exploration, accelerate learning, and reduce unsafe behaviors during both training and deployment phases.

Ultimately, resilience in CPS depends not only on the system’s internal capabilities but also on its ability to incorporate and respond to knowledge from human operators, engineers, and domain experts. By tightly coupling symbolic reasoning, formal safety models, and expert-guided ML, modern CPS can achieve higher assurance, faster recovery from faults, and improved adaptation to new or adversarial environments.

\subsection{Trust and Switching Between Human and CPS Control.}
Trust is a cornerstone of effective human-CPS collaboration, particularly in high-stakes environments such as autonomous vehicles, healthcare robotics, and industrial control systems. The degree of trust between humans and CPS influences system resilience by shaping how and when control is shared or switched between human and machine agents \cite{lee2004trust, sheridan2014humans}. Trust governs reliance, influences operator engagement, and ultimately affects the system’s ability to recover or adapt in the face of faults or adversarial disruptions.
Trust in automation is dynamic and contextual. Users may exhibit overtrust, resulting in uncritical reliance on potentially faulty automation, or undertrust, leading to disuse even when the system is competent \cite{lee2004trust, tenhundfeld2022assessment}. To promote resilience, CPS must calibrate trust, maintaining user confidence while encouraging appropriate skepticism when necessary. This can be achieved through transparent system design, performance feedback, and explainability mechanisms \cite{mhapsekar2024building}.

Recent advances emphasize the dynamic modeling of trust and the development of adaptive trust-aware architectures. Wang \etal \cite{wang2024enhancing} proposed trust-aware control strategies that adjust trajectory planning based on real-time trust metrics, thereby reducing operator workload in teleoperation tasks. Similarly, recent frameworks for human-robot collaboration have explored dynamic role allocation, where the system adjusts its behavior, such as switching between autonomous and compliant modes, based on the operator's inferred trust and cognitive state \cite{li2025trusttriggered, wildman2024trust}. These mechanisms are particularly vital in collaborative settings like mixed-initiative control of UAVs, where seamless transitions between human-led and autonomous operations are required to maintain safety under uncertainty \cite{wang2024trustreflective}.

A critical enabler of this calibrated trust is explainability, the system’s ability to communicate the rationale behind its behaviors. While explainability will be discussed in greater detail in Section \ref{sec:explainability}, it is worth noting here that Explainable AI (XAI) techniques are essential for preventing both over-trust and distrust \cite{stray2025explainability}. Quantitative studies demonstrate that interactive explanations significantly improve user confidence and decision accuracy, allowing operators to calibrate their trust levels to the system's actual capabilities \cite{lai2025exploring}.

Trust is also bidirectional. Not only must humans trust the CPS, but the system must also assess when it can rely on human input. For example, a distracted driver may not be capable of taking over from an autonomous vehicle in a timely manner. Research in human state estimation and adaptive delegation seeks to address this challenge by monitoring human readiness and context \cite{endsley2017here, faroni2022safety}. These bidirectional trust mechanisms are vital for safe and resilient switching between control regimes.
Recent studies have begun exploring trust formation in multi-agent and distributed CPS, where teams of humans and AI agents collaborate under uncertainty \cite{fan2008influence}. The dynamics of inter-agent trust, conflict resolution, and shared situational awareness present new challenges and opportunities for CPS resilience.

\subsection{CPS Resilience through Explainability-enabled Human-CPS Teaming}
\label{sec:explainability}

Despite significant technological advancements in CPS, stakeholders remain hesitant to fully accept and rely on these systems for decision-making~\cite{zheng2018smart, deguchi2020smart, tian2019smart}. One fundamental reason for this reluctance is the lack of explainability, which leads to unknown reliability and trustworthiness. This hesitation is particularly acute when CPS decisions directly impact public safety, urban planning, and critical infrastructure. When a CPS decision results in negative outcomes, city officials and decision-makers must be able to explain and justify how that decision was made. Additionally, many urban operations are tightly regulated. If CPS cannot demonstrate compliance with regulatory requirements due to opaque decision-making processes, its deployment may not be legally permissible.

Explainable AI is a prerequisite for resilient CPS operations, particularly when human operators must validate autonomous countermeasures against cyber-attacks. Explainable AI research has its roots in the need to comprehend, trust, and enhance AI algorithms~\cite{arrieta2020explainable}. Recent studies have highlighted the importance of assisting human users in trusting decisions made by models like neural networks (e.g.~\cite{ma2021predictive,ma2019data}), a critical factor when these models control safety-critical physical infrastructure.

The first category, result-oriented explanations, is vital for post-incident resilience and forensics. Here, the objective is to reveal the rationale behind a decision post hoc. Previous work, such as Langley~\cite{langley2016explainable}, discusses behaviors encompassing explaining the objectives of a planning task. In a CPS context, this capability allows operators to verify if a system's \textit{resilient} response (e.g., shutting down a valve) was triggered by a legitimate threat or a sensor error. A comprehensive survey by Sreedharan \etal~\cite{sreedharan2020emerging} defines considerations for the control operator in resilient systems. Fox \etal~\cite{fox2017explainable} identify key questions an explanation system must answer; answering these questions in real-time is essential to reducing the mean time to recovery (MTTR) in compromised systems.
The second category, generating explanations as a sub-goal of planning, directly supports human-in-the-loop resilience. For instance, Chakraborti \etal~\cite{chakraborti2017plan} views the planning problem as a system where the AI provides suggestions to the human. This aligns with resilient CPS architectures where AI detects anomalies but requires human authorization to execute high-risk mitigation strategies. Boggess \etal~\cite{boggess2022toward} focus on explaining MARL policies; such explanations are necessary to ensure that distributed agents (e.g., in a smart grid) act cohesively during a cyber-physical attack.

\noindent \textit{Explainability through runtime verification of sanity checks}
One method of explainability for trustworthiness of CPS is using runtime verification to monitor properties representing \emph{sanity checks}. The efficient reporting of such sanity checks in a human-understandable way enables human or automated intervention. One benefit of this method is that the runtime verification engine can be very rigorously verified and run efficiently, on-board CPS in real time, unlike AI-based solutions. For example, the \textsc{Realizable Responsive Unobtrusive Unit} (R2U2)\cite{JJKRZ23} is a real-time, online, stream-based runtime verification engine that takes as input a set of MLTL specifications to monitor and a description of the CPS hardware resources. R2U2 embeds on-board CPS during operation to continuously monitor that the current mission adheres to the operational requirements, while obeying provable guarantees with respect to the particular CPS hardware. There are currently three implementations of R2U2, in VHDL, C, and embedded Rust; each implements the same core algorithm but tailored to different CPS platforms. R2U2 has successfully verified many real CPS that were specified using MLTL, including NASA's Robonaut2 \cite{KZJZR20}, the JAXA OPS-SAT autonomous satellite \cite{JAXA}, a UAS Traffic Management (UTM) system involving Collins and Mosaic Aerospace \cite{HCHJR21}, a sounding rocket \cite{HLR21}, and multiple small satellites \cite{LLR21, AJR22}.
We can use tools like R2U2 to monitor for different types of sanity checks, such as that we are not using data from a sensor that is out of range, or whether a control action resulted in the expected temporal reaction; see \cite{Roz16} for a more complete list of sanity checks. Reporting the result of sanity checks in real time can enable more confident intervention. R2U2 has GUI interfaces to enable easy human understandability of its outputs \cite{JJKRZ23}. 

%% file: Sections/look-ahead.tex
\section{Looking Ahead: Five Mid-term and Long-term Goals}
\label{sec:look-ahead}

Here, we outline five of the most important open challenges that the community must address to enable widespread adoption of CPS in critical application domains. These challenges organically arise from our discussion in the article on the five themes for resilient CPS. For each challenge, we lay out some mid-term goals, defined roughly as 2--3 years out, and some long-term goals, defined as 5+ years out. 

\medskip
\noindent \textbf{1. Compositional and Adaptive Foundation Models for Edge CPS}

Learning-enabled CPS relies on making CPS applications runnable on the edge, which will comprise primarily resource-constrained devices. Current trends involve adapting general-purpose foundation models (FMs) or building ``micro'' domain-specific models~\cite{tang2025towards}. An open challenge is to create a workflow to achieve a higher-level task, such as in industrial CPS. For example, a pipeline of activities in autonomous warehouse robotics could involve perception and sensing, prediction and reasoning, high-level planning, and human-in-the-loop oversight. The challenge is to shape this pipeline of multiple task-specific adapters into a unified, lightweight CPS application that can run on the edge.

\begin{itemize}

\item \textbf{Mid-term Goals (2–3 years):} One goal is to develop \textit{parameter-efficient "functional patches"} (e.g., advanced LoRA variants) that allow a base CPS model to rapidly switch between distinct operational modes (e.g., from nominal driving to emergency hazard mitigation) without catastrophic forgetting. A promising direction is zero-shot adaptation to heterogeneous sensor resolutions and sampling rates, bridging the gap between pre-training and deployment. Rapidly adapting to environmental changes may also require data collection, so a promising direction is to explore planning and control models that can safely collect data in the presence of environmental changes.

\item \textbf{Long-term Goals ($\geq$ 5 years):} A worthy if challenging goal is to achieve \textit{fully autonomous model evolution}, where the CPS identifies out-of-distribution (OOD) environments and self-synthesizes new adapters using local data and synthetic simulations. The ultimate goal is a \textit{modular "plug-and-play" architecture} in which distributed CPS components can exchange and compose learned adapters to handle emergent, complex, multi-modal tasks in real-time.

\end{itemize}

\noindent \textbf{2. Bidirectional Trust and Learning with Cognitive-Aware Human-CPS Teaming}

Typically, trust in a resilient CPS is assumed to be a binary property: the human either trusts the system or does not. However, in the future, as resilient CPS are deployed more widely, there will be increasingly frequent settings in which the human user or operator is not an expert and will experience low-to-high cognitive failures while interacting with the system. Therefore, such systems should be able to accurately estimate human readiness and calibrate trust levels, thereby making trust a bi-directional property. This angle is intended to complement, rather than replace, ongoing research efforts to enable humans to assess the trustworthiness of the system under various scenarios. The same argument applies to learning: CPS should be designed so that the human operator can improve decision-making by interacting with the system, and vice versa, with the system improving by capturing tacit human (expert) knowledge and adapting over time. This is a critically required attribute for resilient CPS and will continue to grow in importance. It is only through such a calibrated and verified trust relationship that the human (or other interacting systems) can make decisions needed for the resilient operation of the CPS. 

\begin{itemize}

\item \textbf{Mid-term Goals (2–3 years):} A goal is to design \textit{real-time trust calibration frameworks} that use Explainable AI (XAI) to provide result-oriented explanations for tasks at various levels of autonomy. These should move beyond simple post-hoc reasoning to real-time explanations in the event of failures, allowing human operators to authorize effective, and occasionally high-risk, recovery strategies with full situational awareness and querying counterfactuals. The CPS must, over time, adapt through interactions with human experts, e.g., by adjusting its objective to capture implicit parameters not captured in the design.

\item \textbf{Long-term Goals ($\geq$ 5 years):} A long-term goal is to develop \textit{cognitive-state-aware control regimes} where the CPS can autonomously delegate or reclaim authority based on inferred operator stress, fatigue, or low trust value for whatever reason. This includes resolving conflicts in multi-agent environments where teams of humans and AI agents must maintain a shared mental model, including during critical operational periods such as cascading security attacks.

\end{itemize}

\noindent \textbf{3. Provably Safe ``Good Enough'' Recovery Strategies}

Resilience necessitates the ability to handle perturbations beyond pre-specified fault models, often requiring the system to settle for just ``good enough'' safety-critical functionality. This is a nod to the reality of complex CPS, where theoretically optimal recovery actions are not possible due to real-world constraints, such as policy or competitive constraints that disallow cooperation between some stakeholders, transient degraded connectivity between certain elements of the CPS, or the lack of capable human operators. Hence, we need to systematically develop the foundations for just-good-enough recovery. This should encompass several properties such as a notion of outcomes that absolutely must be avoided and then a graded degree of undesirability of other outcomes. 

\begin{itemize}

\item \textbf{Mid-term Goals (2–3 years):} We have to refine \textit{online safety verification components} (like Gatekeeper~\cite{DAgrawal_TRO24}) to handle highly nonlinear systems in rapidly changing environments. Research should focus on trajectories that guarantee a path to a safe state even when the planner explores unknown or unsafe state spaces.

\item \textbf{Long-term Goals ($\geq$ 5 years):} We have to develop \textit{deep recovery controllers} that can autonomously re-synthesize provable control logic in real-time in response to altered system dynamics. Such alterations may be caused by cyber or physical, failures or attacks. These strategies must be tamper-proof, utilizing hardware roots of trust (such as the ARM TrustZone), to ensure that the recovery process itself cannot be hijacked.

\end{itemize}

\noindent \textbf{4. Verification of Neuro-Symbolic and Generative World Models}

With the revolution of generative AI, it is inevitable that it will play a central role in CPS applications as well. To ensure that this can be used in critical CPS applications as well, it is important to ground generative AI in physical reality to prevent hallucinations in safety-critical settings. There is growing interest and promising results in neuro-symbolic models for CPS~\cite{munir2023neuro, lu2024surveying}; neuro-symbolic models combine the learning power of neural networks with the structured reasoning of symbolic logic. Generative world models are used to predict how a physical environment (the ``world'') will evolve based on past observations and actions. There is a synergy between the two: neuro-symbolic models can be used to generate world representations due to their combined strengths of neural networks and symbolic logic. In that case, it would be important that the components of such models operate within provable safety and reliability bounds. 

\begin{itemize}

\item \textbf{Mid-term Goals (2–3 years):} We have to advance neuro-symbolic specification languages (like NeSaL~\cite{xie_neuro-symbolic_2022}) that allow verification engines to reason about high-dimensional pixel spaces, such as identifying a pedestrian in a blurry camera feed. This includes establishing probabilistic abstractions of perception that provide formal confidence intervals for learning-enabled CPS.

\item \textbf{Long-term Goals ($\geq$ 5 years):} We have to create \textit{physically interpretable world models}~\cite{peper_four_2025} that natively incorporate the underlying physics of the CPS components as well as the cyber request-response characteristics. We have to then enable closed-loop verification of such world models, where the system can prove its own resilience against real-world distribution shifts or other perturbations. A possible approach is coupling statistical models of perception with deterministic reachability analysis.

\end{itemize}

\noindent \textbf{5. Resilient Interconnection for Large-Scale Infrastructure CPS}

For infrastructure-scale CPS, resilience moves beyond individual component reliability to the system's capacity to serve society at a large scale. This requires managing widely varying connectivity among the different system elements. On a human level, this requires managing partially aligned incentives among different stakeholders and resulting in partial connectivity and cooperation among these stakeholders.

\begin{itemize}

\item \textbf{Mid-term Goals (2–3 years):} We have to develop \textit{graph-theoretic measures} (e.g., algebraic connectivity, r-robustness, strong r-robustness) to design motion and communication controllers that preserve network integrity during coordinated movement or node failures. Research should focus on balancing planned network redundancy with cost and performance constraints, both during normal operation and under perturbations.

\item \textbf{Long-term Goals ($\geq$ 5 years):}
We have to develop methodologies to strengthen sub-systems within the
CPS, taking into account the risks and the interdependencies among the assets. Models of the CPS system can help identify cost-effective resilience measures. However, these must not only model the technological
elements, but also the economic (who are the stakeholders and what are their economic drivers) and the
policy (what controls can each stakeholder implement and how can they collaborate) factors that guide CPS
operational controls. These models, when instantiated with parameters from the real system, should enable
rational and distributed decision-making among the multiple stakeholders on how to protect the CPS. Then,
at runtime, based on inputs from sensors, the system can determine if a perturbation is currently underway
and, if so, the optimal response to deploy. The goal is to ensure \textit{system-wide resilience} (rather than individual subsystem or component-wise resilience) that can detect and adaptively react to both internal and exogenous perturbations.

\end{itemize}

\lettrine{I}{n summary}, when the five themes that we have described work together cohesively, they hold the greatest potential for us to achieve resilient CPS. Each theme has already delivered a rich array of solutions --- and also presents to us a challenging set of questions that we, as an energized and well-coordinated community, will embark on and solve. We look forward to sustaining a vibrant field of resilient CPS research, working hand-in-hand with robust practical implementations and strong policy, all leading to widespread use of CPS in highly critical applications.

%% file: Sections/acknowledgments.tex
\begin{acks}
This material is based in part upon work supported by: 
(i) {\bf S. Bagchi and H. Kim}: the DEVCOM ARL Army Research Office under Contract number W911NF-2020-221, and the National Science Foundation under Grant Numbers CNS-2333487 (CPS Frontier) and CNS-2038986; 
(ii) {\bf J. Li and N. Li}: Cisco Research and the National Science Foundation under Grant Numbers CNS-2247794 and CNS-2207204;
(iii) \textbf{S. Z. Yong}: National Science Foundation under Grant Numbers CNS-2312007 and CNS-2313814 and Office of Naval Research grant N00014-23-1-2093;
(iv) \textbf{D. Panagou}: NSF Grant Numbers 1942907 (NSF CAREER), 2137195 (NSF I/UCRC), 2223845 (NSF CPS), and AFOSR Grant Number FA9550-23-1-0557 (Complex Networks);
(v) \textbf{Y. Li}: DEVCOM ARL Army Research Office under Contract number W911NF-2020-221, NSF Grant Numbers IIS-2442739 (NSF CAREER), and CNS-2333491 (NSF-FRONTIER);
(vi) \textbf{F. Kong}: the NSF Grant Numbers CNS-2442914 (NSF CAREER) and CNS-2333980 (NSF CPS);
(vii) \textbf{H. Alemzadeh}: the NSF Grant Numbers CNS-2146295 (NSF CAREER) and CCF-2402941 (NSF SHF);
(viii) \textbf{I. Ruchkin}: the NSF Grant Numbers CNS-2440920 (NSF CAREER) and CNS-2513076 (NSF CPS); (ix) \textbf{M. Ornik}: AFOSR Grant Number FA9550-23-1-0131, NASA Grant Number 80NSSC22M0070, and ONR Grant Numbers N00014-23-1-2651, N00014-23-1-2505, N00014-25-1-2369 and~
(x) \textbf{S. Mohan}: U.S. National Science Foundation grant CPS 2246937 (NSF CAREER).
(xi) \textbf{A. Mukhopadhyay}: U.S. National Science Foundation grant CNS-2531369.
(xii) \textbf{M.D. Lemmon and Y. Duan:} National Science Foundation CNS-2228092. 
(xiii) \textbf{M. Ma}: NSF Grant Numbers 2443803 (NSF CAREER) and 2427711 (NSF ReDDDoT).
(xiv) \textbf{W. Luo}: NSF Grant Number 2528997 (NSF ERI).
(xv) \textbf{Chaterji}: NSF Grant Numbers 2333487 (CPS Frontier) and 2146449 (CPS CAREER).
Any opinions, findings, and conclusions or recommendations expressed in this material are those of the authors and do not necessarily reflect the views of the sponsors.
\end{acks}

%% file: refs/KYR.bib
@inproceedings{DBR21,
author = {James B. Dabney and Julia M. Badger and Pavan Rajagopal},
title = {Adding a Verification View for an Autonomous Real-Time System Architecture},
booktitle = {Proceedings of SciTech Forum},
series = {2021-0566},
publisher = {{AIAA}},
pages = {Online},
month = {January},
year = {2021},
doi = {https://doi.org/10.2514/6.2021-0566},
}

@misc{Dab21,
title = {Using Assume-Guarantee Contracts in Autonomous Spacecraft},
author = {James Bruster Dabney},
howpublished = {Flight Software Workshop (FSW) Online: \url{https://www.youtube.com/watch?v=zrtyiyNf674}},
month = {February},
year = {2021},
}

@misc{DRB22,
title = {Using Assume-Guarantee Contracts for Developmental Verification of Autonomous Spacecraft},
author = {James Bruster Dabney and Pavan Rajagopal and Julia M. Badger},
howpublished = {Flight Software Workshop (FSW) Online: \url{https://www.youtube.com/watch?v=HFnn6TzblPg}},
month = {February},
year = {2022},
}

@inproceedings{DBR23,
  title={Trustworthy Autonomy for Gateway Vehicle System Manager},
  author={Dabney, James B and Badger, Julia M and Rajagopal, Pavan},
  booktitle={2023 IEEE Space Computing Conference (SCC)},
  pages={57--62},
  year={2023},
  organization={IEEE}
}

@inproceedings{AH90,
	Author = {Alur, R. and Henzinger, T.~A.},
	Booktitle = {LICS},
	Pages = {390--401},
	Publisher = {IEEE},
	Title = {{R}eal-time {L}ogics: {C}omplexity and {E}xpressiveness},
	Year = {1990}
}

@inproceedings{MN04,
author = {O.~Maler and D.~Nickovic},
title = {Monitoring temporal properties of continuous signals},
Booktitle = {Proceedings of Formal Techniques, Modelling and Analysis of
  Timed and Fault-Tolerant Systems (FORMATS)},
pages =  {152--166},
year = {2004},
}

@inproceedings{RRS14,
  author = "Thomas Reinbacher and Kristin Y.~Rozier and Johann Schumann",
  title = "Temporal-Logic Based Runtime Observer Pairs for System Health Management of Real-Time Systems",
  editors     =   "Erika Abraham and Klaus Havelund",
  booktitle = "Proceedings of the 20th International Conference on Tools and Algorithms for the Construction and Analysis of Systems (TACAS)",
    series      =   "Lecture Notes in Computer Science (LNCS)",
    volume      =   "8413",
    publisher   =   "Springer-Verlag",
    pages       =   "357--372",
  month = "April",
  year = "2014", 
}

@article{EH10,
author = {Erzberger, H and Heere, K}, 
title = {Algorithm and operational concept for resolving short-range conflicts}, 
volume = {224}, 
number = {2}, 
pages = {225-243}, 
year = {2010}, 
doi = {10.1243/09544100JAERO546}, 
URL = {http://pig.sagepub.com/content/224/2/225.abstract}, 
eprint = {http://pig.sagepub.com/content/224/2/225.full.pdf+html}, 
journal = {Proc. {IMechE} G J. Aerosp. Eng.} 
}

@techreport{GPM,
author = {Shirley Dion},
title = {{Global Precipitation Measurement (GPM) Safety Inhibit Timeline Tool}},
institution = {{NASA Goddard Space Flight Center}},
number = {GSFC.ABS.7501.2012},
year = {2013},
address = {Greenbelt, MD, United States},
note = {\url{https://ntrs.nasa.gov/citations/20130000831}},
}

@inproceedings{RCRBS11,
  title={Designing an interactive local and global decision support system for aircraft carrier deck scheduling},
  author={Ryan, Jason and Cummings, Mary and Roy, Nick and Banerjee, Ashis and Schulte, Axel},
  booktitle={Infotech@Aerospace},
  publisher = {{AIAA}},
  year={2011}
}

@inproceedings{FMRSWY21b,
author = {Michael Fisher and Viviana Mascardi and Kristin Yvonne Rozier and Holger Schlingloff and Michael Winikoff and Neil Yorke-Smith},
title = {Towards a Framework for Certification of Reliable Autonomous Systems},
booktitle = {{20th International Conference on Autonomous Agents and Multiagent Systems (AAMAS)}},
series = {{Journal-first (JAAMAS) track}},
publisher = {Springer},
year = {2021},
month = {May},
}

@inproceedings{LVR19,
    author = {Jianwen Li and Moshe Y.~Vardi and Kristin Y.~ Rozier},
    title = {Satisfiability Checking for {M}ission-Time {LTL}},
    booktitle = {{Proceedings of 31st International Conference on Computer Aided Verification (CAV)}},
    publisher = {Springer},
    series = {{LNCS}},
    volume = {11562},
    address = {New York, NY, USA},
    editors = {Isil Dillig and Serdar Tasiran},
    pages={3--22},
    month = {July},
    year = {2019},
}

@InProceedings{OW08,
author="J. Ouaknine
and J. Worrell",
editor="Cassez, Franck
and Jard, Claude",
title="Some Recent Results in Metric Temporal Logic",
booktitle="Formal Modeling and Analysis of Timed Systems",
year="2008",
publisher="Springer Berlin Heidelberg",
address="Berlin, Heidelberg",
pages="1--13",
}

@InProceedings{Ale13,
author="Donz{\'e}, Alexandre",
editor="Legay, Axel
and Bensalem, Saddek",
title="On Signal Temporal Logic",
booktitle="Runtime Verification",
year="2013",
publisher="Springer Berlin Heidelberg",
address="Berlin, Heidelberg",
pages="382--383",
isbn="978-3-642-40787-1"
}

@inproceedings{Roz16,
    author = {Kristin Yvonne Rozier},
    title = {Specification: The Biggest Bottleneck in Formal Methods and Autonomy},
    booktitle = {{Proceedings of 8th Working Conference on Verified Software: Theories, Tools, and Experiments (VSTTE 2016)}},
    publisher = {Springer-Verlag},
    series = {{LNCS}},
    volume = {9971},
    address = {Toronto, ON, Canada},
    editors = {Marsha Chechik and Sandrine Blazy},
    pages = {1--19},
    doi = {10.1007/978-3-319-48869-1_2},
    month = {July},
    year = {2016},
}

@inproceedings{AJR22,
author = {Alexis Aurandt and Phillip Jones and Kristin Yvonne Rozier},
title = {{Runtime Verification Triggers Real-time, Autonomous Fault Recovery on the CySat-I}},
booktitle = {{Proceedings of the 14th NASA Formal Methods Symposium (NFM 2022)}},
publisher = {Springer, Cham},
series = {Lecture Notes in Computer Science (LNCS)},
volume = {13260},
address = {Caltech, California, USA},
month = {May},
year = {2022},
isbn = {978-3-031-06772-3},
}

@article{HCHJR21,
    author = {Abigail Hammer and Matthew Cauwels and Benjamin Hertz and Phillip Jones and Kristin Yvonne Rozier},
    title = {Integrating Runtime Verification into an Automated UAS Traffic Management System},
    journal = {{Innovations in Systems and Software Engineering: A NASA Journal}},
    publisher = {Springer},
    month = {July},
    year = {2021},
    doi = {10.1007/s11334-021-00407-5},
}

@InProceedings{JJKRZ23,
author="Johannsen, Chris
and Jones, Phillip
and Kempa, Brian
and Rozier, Kristin Yvonne
and Zhang, Pei",
editor="Enea, Constantin
and Lal, Akash",
title={{R2U2 Version 3.0: Re-Imagining a Toolchain for Specification, Resource Estimation, and Optimized Observer Generation for Runtime Verification in Hardware and Software}},
booktitle="Computer Aided Verification",
year="2023",
publisher="Springer Nature Switzerland",
address="Cham",
pages="483--497",
isbn="978-3-031-37709-9"
}

@inproceedings{KZJZR20,
    author = {Brian Kempa and Pei Zhang and Phillip H.~Jones and Joseph Zambreno and Kristin Yvonne Rozier},
    title = {{Embedding Online Runtime Verification for Fault Disambiguation on Robonaut2}},
    booktitle = {{Proceedings of the 18th International Conference on Formal Modeling and Analysis of Timed Systems (FORMATS)}},
    publisher = {Springer},
    series = {{Lecture Notes in Computer Science (LNCS)}},
    address = {Vienna, Austria},
    editors = {Nathalie Bertrand and Nils Jansen},
    url = {http://research.temporallogic.org/papers/KZJZR20.pdf},
    month = {September},
    year = {2020},
    pages = {196-214},
}

@inproceedings{HLR21,
  author       = {Benjamin Hertz and
                  Zachary Luppen and
                  Kristin Yvonne Rozier},
  editor       = {Aaron Dutle and
                  Mariano M. Moscato and
                  Laura Titolo and
                  C{\'{e}}sar A. Mu{\~{n}}oz and
                  Ivan Perez},
  title        = {Integrating Runtime Verification into a Sounding Rocket Control System},
  booktitle    = {{NASA} Formal Methods - 13th International Symposium, {NFM} 2021,
                  Virtual Event, May 24-28, 2021, Proceedings},
  series       = {Lecture Notes in Computer Science},
  volume       = {12673},
  pages        = {151--159},
  publisher    = {Springer},
  year         = {2021},
  doi          = {10.1007/978-3-030-76384-8\_10},
  bibsource    = {dblp computer science bibliography, https://dblp.org}
}

@inproceedings{LLR21,
author = {Zachary A. Luppen and Dae Young Lee and Kristin Yvonne Rozier},
title = {{A Case Study in Formal Specification and Runtime Verification of a CubeSat Communications System}},
booktitle = {{SciTech}},
publisher = {{AIAA}},
address = {Nashville, TN, USA},
month = {January},
year = {2021},
doi = {10.2514/6.2021-0997.c1},
}

@misc{JAXA,
author = {Naoko Okubo and Tsutomu Kobayashi},
title = {{Using R2U2 in JAXA program}},
howpublished = {Electronic correspondence; \url{https://www.esa.int/Enabling_Support/Operations/OPS-SAT}},
month = {November--December},
year = {2020},
note = {Series of emails and zoom call from {JAXA} to {PI} with technical questions about embedding {R2U2} into an autonomous satellite mission with a provable memory bound of 200KB},
}


%% file: refs/mybibfile.bib
@book{DEDSBook,
	Author = {R. Kumar and V.K. Garg},
	Publisher = {Kluwer Academic Publishers},
	Title = {Modeling and Control of Logical Discrete Event Systems},
	Year = {1995}}

@book{CassandrasBook,
	Address = {Boston, MA},
	Author = {C. Cassandras and S. Lafortune},
	Publisher = {Kluwer Academic Publishers},
	Title = {Introduction to discrete event systems},
	Year = {1999}}

@article{Raisch98,
	Author = {J. Raisch and S. D. O'Young},
	Journal = {IEEE Transactions on Automatic Control},
	Month = {April},
	Number = {4},
	Pages = {569-573},
	Title = {Discrete approximation and supervisory control of continuous systems},
	Volume = {43},
	Year = {1998}}

@article{AHLP00,
	Author = {R. Alur and T. Henzinger and G. Lafferriere and G. J. Pappas},
	Journal = {Proceedings of the IEEE},
	Month = {July},
	Number = {7},
	Pages = {971-984},
	Title = {Discrete abstractions of hybrid systems},
	Volume = {88},
	Year = {2000}}

@article{asarin,
	Author = {E. Asarin and O. Maler and A. Pnueli},
	Journal = {Theoretical Computer Science},
	Pages = {35-66},
	Title = {Reachability analysis of dynamical systems having piecewise-constant derivatives},
	Volume = {138},
	Year = {1995}}

@article{majid,
	Author = {M. Zamani and G. Pola and M. Mazo Jr. and P. Tabuada},
	Journal = {IEEE Transaction on Automatic Control},
	Month = {July},
	Number = {7},
	Pages = {1804-1809},
	Title = {Symbolic models for nonlinear control systems without stability assumptions},
	Volume = {57},
	Year = {2012}}

@article{girard2,
	Author = {A. Girard and G. Pola and P. Tabuada},
	Journal = {IEEE Transactions on Automatic Control},
	Month = {January},
	Number = {1},
	Pages = {116-126},
	Title = {Approximately bisimilar symbolic models for incrementally stable switched systems},
	Volume = {55},
	Year = {2010}}

@article{pola,
	Author = {G. Pola and A. Girard and P. Tabuada},
	Journal = {Automatica},
	Month = {October},
	Number = {10},
	Pages = {2508-2516},
	Title = {Approximately bisimilar symbolic models for nonlinear control systems},
	Volume = {44},
	Year = {2008}}

@book{paulo,
	Author = {P. Tabuada},
	Publisher = {Springer},
	Title = {Verification and Control of Hybrid Systems: A symbolic approach},
	Year = {2009}}

@article{girard,
	Author = {A. Girard and G. J. Pappas},
	Journal = {IEEE Transactions on Automatic Control},
	Month = {May},
	Number = {5},
	Pages = {782-798},
	Title = {Approximation metrics for discrete and continuous systems},
	Volume = {25},
	Year = {2007}}

@article{KloetzerBelta08,
	Author = {Kloetzer, M. and Belta, C.},
	Journal = {IEEE Transactions on Automatic Control},
	Number = {1},
	Pages = {287--297},
	Title = {A fully automated framework for control of linear systems from temporal logic specifications},
	Volume = {53},
	Year = {2008}}

@book{milner,
	Author = {R. Milner},
	Publisher = {Prentice-Hall, Inc.},
	Title = {Communication and Concurrency},
	Year = {1989}}

@article{park,
	Author = {D. M. R. Park},
	Journal = {in Proceedings of the 5th GI Conference on Theoretical Computer Science, LNCS},
	Pages = {167-183},
	Title = {Concurrency and automata on infinite sequences},
	Volume = {104},
	Year = {London, 1981}}

@article{MoorRaisch99,
	Author = {Moor, T. and Raisch, J.},
	Journal = {Systems \& Control Letters},
	Number = {3},
	Pages = {157--166},
	Title = {Supervisory control of hybrid systems within a behavioural framework},
	Volume = {38},
	Year = {1999}}

@inproceedings{Thomas95,
	Author = {W. Thomas},
	Booktitle = {Proceedings of the 12th Annual Symposium on Theoretical Aspects of Computer Science},
	Editor = {E. W. Mayr and C. Puech},
	Month = {March},
	Pages = {1-13},
	Publisher = {Springer Berlin Heidelberg},
	Series = {LNCS},
	Title = {On the synthesis of strategies in infinite games},
	Volume = {900},
	Year = {1995}}

@inproceedings{MalerPnueliSifakis95,
	Author = {O. Maler and A. Pnueli and J. Sifakis},
	Booktitle = {Symposium on Theoretical Aspects of Computer Science},
	Editor = {E. W. Mayr and C. Puech},
	Pages = {229-242},
	Publisher = {Springer-Verlag},
	Series = {LNCS},
	Title = {On the synthesis of discrete controllers for timed systems},
	Volume = {900},
	Year = {1995}}

@article{ComputeGames,
	Author = {P. Madhusudan and W. Nam and R. Alur},
	Journal = {Electronic Notes in Theoretical Computer Science},
	Number = {4},
	Title = {Symbolic Computational Techniques for Solving Games},
	Volume = {89},
	Year = {2003}}

@inproceedings{ACZ06,
	Author = {Alur, R. and {\v{C}}ern{\'y}, P. and Zdancewic, S.},
	Booktitle = {Automata, Languages and Programming},
	Pages = {107--118},
	Publisher = {Springer Berlin Heidelberg},
	Title = {Preserving Secrecy Under Refinement},
	Year = {2006}}

@article{Clark10,
  title={Hyperproperties},
  author={Clarkson, Michael R and Schneider, Fred B},
  journal={Journal of Computer Security},
  volume={18},
  number={6},
  pages={1157--1210},
  year={2010}}

@article{gunther2,
	Author = {G. Rei{\ss}ig and A. Weber and M. Rungger},
	Journal = {IEEE Transactions on Automatic Control},
	Number = {4},
	Pages = {1781-1796},
	Title = {Feedback refinement relations for the synthesis of symbolic controllers},
	Volume = {62},
	Year = {2017}}

@article{LTYZ,
	author = {S. Liu and A. Trivedi and X. Yin and \textbf{M. Zamani}},
	journal = {Annual Reviews in Control},
	title = {Secure-by-construction synthesis of cyber-physical systems},
	year = {2022}}

@article{Yinapproximate,
	Author = {Yin, Xiang and Zamani, Majid and Liu, Siyuan},
	Journal = {IEEE Transactions on Automatic Control},
	Number = {4},
	Pages = {1630--1645},
	Title = {On approximate opacity of cyber-physical systems},
	Volume = {66},
	Year = {2021}}

@article{Zhang2018OpacitySimilation,
	Author = {K. Zhang and X. Yin and M. Zamani},
	Journal = {IEEE Transactions on Automatic Control},
	Number = {12},
	Pages = {5116--5123},
	Title = {Opacity of nondeterministic transition systems: A (bi)simulation relation approach},
	Volume = {64},
	Year = {2019}}

@article{majid8,
	Author = {M. Zamani and P. Mohajerin Esfahani and R. Majumdar and A. Abate and J. Lygeros},
	Journal = {IEEE Transactions on Automatic Control, \textbf{Special Issue on Control of Cyber-Physical Systems}},
	Month = {November},
	Number = {12},
	Pages = {3135-3150},
	Title = {Symbolic control of stochastic systems via approximately bisimilar finite abstractions},
	Volume = {59},
	Year = {2014}}

@article{willems,
	Author = {J. C. Willems},
	Journal = {IEEE Control Systems Magazine},
	Number = {6},
	Pages = {46--99},
	Title = {The behavioral approach to open and interconnected systems},
	Volume = {27},
	Year = {2007}}

@article{Lunze99,
	Author = {J. Lunze and B. Nixdorf and J. Schr\"oder},
	Journal = {Automatica},
	Month = {March},
	Pages = {395-406},
	Title = {Deterministic discrete-event representation of continuous-variable systems},
	Volume = {35},
	Year = {1999}}

@book{katoen08,
	Author = {C. Baier and J. P. Katoen},
	Publisher = {The MIT Press},
	Title = {Principles of model checking},
	Year = {2008}}

@inproceedings{DeGia13,
	Author = {De Giacomo, Giuseppe and Vardi, Moshe Y.},
	Booktitle = {23rd International Joint Conference on Artificial Intelligence (IJCAI)},
	Numpages = {7},
	Pages = {854--860},
	Publisher = {AAAI Press},
	Title = {Linear Temporal Logic and Linear Dynamic Logic on Finite Traces},
	Year = {2013}}

@article{yu2016smart,
  title={Smart grids: A cyber--physical systems perspective},
  author={Yu, Xinghuo and Xue, Yusheng},
  journal={Proceedings of the IEEE},
  volume={104},
  number={5},
  pages={1058--1070},
  year={2016}
}

@inproceedings{prajna_safety_2004,
	author = {Prajna, Stephen and Jadbabaie, Ali},
	booktitle = {Hybrid {Systems}: {Computation} and {Control}},
	doi = {10.1007/978-3-540-24743-2\_32},
	language = {en},
	pages = {477--492},
	series = {Lecture {Notes} in {Computer} {Science}},
	title = {Safety {verification} of {hybrid} {systems} {using} {barrier} {certificates}},
	year = {2004}}

@article{Parrilo_2003,
	Author = {Parrilo, Pablo A.},
	Journal = {Mathematical Programming},
	Pages = {293--320},
	Title = {Semidefinite programming relaxations for semialgebraic problems},
	Volume = {96},
	Year = {2003},
	}

@article{murali2023co,
  title={Co-Buchi Barrier Certificates for Discrete-time Dynamical Systems},
  author={Murali, Vishnu and Trivedi, Ashutosh and Zamani, Majid},
  journal={arXiv preprint arXiv:2311.07695},
  year={2023}
}


%% file: refs/ref-Huajie.bib
@article{sinha2024real,
  title={Real-time anomaly detection and reactive planning with large language models},
  author={Sinha, Rohan and Elhafsi, Amine and Agia, Christopher and Foutter, Matthew and Schmerling, Edward and Pavone, Marco},
  journal={arXiv preprint arXiv:2407.08735},
  year={2024}
}

@article{sun2020real,
  title={Real-time fusion network for RGB-D semantic segmentation incorporating unexpected obstacle detection for road-driving images},
  author={Sun, Lei and Yang, Kailun and Hu, Xinxin and Hu, Weijian and Wang, Kaiwei},
  journal={IEEE Robotics and Automation Letters},
  volume={5},
  number={4},
  pages={5558--5565},
  year={2020},
  publisher={IEEE}
}

@article{wang2021radar,
  title={Radar ghost target detection via multimodal transformers},
  author={Wang, LeiChen and Giebenhain, Simon and Anklam, Carsten and Goldluecke, Bastian},
  journal={IEEE Robotics and Automation Letters},
  volume={6},
  number={4},
  pages={7758--7765},
  year={2021},
  publisher={IEEE}
}

@article{qiu2022unsupervised,
  title={Unsupervised Scalable Multimodal Driving Anomaly Detection},
  author={Qiu, Yuning and Misu, Teruhisa and Busso, Carlos},
  journal={IEEE Transactions on Intelligent Vehicles},
  year={2022},
  publisher={IEEE}
}

@article{duonggeneral2024,
  title={A General-Purpose Multi-Modal OOD Detection Framework},
  author={Duong, Viet Quoc and Wu, Qiong and Zhou, Zhengyi and Zavesky, Eric and Hsu, WenLing and Zhao, Han and Shao, Huajie},
  journal={Transactions on Machine Learning Research},
  year ={2024}
}

@article{zhang2024holmes,
  title={Holmes-vad: Towards unbiased and explainable video anomaly detection via multi-modal llm},
  author={Zhang, Huaxin and Xu, Xiaohao and Wang, Xiang and Zuo, Jialong and Han, Chuchu and Huang, Xiaonan and Gao, Changxin and Wang, Yuehuan and Sang, Nong},
  journal={arXiv preprint arXiv:2406.12235},
  year={2024}
}

@inproceedings{zhang2024gpt,
  title={Gpt-4v-ad: Exploring grounding potential of vqa-oriented gpt-4v for zero-shot anomaly detection},
  author={Zhang, Jiangning and He, Haoyang and Chen, Xuhai and Xue, Zhucun and Wang, Yabiao and Wang, Chengjie and Xie, Lei and Liu, Yong},
  booktitle={International Joint Conference on Artificial Intelligence},
  pages={3--16},
  year={2024},
  organization={Springer}
}

@article{ramakrishna2022efficient,
  title={Efficient out-of-distribution detection using latent space of $\beta$-vae for cyber-physical systems},
  author={Ramakrishna, Shreyas and Rahiminasab, Zahra and Karsai, Gabor and Easwaran, Arvind and Dubey, Abhishek},
  journal={ACM Transactions on Cyber-Physical Systems (TCPS)},
  volume={6},
  number={2},
  pages={1--34},
  year={2022},
  publisher={ACM New York, NY}
}

@article{mohammadi2020anomaly,
  title={Anomaly detection in cyber-physical systems using machine learning},
  author={Mohammadi Rouzbahani, Hossein and Karimipour, Hadis and Rahimnejad, Abolfazl and Dehghantanha, Ali and Srivastava, Gautam},
  journal={Handbook of big data privacy},
  pages={219--235},
  year={2020},
  publisher={Springer}
}

@inproceedings{kaur2023codit,
  title={CODiT: Conformal out-of-distribution Detection in time-series data for cyber-physical systems},
  author={Kaur, Ramneet and Sridhar, Kaustubh and Park, Sangdon and Yang, Yahan and Jha, Susmit and Roy, Anirban and Sokolsky, Oleg and Lee, Insup},
  booktitle={Proceedings of the ACM/IEEE 14th International Conference on Cyber-Physical Systems (with CPS-IoT Week 2023)},
  pages={120--131},
  year={2023}
}


%% file: refs/ref-ayan.bib
@article{ackerson1970state,
  title={On state estimation in switching environments},
  author={Ackerson, Guy and Fu, K},
  journal={IEEE transactions on automatic control},
  volume={15},
  number={1},
  pages={10--17},
  year={1970},
  publisher={IEEE}
}

@article{campo1991state,
  title={State estimation for systems with sojourn-time-dependent Markov model switching},
  author={Campo, L and Mookerjee, P and Bar-Shalom, Y},
  journal={IEEE Transactions on Automatic Control},
  volume={36},
  number={2},
  pages={238--243},
  year={1991},
  publisher={IEEE}
}

@article{lecarpentier2019non,
  title={Non-stationary Markov decision processes, a worst-case approach using model-based reinforcement learning},
  author={Lecarpentier, Erwan and Rachelson, Emmanuel},
  journal={Advances in neural information processing systems},
  volume={32},
  year={2019}
}

@article{keplinger2025ns,
  title={NS-Gym: Open-Source Simulation Environments and Benchmarks for Non-Stationary Markov Decision Processes},
  author={Keplinger, Nathaniel S and Luo, Baiting and Bektas, Iliyas and Zhang, Yunuo and Wray, Kyle Hollins and Laszka, Aron and Dubey, Abhishek and Mukhopadhyay, Ayan},
  journal={arXiv preprint arXiv:2501.09646},
  year={2025}
}

@article{luo2024act,
  title={Act as you learn: Adaptive decision-making in non-stationary markov decision processes},
  author={Luo, Baiting and Zhang, Yunuo and Dubey, Abhishek and Mukhopadhyay, Ayan},
  journal={International Conference on Autonomous Agents and Multi-Agent Systems},
  year={2024}
}

@article{chandak2020optimizing,
  title={Optimizing for the Future in Non-Stationary MDPs},
  author={Chandak, Yash and Theocharous, Georgios and Shankar, Shiv and Mahadevan, Sridhar and White, Martha and Thomas, Philip S},
  journal={Thirty-seventh International Conference on Machine Learning (ICML)},
  year={2020}
}

@article{pettet2024decision,
  title={Decision Making in Non-Stationary Environments with Policy-Augmented Search},
  author={Pettet, Ava and Zhang, Yunuo and Luo, Baiting and Wray, Kyle and Baier, Hendrik and Laszka, Aron and Dubey, Abhishek and Mukhopadhyay, Ayan},
  journal={arXiv preprint arXiv:2401.03197},
  year={2024}
}


%% file: refs/ref-chaterji.bib
@inproceedings{agile3D,
author = {Xu, Ran and Lee, Jayoung and Wang, Pengcheng and Bagchi, Saurabh and Li, Yin and Chaterji, Somali},
title = {LiteReconfig: cost and content aware reconfiguration of video object detection systems for mobile GPUs},
year = {2022},
isbn = {9781450391627},
publisher = {Association for Computing Machinery},
address = {New York, NY, USA},
url = {https://doi.org/10.1145/3492321.3519577},
doi = {10.1145/3492321.3519577},
booktitle = {Proceedings of the Seventeenth European Conference on Computer Systems},
pages = {334–351},
numpages = {18},
location = {Rennes, France},
series = {EuroSys '22}
}

@inproceedings{litereconfig,
author = {Xu, Ran and Lee, Jayoung and Wang, Pengcheng and Bagchi, Saurabh and Li, Yin and Chaterji, Somali},
title = {LiteReconfig: cost and content aware reconfiguration of video object detection systems for mobile GPUs},
year = {2022},
isbn = {9781450391627},
publisher = {Association for Computing Machinery},
address = {New York, NY, USA},
url = {https://doi.org/10.1145/3492321.3519577},
doi = {10.1145/3492321.3519577},
booktitle = {Proceedings of the Seventeenth European Conference on Computer Systems},
pages = {334–351},
numpages = {18},
location = {Rennes, France},
series = {EuroSys '22}
}

@inproceedings{SWSEG,
  title={Improving semi-supervised semantic segmentation with sliced-Wasserstein feature alignment and uniformity},
  author={Lu, Chen-Yi and Derakhshandeh, Kasra and Chaterji, Somali},
  booktitle={Proceedings of the Computer Vision and Pattern Recognition Conference},
  pages={20233--20243},
  year={2025}
}

@inproceedings{mahadik2016sarvavid,
  title={Sarvavid: a domain specific language for developing scalable computational genomics applications},
  author={Mahadik, Kanak and Wright, Christopher and Zhang, Jinyi and Kulkarni, Milind and Bagchi, Saurabh and Chaterji, Somali},
  booktitle={Proceedings of the 2016 International Conference on Supercomputing (ICS)},
  pages={1--12},
  year={2016}
}


%% file: refs/ref-fanxinmengyu.bib
@inproceedings{abad2016reset,
  title={Reset-based recovery for real-time cyber-physical systems with temporal safety constraints},
  author={Abad, Fardin Abdi Taghi and Mancuso, Renato and Bak, Stanley and Dantsker, Or and Caccamo, Marco},
  booktitle={2016 IEEE 21st International Conference on Emerging Technologies and Factory Automation (ETFA)},
  pages={1--8},
  year={2016},
  organization={IEEE}
}

@inproceedings{zhang2024fast,
  title={Fast Attack Recovery for Stochastic Cyber-Physical Systems},
  author={Zhang, Lin and Burbano, Luis and Chen, Xin and Cardenas, Alvaro A and Drager, Steven and Anderson, Matthew and Kong, Fanxin},
  booktitle={2024 IEEE 30th Real-Time and Embedded Technology and Applications Symposium (RTAS)},
  pages={280--293},
  year={2024},
  organization={IEEE}
}

@inproceedings{zhang2020real,
  title={Real-time attack-recovery for cyber-physical systems using linear approximations},
  author={Zhang, Lin and Chen, Xin and Kong, Fanxin and Cardenas, Alvaro A},
  booktitle={2020 IEEE Real-Time Systems Symposium (RTSS)},
  pages={205--217},
  year={2020},
  organization={IEEE}
}

@article{zhang2021real,
  title={Real-time attack-recovery for cyber-physical systems using linear-quadratic regulator},
  author={Zhang, Lin and Lu, Pengyuan and Kong, Fanxin and Chen, Xin and Sokolsky, Oleg and Lee, Insup},
  journal={ACM Transactions on Embedded Computing Systems (TECS)},
  volume={20},
  number={5s},
  pages={1--24},
  year={2021},
  publisher={ACM New York, NY}
}

@inproceedings{liu2023learn,
  title={Learn-to-respond: Sequence-predictive recovery from sensor attacks in cyber-physical systems},
  author={Liu, Mengyu and Zhang, Lin and Phoha, Vir V and Kong, Fanxin},
  booktitle={2023 IEEE Real-Time Systems Symposium (RTSS)},
  pages={78--91},
  year={2023},
  organization={IEEE}
}

@article{lu2024recovery,
  title={Recovery from adversarial attacks in cyber-physical systems: Shallow, deep, and exploratory works},
  author={Lu, Pengyuan and Zhang, Lin and Liu, Mengyu and Sridhar, Kaustubh and Sokolsky, Oleg and Kong, Fanxin and Lee, Insup},
  journal={ACM Computing Surveys},
  volume={56},
  number={8},
  pages={1--31},
  year={2024},
  publisher={ACM New York, NY}
}


%% file: refs/ref-glen.bib
@inproceedings{DBLP:conf/icra/PanCB23,
  author       = {Jiayi Pan and
                  Glen Chou and
                  Dmitry Berenson},
  title        = {Data-Efficient Learning of Natural Language to Linear Temporal Logic
                  Translators for Robot Task Specification},
  booktitle    = {{IEEE} International Conference on Robotics and Automation, {ICRA}
                  2023, London, UK, May 29 - June 2, 2023},
  pages        = {11554--11561},
  publisher    = {{IEEE}},
  year         = {2023},
  url          = {https://doi.org/10.1109/ICRA48891.2023.10161125},
  doi          = {10.1109/ICRA48891.2023.10161125},
  bibsource    = {dblp computer science bibliography, https://dblp.org}
}

@article{DBLP:journals/ijrr/ChouBO21,
  author       = {Glen Chou and
                  Dmitry Berenson and
                  Necmiye Ozay},
  title        = {Learning constraints from demonstrations with grid and parametric
                  representations},
  journal      = {Int. J. Robotics Res.},
  volume       = {40},
  number       = {10-11},
  year         = {2021},
  url          = {https://doi.org/10.1177/02783649211035177},
  doi          = {10.1177/02783649211035177},
  bibsource    = {dblp computer science bibliography, https://dblp.org}
}

@inproceedings{DBLP:conf/corl/ChouBO20,
  author       = {Glen Chou and
                  Dmitry Berenson and
                  Necmiye Ozay},
  editor       = {Jens Kober and
                  Fabio Ramos and
                  Claire J. Tomlin},
  title        = {Uncertainty-Aware Constraint Learning for Adaptive Safe Motion Planning
                  from Demonstrations},
  booktitle    = {4th Conference on Robot Learning, CoRL 2020, 16-18 November 2020,
                  Virtual Event / Cambridge, MA, {USA}},
  series       = {Proceedings of Machine Learning Research},
  volume       = {155},
  pages        = {1612--1639},
  publisher    = {{PMLR}},
  year         = {2020},
  url          = {https://proceedings.mlr.press/v155/chou21a.html},
  bibsource    = {dblp computer science bibliography, https://dblp.org}
}

@inproceedings{DBLP:conf/wafr/ChouOB22,
  author       = {Glen Chou and
                  Necmiye Ozay and
                  Dmitry Berenson},
  title        = {Safe Output Feedback Motion Planning from Images via Learned Perception
                  Modules and Contraction Theory},
  booktitle    = {Algorithmic Foundations of Robotics {XV} - Proceedings of the Fifteenth
                  Workshop on the Algorithmic Foundations of Robotics, {WAFR} 2022,
                  College Park, MD, USA, 22-24 June, 2022},
  series       = {Springer Proceedings in Advanced Robotics},
  volume       = {25},
  pages        = {349--367},
  publisher    = {Springer},
  year         = {2022}
}


%% file: refs/ref-homa.bib
@techreport{NIST_SP_1500_202_2017,
  author       = {{Cyber‑Physical Systems Public Working Group}},
  title        = {Framework for Cyber‑Physical Systems: Volume 2, Working Group Reports},
  institution  = {National Institute of Standards and Technology (NIST), U.S. Department of Commerce},
  type         = {NIST Special Publication},
  number       = {1500‑202},
  version      = {1.0},
  month        = jun,
  year         = {2017},
  url          = {https://doi.org/10.6028/NIST.SP.1500-202},
  note         = {163 pp.}
}

@inproceedings{chen2022stl,
  title={An STL-based formulation of resilience in cyber-physical systems},
  author={Chen, Hongkai and Lin, Shan and Smolka, Scott A and Paoletti, Nicola},
  booktitle={International Conference on Formal Modeling and Analysis of Timed Systems},
  pages={117--135},
  year={2022},
  organization={Springer}
}

@book{hollnagel2006resilience,
  title={Resilience engineering: Concepts and precepts},
  author={Hollnagel, Erik and Woods, David D and Leveson, Nancy},
  year={2006},
  publisher={Ashgate Publishing, Ltd.}
}

@article{leveson2004new,
  title={A new accident model for engineering safer systems},
  author={Leveson, Nancy},
  journal={Safety science},
  volume={42},
  number={4},
  pages={237--270},
  year={2004},
  publisher={Elsevier}
}


%% file: refs/ref-hyunseung.bib
@article{segovia2024survey,
  title={A survey on cyber-resilience approaches for cyber-physical systems},
  author={Segovia-Ferreira, Mariana and Rubio-Hernan, Jose and Cavalli, Ana and Garcia-Alfaro, Joaquin},
  journal={ACM Computing Surveys},
  volume={56},
  number={8},
  pages={1--37},
  year={2024},
  publisher={ACM New York, NY}
}

@article{ratasich2019roadmap,
  title={A roadmap toward the resilient internet of things for cyber-physical systems},
  author={Ratasich, Denise and Khalid, Faiq and Geissler, Florian and Grosu, Radu and Shafique, Muhammad and Bartocci, Ezio},
  journal={IEEE Access},
  volume={7},
  pages={13260--13283},
  year={2019},
  publisher={IEEE}
}

@article{cassottana2023resilience,
  title={Resilience analysis of cyber-physical systems: A review of models and methods},
  author={Cassottana, Beatrice and Roomi, Muhammad M and Mashima, Daisuke and Sansavini, Giovanni},
  journal={Risk Analysis},
  volume={43},
  number={11},
  pages={2359--2379},
  year={2023},
  publisher={Wiley Online Library}
}

@article{kim2022survey,
  title={A survey on network security for cyber--physical systems: From threats to resilient design},
  author={Kim, Sangjun and Park, Kyung-Joon and Lu, Chenyang},
  journal={IEEE Communications Surveys \& Tutorials},
  volume={24},
  number={3},
  pages={1534--1573},
  year={2022},
  publisher={IEEE}
}

@article{yu2023survey,
  title={A survey on cyber--physical systems security},
  author={Yu, Zhenhua and Gao, Hongxia and Cong, Xuya and Wu, Naiqi and Song, Houbing Herbert},
  journal={IEEE Internet of Things Journal},
  volume={10},
  number={24},
  pages={21670--21686},
  year={2023},
  publisher={IEEE}
}


%% file: refs/ref-ivan.bib
@inproceedings{ha_world_2018,
	title = {World {Models}},
	url = {http://arxiv.org/abs/1803.10122},
	doi = {10.5281/zenodo.1207631},
	urldate = {2022-12-06},
	booktitle = {Proc. of {NeurIPS}},
	author = {Ha, David and Schmidhuber, Jürgen},
	month = mar,
	year = {2018},
	note = {arXiv:1803.10122 [cs, stat]},
}

@article{katz_verification_2022,
	title = {Verification of {Image}-{Based} {Neural} {Network} {Controllers} {Using} {Generative} {Models}},
	volume = {19},
	issn = {1940-3151},
	url = {https://doi.org/10.2514/1.I011071},
	doi = {10.2514/1.I011071},
	number = {9},
	urldate = {2023-01-04},
	journal = {Journal of Aerospace Information Systems},
	author = {Katz, Sydney M. and Corso, Anthony L. and Strong, Christopher A. and Kochenderfer, Mykel J.},
	year = {2022},
	note = {Publisher: American Institute of Aeronautics and Astronautics
\_eprint: https://doi.org/10.2514/1.I011071},
	pages = {574--584},
}

@inproceedings{luo_dynamic_2023,
	address = {New York, NY, USA},
	series = {{ICCPS} '23},
	title = {Dynamic {Simplex}: {Balancing} {Safety} and {Performance} in {Autonomous} {Cyber} {Physical} {Systems}},
	isbn = {979-8-4007-0036-1},
	shorttitle = {Dynamic {Simplex}},
	url = {https://dl.acm.org/doi/10.1145/3576841.3585934},
	doi = {10.1145/3576841.3585934},
	urldate = {2023-05-12},
	booktitle = {Proceedings of the {ACM}/{IEEE} 14th {International} {Conference} on {Cyber}-{Physical} {Systems} (with {CPS}-{IoT} {Week} 2023)},
	publisher = {Association for Computing Machinery},
	author = {Luo, Baiting and Ramakrishna, Shreyas and Pettet, Ava and Kuhn, Christopher and Karsai, Gabor and Mukhopadhyay, Ayan},
	month = may,
	year = {2023},
	pages = {177--186},
}

@inproceedings{acharya_competency_2022,
	title = {Competency {Assessment} for {Autonomous} {Agents} using {Deep} {Generative} {Models}},
	doi = {10.1109/IROS47612.2022.9981991},
	booktitle = {2022 {IEEE}/{RSJ} {International} {Conference} on {Intelligent} {Robots} and {Systems} ({IROS})},
	author = {Acharya, Aastha and Russell, Rebecca and Ahmed, Nisar R.},
	month = oct,
	year = {2022},
	note = {ISSN: 2153-0866},
	pages = {8211--8218},
}

@inproceedings{mao_how_2024,
	title = {How {Safe} {Am} {I} {Given} {What} {I} {See}? {Calibrated} {Prediction} of {Safety} {Chances} for {Image}-{Controlled} {Autonomy}},
	shorttitle = {How {Safe} {Am} {I} {Given} {What} {I} {See}?},
	url = {http://arxiv.org/abs/2308.12252},
	doi = {10.48550/arXiv.2308.12252},
	urldate = {2023-08-29},
	booktitle = {Proc. of the {Annual} {Conference} on {Learning} for {Dynamics} and {Control} ({L4DC})},
	author = {Mao, Zhenjiang and Sobolewski, Carson and Ruchkin, Ivan},
	year = {2024},
	note = {arXiv:2308.12252 [cs]},
}

@inproceedings{geng_bridging_2024,
	address = {Milano, Italy},
	title = {Bridging {Dimensions}: {Confident} {Reachability} for {High}-{Dimensional} {Controllers}},
	shorttitle = {Bridging {Dimensions}},
	doi = {10.48550/arXiv.2311.04843},
	urldate = {2023-11-09},
	booktitle = {Proc. of the {International} {Symposium} on {Formal} {Methods} ({FM})},
	author = {Geng, Yuang and Baldauf, Jake Brandon and Dutta, Souradeep and Huang, Chao and Ruchkin, Ivan},
	year = {2024},
	note = {arXiv:2311.04843 [cs]},
}

@article{wu_toward_2023,
	title = {Toward {Certified} {Robustness} {Against} {Real}-{World} {Distribution} {Shifts}},
	url = {https://ieeexplore.ieee.org/document/10136136/},
	doi = {10.1109/SaTML54575.2023.00042},
	urldate = {2024-01-03},
	journal = {2023 IEEE Conference on Secure and Trustworthy Machine Learning (SaTML)},
	author = {Wu, Haoze and Tagomori, Teruhiro and Robey, Alexander and Yang, Fengjun and Matni, Nikolai and Pappas, George and Hassani, Hamed and Pasareanu, Corina and Barrett, Clark},
	month = feb,
	year = {2023},
	note = {Conference Name: 2023 IEEE Conference on Secure and Trustworthy Machine Learning (SaTML)
ISBN: 9781665462990
Place: Raleigh, NC, USA
Publisher: IEEE},
	pages = {537--553},
}

@inproceedings{xie_neuro-symbolic_2022,
	title = {Neuro-{Symbolic} {Verification} of {Deep} {Neural} {Networks}},
	doi = {10.48550/ARXIV.2203.00938},
	urldate = {2024-01-03},
	booktitle = {Proc. of {IJCAI}-22},
	author = {Xie, Xuan and Kersting, Kristian and Neider, Daniel},
	year = {2022},
}

@inproceedings{pasareanu_closed-loop_2023,
	address = {Cham},
	series = {Lecture {Notes} in {Computer} {Science}},
	title = {Closed-{Loop} {Analysis} of {Vision}-{Based} {Autonomous} {Systems}: {A} {Case} {Study}},
	isbn = {978-3-031-37706-8},
	shorttitle = {Closed-{Loop} {Analysis} of {Vision}-{Based} {Autonomous} {Systems}},
	doi = {10.1007/978-3-031-37706-8_15},
	language = {en},
	booktitle = {Computer {Aided} {Verification}},
	publisher = {Springer Nature Switzerland},
	author = {Pasareanu, Corina S. and Mangal, Ravi and Gopinath, Divya and Getir Yaman, Sinem and Imrie, Calum and Calinescu, Radu and Yu, Huafeng},
	year = {2023},
	pages = {289--303},
}

@inproceedings{waite_state-dependent_2025,
	address = {Philadelphia, PA, USA},
	title = {State-{Dependent} {Conformal} {Perception} {Bounds} for {Neuro}-{Symbolic} {Verification} of {Autonomous} {Systems}},
	url = {http://arxiv.org/abs/2502.21308},
	doi = {10.48550/arXiv.2502.21308},
	urldate = {2025-03-04},
	booktitle = {Proc. of 2nd {International} {Conference} on {Neuro}-symbolic {Systems} ({NeuS})},
	publisher = {PMLR},
	author = {Waite, Thomas and Geng, Yuang and Turnquist, Trevor and Ruchkin, Ivan and Ivanov, Radoslav},
	month = feb,
	year = {2025},
}

@inproceedings{peper_four_2025,
	address = {Philadelphia, PA, USA},
	title = {Four {Principles} for {Physically} {Interpretable} {World} {Models}},
	doi = {10.48550/arXiv.2503.02143},
	urldate = {2025-03-05},
	booktitle = {Proc. of 2nd {International} {Conference} on {Neuro}-symbolic {Systems} ({NeuS})},
	publisher = {PMLR},
	author = {Peper, Jordan and Mao, Zhenjiang and Geng, Yuang and Pan, Siyuan and Ruchkin, Ivan},
	year = {2025},
}

@incollection{mitra_formal_2025,
	address = {Cham},
	title = {Formal {Verification} {Techniques} for {Vision}-{Based} {Autonomous} {Systems} – {A} {Survey}},
	isbn = {978-3-031-75778-5},
	url = {https://doi.org/10.1007/978-3-031-75778-5_5},
	language = {en},
	urldate = {2025-04-15},
	booktitle = {Principles of {Verification}: {Cycling} the {Probabilistic} {Landscape} : {Essays} {Dedicated} to {Joost}-{Pieter} {Katoen} on the {Occasion} of {His} 60th {Birthday}, {Part} {III}},
	publisher = {Springer Nature Switzerland},
	author = {Mitra, Sayan and Păsăreanu, Corina and Prabhakar, Pavithra and Seshia, Sanjit A. and Mangal, Ravi and Li, Yangge and Watson, Christopher and Gopinath, Divya and Yu, Huafeng},
	editor = {Jansen, Nils and Junges, Sebastian and Kaminski, Benjamin Lucien and Matheja, Christoph and Noll, Thomas and Quatmann, Tim and Stoelinga, Mariëlle and Volk, Matthias},
	year = {2025},
	doi = {10.1007/978-3-031-75778-5_5},
	pages = {89--108},
}

@misc{cleaveland_conservative_2025,
	title = {Conservative {Perception} {Models} for {Probabilistic} {Verification}},
	url = {http://arxiv.org/abs/2503.18077},
	doi = {10.48550/arXiv.2503.18077},
	urldate = {2025-04-16},
	publisher = {arXiv},
	author = {Cleaveland, Matthew and Lu, Pengyuan and Sokolsky, Oleg and Lee, Insup and Ruchkin, Ivan},
	month = apr,
	year = {2025},
	note = {arXiv:2503.18077 [cs]
version: 2},
}

@inproceedings{peper_towards_2025,
	address = {Cham},
	title = {Towards {Unified} {Probabilistic} {Verification} and {Validation} of {Vision}-{Based} {Autonomy}},
	isbn = {978-3-032-08707-2},
	doi = {10.1007/978-3-032-08707-2_11},
	language = {en},
	booktitle = {Automated {Technology} for {Verification} and {Analysis}},
	publisher = {Springer Nature Switzerland},
	author = {Peper, Jordan and Miao, Yan and Mitra, Sayan and Ruchkin, Ivan},
	year = {2025},
	pages = {231--259},
}


%% file: refs/ref-lemmon.bib
@article{pare2007systematic,
	author = {Par{\'e}, Guy and Jaana, Mirou and Sicotte, Claude},
	journal = {Journal of the American Medical Informatics Association},
	number = {3},
	pages = {269--277},
	title = {Systematic review of home telemonitoring for chronic diseases: the evidence base},
	volume = {14},
	year = {2007}}

@article{lee2011challenges,
	author = {Lee, Insup and Sokolsky, Oleg and Chen, Sanjian and Hatcliff, John and Jee, Eunkyoung and Kim, BaekGyu and King, Andrew and Mullen-Fortino, Margaret and Park, Soojin and Roederer, Alexander and others},
	journal = {Proceedings of the IEEE},
	number = {1},
	pages = {75--90},
	title = {Challenges and research directions in medical cyber--physical systems},
	volume = {100},
	year = {2011}}

@article{varma2023remote,
	author = {Varma, Niraj and Braunschweig, Frieder and Burri, Haran and Hindricks, Gerhard and Linz, Dominik and Michowitz, Yoav and Ricci, Renato Pietro and Nielsen, Jens Cosedis},
	journal = {Europace},
	number = {9},
	pages = {euad233},
	publisher = {Oxford University Press US},
	title = {Remote monitoring of cardiac implantable electronic devices and disease management},
	volume = {25},
	year = {2023}}

@article{padros2023smart,
	author = {Padr{\'o}s, Manuel Ram{\'o}n Castillo and Pastor, Nuria and Paracolls, J{\'u}lia Altarriba and Pe{\~n}a, Marcelino Mosquera and Pergolizzi, Denise and Verg{\`e}s, {\`A}ngels Salvador},
	journal = {JMIR formative research},
	number = {1},
	pages = {e45654},
	publisher = {JMIR Publications Inc., Toronto, Canada},
	title = {A smart system for remote monitoring of patients in palliative care (humanITcare platform): mixed methods study},
	volume = {7},
	year = {2023}}

@article{atilgan2021remote,
	author = {Atilgan, K{\i}van{\c{c}} and Onuk, Burak E and K{\"o}ksal Co{\c{s}}kun, P{\i}nar and Ye{\c{s}}il, Fahri G and Aslan, Cemal and {\c{C}}olak, Abdullah and {\c{C}}elebi, Aks{\"u}yek S and Bozba{\c{s}}, H{\"u}seyin},
	journal = {Journal of Cardiac Surgery},
	number = {11},
	pages = {4226--4234},
	publisher = {Wiley Online Library},
	title = {Remote patient monitoring after cardiac surgery: the utility of a novel telemedicine system},
	volume = {36},
	year = {2021}}

@article{salehi2020assessment,
	author = {Salehi, Sahar and Olyaeemanesh, Alireza and Mobinizadeh, Mohammadreza and Nasli-Esfahani, Ensieh and Riazi, Hossein},
	journal = {Journal of Diabetes \& Metabolic Disorders},
	pages = {115--127},
	publisher = {Springer},
	title = {Assessment of remote patient monitoring (RPM) systems for patients with type 2 diabetes: a systematic review and meta-analysis},
	volume = {19},
	year = {2020}}

@inproceedings{10.1145/1837274.1837463,
	address = {New York, NY, USA},
	author = {Lee, Insup and Sokolsky, Oleg},
	booktitle = {Proceedings of the 47th Design Automation Conference},
	doi = {10.1145/1837274.1837463},
	isbn = {9781450300025},
	location = {Anaheim, California},
	numpages = {6},
	pages = {743--748},
	publisher = {Association for Computing Machinery},
	series = {DAC '10},
	title = {Medical cyber physical systems},
	url = {https://doi.org/10.1145/1837274.1837463},
	year = {2010}}

@inproceedings{montestruque2015globally,
  title={Globally coordinated distributed storm water management system},
  author={Montestruque, Luis and Lemmon, MD},
  booktitle={Proceedings of the 1st ACM International Workshop on Cyber-Physical Systems for Smart Water Networks},
  pages={1--6},
  year={2015}
}

@INPROCEEDINGS{7535338,
  author={Möller, Dietmar P.F. and Vakilzadian, Hamid},
  booktitle={2016 IEEE International Conference on Electro Information Technology (EIT)}, 
  title={Cyber-physical systems in smart transportation}, 
  year={2016},
  volume={},
  number={},
  pages={0776-0781},
  doi={10.1109/EIT.2016.7535338}}


%% file: refs/ref-meiyi.bib
@article{deguchi2020smart,
  title={From smart city to society 5.0},
  author={Deguchi, Atsushi and others},
  journal={Society},
  volume={5},
  pages={43--65},
  year={2020}
}

@article{tian2019smart,
  title={Smart healthcare: making medical care more intelligent},
  author={Tian, Shuo and Yang, Wenbo and Le Grange, Jehane Michael and Wang, Peng and Huang, Wei and Ye, Zhewei},
  journal={Global Health Journal},
  volume={3},
  number={3},
  pages={62--65},
  year={2019},
  publisher={Elsevier}
}

@article{zheng2018smart,
  title={Smart manufacturing systems for Industry 4.0: Conceptual framework, scenarios, and future perspectives},
  author={Zheng, Pai and Wang, Honghui and Sang, Zhiqian and Zhong, Ray Y and Liu, Yongkui and Liu, Chao and Mubarok, Khamdi and Yu, Shiqiang and Xu, Xun},
  journal={Frontiers of Mechanical Engineering},
  volume={13},
  pages={137--150},
  year={2018},
  publisher={Springer}
}

@article{arrieta2020explainable,
  title={Explainable Artificial Intelligence (XAI): Concepts, taxonomies, opportunities and challenges toward responsible AI},
  author={Arrieta, Alejandro Barredo and D{\'\i}az-Rodr{\'\i}guez, Natalia and Del Ser, Javier and Bennetot, Adrien and Tabik, Siham and Barbado, Alberto and Garc{\'\i}a, Salvador and Gil-L{\'o}pez, Sergio and Molina, Daniel and Benjamins, Richard and others},
  journal={Information fusion},
  volume={58},
  pages={82--115},
  year={2020},
  publisher={Elsevier}
}

@inproceedings{sreedharan2020emerging,
  title={The emerging landscape of explainable automated planning \& decision making},
  author={Sreedharan, S and Chakraborti, T and Kambhampati, S},
  year={2020},
  organization={IJCAI}
}

@article{fox2017explainable,
  title={Explainable planning},
  author={Fox, Maria and Long, Derek and Magazzeni, Daniele},
  journal={arXiv preprint arXiv:1709.10256},
  year={2017}
}

@article{boggess2022toward,
  title={Toward policy explanations for multi-agent reinforcement learning},
  author={Boggess, Kayla and Kraus, Sarit and Feng, Lu},
  journal={arXiv preprint arXiv:2204.12568},
  year={2022}
}

@inproceedings{langley2016explainable,
  title={Explainable agency in human-robot interaction},
  author={Langley, Pat},
  booktitle={AAAI fall symposium series},
  year={2016}
}

@article{chakraborti2017plan,
  title={Plan explanations as model reconciliation: Moving beyond explanation as soliloquy},
  author={Chakraborti, Tathagata and Sreedharan, Sarath and Zhang, Yu and Kambhampati, Subbarao},
  journal={arXiv preprint arXiv:1701.08317},
  year={2017}
}

@inproceedings{kocsis2006bandit,
  title={Bandit based monte-carlo planning},
  author={Kocsis, Levente and Szepesv{\'a}ri, Csaba},
  booktitle={European conference on machine learning},
  pages={282--293},
  year={2006},
  organization={Springer}
}

@article{ma2019data,
  title={Data sets, modeling, and decision making in smart cities: A survey},
  author={Ma, Meiyi and Preum, Sarah and Ahmed, Mohsin and T{\"a}rneberg, William and Hendawi, Abdeltawab and Stankovic, John },
  journal={ACM Transactions on Cyber-Physical Systems},
  volume={4},
  number={2},
  pages={1--28},
  year={2019},
  publisher={ACM},
  doi={10.1145/3355283}
}

@article{ma2021predictive,
  title={Predictive monitoring with logic-calibrated uncertainty for cyber-physical systems},
  author={Ma, Meiyi and Stankovic, John and Bartocci, Ezio and Feng, Lu},
  journal={ACM Transactions on Embedded Computing Systems},
  volume={20},
  number={5s},
  pages={1--25},
  year={2021},
  publisher={ACM New York, NY}
}


%% file: refs/ref-panagou.bib
@ARTICLE{DAgrawal_TRO24,
  author={Agrawal, Devansh Ramgopal and Chen, Ruichang and Panagou, Dimitra},
  journal={IEEE Transactions on Robotics}, 
  title={gatekeeper: Online Safety Verification and Control for Nonlinear Systems in Dynamic Environments}, 
  year={2024},
  volume={40},
  number={},
  pages={4358-4375},
  doi={10.1109/TRO.2024.3454415}}

@article{cherenson2025autonomy,
  title={Autonomy Architectures for Safe Planning in Unknown Environments Under Budget Constraints},
  author={Cherenson, Daniel M and Agrawal, Devansh R and Panagou, Dimitra},
  journal={arXiv preprint arXiv:2504.03001},
  year={2025}
}

@inproceedings{kim2024learning,
  title={Learning to Refine Input Constrained Control Barrier Functions via Uncertainty-Aware Online Parameter Adaptation},
  author={Kim, Taekyung and Kee, Robin Inho and Panagou, Dimitra},
  booktitle={2025 IEEE International Conference on Robotics and Automation},
  year={2025}
}

@article{naveed2024mesch,
  title={meSch: Multi-Agent Energy-Aware Scheduling for Task Persistence},
  author={Naveed, Kaleb Ben and Dang, An and Kumar, Rahul and Panagou, Dimitra},
  journal={arXiv preprint arXiv:2406.04560, to be presented in IROS 2025},
  year={2024}
}

@inproceedings{black2023safety,
  title={Safety under uncertainty: Tight bounds with risk-aware control barrier functions},
  author={Black, Mitchell and Fainekos, Georgios and Hoxha, Bardh and Prokhorov, Danil and Panagou, Dimitra},
  booktitle={2023 IEEE International Conference on Robotics and Automation (ICRA)},
  pages={12686--12692},
  year={2023},
  organization={IEEE}
}

@article{garg2024advances,
  title={Advances in the theory of control barrier functions: Addressing practical challenges in safe control synthesis for autonomous and robotic systems},
  author={Garg, Kunal and Usevitch, James and Breeden, Joseph and Black, Mitchell and Agrawal, Devansh and Parwana, Hardik and Panagou, Dimitra},
  journal={Annual Reviews in Control},
  volume={57},
  pages={100945},
  year={2024},
  publisher={Pergamon}
}

@inproceedings{wang2025safe,
  title={Safe Navigation in Uncertain Crowded Environments Using Risk Adaptive CVaR Barrier Functions},
  author={Wang, Xinyi and Kim, Taekyung and Hoxha, Bardh and Fainekos, Georgios and Panagou, Dimitra},
  booktitle={2025 International Conference on Intelligent Robots and Systems},
  year={2025}
}

@inproceedings{black2021fixed,
  title={A fixed-time stable adaptation law for safety-critical control under parametric uncertainty},
  author={Black, Mitchell and Arabi, Ehsan and Panagou, Dimitra},
  booktitle={2021 European Control Conference (ECC)},
  pages={1328--1333},
  year={2021},
  organization={IEEE}
}

@inproceedings{agrawal2021safe,
  title={Safe control synthesis via input constrained control barrier functions},
  author={Agrawal, Devansh R and Panagou, Dimitra},
  booktitle={2021 60th IEEE Conference on Decision and Control},
  pages={6113--6118},
  year={2021}
}

@article{kim2024visibility,
  title={Visibility-Aware RRT* for Safety-Critical Navigation of Perception-Limited Robots in Unknown Environments},
  author={Kim, Taekyung and Panagou, Dimitra},
  journal={arXiv preprint arXiv:2406.07728, accepted in RA-L, in press; will be presented in IROS 2025},
  year={2024}
}

@inproceedings{breeden2021high,
  title={High relative degree control barrier functions under input constraints},
  author={Breeden, Joseph and Panagou, Dimitra},
  booktitle={2021 60th IEEE Conference on Decision and Control},
  pages={6119--6124},
  year={2021},
  organization={IEEE}
}

@article{agrawal2021constructive,
  title={A constructive method for designing safe multirate controllers for differentially-flat systems},
  author={Agrawal, Devansh R and Parwana, Hardik and Cosner, Ryan K and Rosolia, Ugo and Ames, Aaron D and Panagou, Dimitra},
  journal={IEEE Control Systems Letters},
  volume={6},
  pages={2138--2143},
  year={2021}
}

@article{usevitch2021resilient,
  title={Resilient trajectory propagation in multirobot networks},
  author={Usevitch, James and Panagou, Dimitra},
  journal={IEEE Trans. on Robotics},
  volume={38},
  number={1},
  pages={42--56},
  year={2021}
}

@inproceedings{lee2024maintaining,
  title={Maintaining Strong $ r $-Robustness in Reconfigurable Multi-Robot Networks using Control Barrier Functions},
  author={Lee, Haejoon and Panagou, Dimitra},
  booktitle={2025 IEEE International Conference on Robotics and Automation},
  year={2025}
}

@article{PIRANI2023111264,
title = {Graph-theoretic approaches for analyzing the resilience of distributed control systems: A tutorial and survey},
journal = {Automatica},
volume = {157},
pages = {111264},
year = {2023},
issn = {0005-1098},
doi = {https://doi.org/10.1016/j.automatica.2023.111264},
url = {https://www.sciencedirect.com/science/article/pii/S0005109823004259},
author = {Mohammad Pirani and Aritra Mitra and Shreyas Sundaram}
}

@article{LeBlanc2013,
  title={Resilient asymptotic consensus in robust networks},
  author={LeBlanc, Heath J and Zhang, Haotian and Koutsoukos, Xenofon and Sundaram, Shreyas},
  journal={IEEE Journal on Selected Areas in Communications},
  volume={31},
  number={4},
  pages={766--781},
  year={2013},
  publisher={IEEE}
}


%% file: refs/ref-sample-base.bib
@String{Computing = "Computing" }

@String{Computer = "{IEEE} Computer" }

@String{Academic = "Academic Press" }

@String{Chelsea = "Chelsea" }

@String{Springer = "Springer-Verlag" }

@BOOK{test,
   author = "Donald E. Knuth",
   title = "Seminumerical Algorithms",
   volume = 2,
   series = "The Art of Computer Programming",
   publisher = "Addison-Wesley",
   address = "Reading, MA",
   edition = "2nd",
   month = "10~" # jan,
   year = "1981",
}

@ArtifactSoftware{R,
    title = {R: A Language and Environment for Statistical Computing},
    author = {{R Core Team}},
    organization = {R Foundation for Statistical Computing},
    address = {Vienna, Austria},
    year = {2019},
    url = {https://www.R-project.org/},
}


%% file: refs/ref-saurabh.bib
@article{rana2023connected,
  title={Connected and autonomous vehicles and infrastructures: A literature review},
  author={Rana, Md Masud and Hossain, Kamal},
  journal={International Journal of Pavement Research and Technology},
  volume={16},
  number={2},
  pages={264--284},
  year={2023},
  publisher={Springer}
}

@inproceedings{zhang2021emp,
  title={Emp: Edge-assisted multi-vehicle perception},
  author={Zhang, Xumiao and Zhang, Anlan and Sun, Jiachen and Zhu, Xiao and Guo, Y Ethan and Qian, Feng and Mao, Z Morley},
  booktitle={Proceedings of the 27th Annual International Conference on Mobile Computing and Networking (Mobicom)},
  pages={545--558},
  year={2021}
}

@inproceedings{qiu2022autocast,
  title={AutoCast: scalable infrastructure-less cooperative perception for distributed collaborative driving},
  author={Qiu, Hang and Huang, Po-Han and Asavisanu, Namo and Liu, Xiaochen and Psounis, Konstantinos and Govindan, Ramesh},
  booktitle={Proceedings of the 20th Annual International Conference on Mobile Systems, Applications and Services (MobiSys)},
  pages={128--141},
  year={2022}
}

@inproceedings{shi2022vips,
  title={VIPS: Real-time perception fusion for infrastructure-assisted autonomous driving},
  author={Shi, Shuyao and Cui, Jiahe and Jiang, Zhehao and Yan, Zhenyu and Xing, Guoliang and Niu, Jianwei and Ouyang, Zhenchao},
  booktitle={Proceedings of the 28th annual international conference on mobile computing and networking (MobiCom)},
  pages={133--146},
  year={2022}
}

@inproceedings{he2021vi,
  title={VI-eye: semantic-based 3D point cloud registration for infrastructure-assisted autonomous driving},
  author={He, Yuze and Ma, Li and Jiang, Zhehao and Tang, Yi and Xing, Guoliang},
  booktitle={Proceedings of the 27th Annual International Conference on Mobile Computing and Networking (MobiCom)},
  pages={573--586},
  year={2021}
}

@article{tang2025towards,
  title={Towards general industrial intelligence: A survey of large models as a service in industrial IoT},
  author={Tang, Jianhua and Chen, Jiao and He, Jiayi and Chen, Fangfang and Lv, Zuohong and Han, Guangjie and Liu, Zuozhu and Yang, Howard H and Li, Weihua},
  journal={IEEE Communications Surveys \& Tutorials},
  year={2025},
  publisher={IEEE}
}

@article{munir2023neuro,
  title={Neuro-symbolic explainable artificial intelligence twin for zero-touch IoE in wireless network},
  author={Munir, Md Shirajum and Kim, Ki Tae and Adhikary, Apurba and Saad, Walid and Shetty, Sachin and Park, Seong-Bae and Hong, Choong Seon},
  journal={IEEE Internet of Things Journal},
  volume={10},
  number={24},
  pages={22451--22468},
  year={2023},
  publisher={IEEE}
}

@article{lu2024surveying,
  title={Surveying neuro-symbolic approaches for reliable artificial intelligence of things},
  author={Lu, Zhen and Afridi, Imran and Kang, Hong Jin and Ruchkin, Ivan and Zheng, Xi},
  journal={Journal of Reliable Intelligent Environments},
  volume={10},
  number={3},
  pages={257--279},
  year={2024},
  publisher={Springer}
}


%% file: refs/ref-sibin.bib
@inproceedings{resecure,
  author    = {Abdi, Fardin and Chen, Chien-Ying and Hasan, Monowar and Liu, Songran and Mohan, Sibin and Caccamo, Marco},
  booktitle = {2018 ACM/IEEE 9th International Conference on Cyber-Physical Systems (ICCPS)},
  title     = {Guaranteed Physical Security with Restart-Based Design for Cyber-Physical Systems},
  year      = {2018},
  volume    = {},
  pages     = {10-21},
  doi       = {10.1109/ICCPS.2018.00010}
}

@article{resecure-iot,
  author   = {Abdi, Fardin and Chen, Chien-Ying and Hasan, Monowar and Liu, Songran and Mohan, Sibin and Caccamo, Marco},
  journal  = {IEEE Internet of Things Journal},
  title    = {Preserving Physical Safety Under Cyber Attacks},
  year     = {2019},
  volume   = {6},
  number   = {4},
  pages    = {6285-6300},
  doi      = {10.1109/JIOT.2018.2889866}
}

@inproceedings{groundhog,
  author    = { Kashinath, Ashish and Agarwala, Disha and Kulp, Gabriel and Das, Sourav and Mohan, Sibin and Venkatagiri, Radha },
  booktitle = { 2025 IEEE Symposium on Security and Privacy (SP) },
  title     = {{ Groundhog: A Restart-Based Systems Framework for Increasing Availability in Threshold Cryptosystems }},
  year      = {2025},
  volume    = {},
  issn      = {},
  pages     = {165-183},
  doi       = {10.1109/SP61157.2025.00056},
  url       = {https://doi.ieeecomputersociety.org/10.1109/SP61157.2025.00056},
  publisher = {IEEE Computer Society},
  address   = {Los Alamitos, CA, USA},
  month     = May
}

@article{arm-trustzone,
  title={Demystifying arm trustzone: A comprehensive survey},
  author={Pinto, Sandro and Santos, Nuno},
  journal={ACM computing surveys (CSUR)},
  volume={51},
  number={6},
  pages={1--36},
  year={2019},
  publisher={ACM New York, NY, USA}
}

@inproceedings{scate-workshop,
  author    = {Hasan, Monowar and Mohan, Sibin},
  title     = {Protecting Actuators in Safety-Critical IoT Systems from Control Spoofing Attacks},
  year      = {2019},
  isbn      = {9781450368384},
  publisher = {Association for Computing Machinery},
  address   = {New York, NY, USA},
  url       = {https://doi.org/10.1145/3338507.3358615},
  doi       = {10.1145/3338507.3358615},
  booktitle = {Proceedings of the 2nd International ACM Workshop on Security and Privacy for the Internet-of-Things},
  pages     = {8–14},
  numpages  = {7},
  location  = {London, United Kingdom},
  series    = {IoT S&P'19}
}

@inproceedings{scate-isorc,
  author    = {Hasan, Monowar and Mohan, Sibin},
  booktitle = {2023 IEEE 26th International Symposium on Real-Time Distributed Computing (ISORC)},
  title     = {You Can’t Always Check What You Wanted: : Selective Checking and Trusted Execution to Prevent False Actuations in Real-Time Internet-of-Things},
  year      = {2023},
  volume    = {},
  number    = {},
  pages     = {42-53},
  doi       = {10.1109/ISORC58943.2023.00017}
}

@inproceedings{indistinguishability,
  author    = {Chen, Chien-Ying and Sanyal, Debopam and Mohan, Sibin},
  title     = {Indistinguishability Prevents Scheduler Side Channels in Real-Time Systems},
  year      = {2021},
  isbn      = {9781450384544},
  publisher = {Association for Computing Machinery},
  address   = {New York, NY, USA},
  url       = {https://doi.org/10.1145/3460120.3484769},
  doi       = {10.1145/3460120.3484769},
  booktitle = {Proceedings of the 2021 ACM SIGSAC Conference on Computer and Communications Security},
  pages     = {666–684},
  numpages  = {19},
  location  = {Virtual Event, Republic of Korea},
  series    = {CCS '21}
}


%% file: refs/ref-tarek.bib
@inproceedings{kimura2024case,
  title={The Case for Micro Foundation Models to Support Robust Edge Intelligence},
  author={Kimura, Tomoyoshi and Misra, Ashitabh and Chen, Yizhuo and Kara, Denizhan and Li, Jinyang and Wang, Tianshi and Wang, Ruijie and Bhattacharyya, Joydeep and Kim, Jae and Shenoy, Prashant and others},
  booktitle={2024 IEEE 6th International Conference on Cognitive Machine Intelligence (CogMI)},
  pages={23--31},
  year={2024},
  organization={IEEE}
}

@inproceedings{kimura2024vibrofm,
  title={Vibrofm: Towards micro foundation models for robust multimodal iot sensing},
  author={Kimura, Tomoyoshi and Li, Jinyang and Wang, Tianshi and Chen, Yizhuo and Wang, Ruijie and Kara, Denizhan and Wigness, Maggie and Bhattacharyya, Joydeep and Srivatsa, Mudhakar and Liu, Shengzhong and others},
  booktitle={2024 IEEE 21st International Conference on Mobile Ad-Hoc and Smart Systems (MASS)},
  pages={10--18},
  year={2024},
  organization={IEEE}
}

@article{baris2025foundation,
  title={Foundation Models for CPS-IoT: Opportunities and Challenges},
  author={Baris, Ozan and Chen, Yizhuo and Dong, Gaofeng and Han, Liying and Kimura, Tomoyoshi and Quan, Pengrui and Wang, Ruijie and Wang, Tianchen and Abdelzaher, Tarek and Berg{\'e}s, Mario and others},
  journal={arXiv preprint arXiv:2501.16368},
  year={2025}
}

@inproceedings{wang2025dynagen,
  title={DynaGen: Conditional Diffusion Models for Enhancing Acoustic and Seismic-Based Vehicle Detection},
  author={Wang, Tianshi and Li, Jinyang and Yang, Qikai and Wang, Ruijie and Chen, Yizhuo and Sun, Dachun and Li, Bohan and Hu, Yigong and Kimura, Tomoyoshi and Kara, Denizhan and Abdelzaher, Tarek},
  booktitle={In Proc. IEEE Conference on Computer Communications (Infocom)},
  location = {London, UK},
  month = {May},
  year={2025}
}

@inproceedings{wang2023sudokusens,
  title={Sudokusens: Enhancing deep learning robustness for iot sensing applications using a generative approach},
  author={Wang, Tianshi and Li, Jinyang and Wang, Ruijie and Kara, Denizhan and Liu, Shengzhong and Wertheimer, Davis and Viros i Martin, Antoni and Ganti, Raghu and Srivatsa, Mudhakar and Abdelzaher, Tarek},
  booktitle={Proceedings of the 21st ACM Conference on Embedded Networked Sensor Systems},
  pages={15--27},
  year={2023}
}

@inproceedings{wang2024data,
  title={Data augmentation for human activity recognition via condition space interpolation within a generative model},
  author={Wang, Tianshi and Chen, Yizhuo and Yang, Qikai and Sun, Dachun and Wang, Ruijie and Li, Jinyang and Kimura, Tomoyoshi and Abdelzaher, Tarek},
  booktitle={2024 33rd International Conference on Computer Communications and Networks (ICCCN)},
  pages={1--9},
  year={2024},
  organization={IEEE}
}

@inproceedings{kara2024phymask,
  title={PhyMask: An Adaptive Masking Paradigm for Efficient Self-Supervised Learning in IoT},
  author={Kara, Denizhan and Kimura, Tomoyoshi and Chen, Yatong and Li, Jinyang and Wang, Ruijie and Chen, Yizhuo and Wang, Tianshi and Liu, Shengzhong and Abdelzaher, Tarek},
  booktitle={Proceedings of the 22nd ACM Conference on Embedded Networked Sensor Systems},
  pages={97--111},
  year={2024}
}

@inproceedings{liu2021contrastive,
  title={Contrastive self-supervised representation learning for sensing signals from the time-frequency perspective},
  author={Liu, Dongxin and Wang, Tianshi and Liu, Shengzhong and Wang, Ruijie and Yao, Shuochao and Abdelzaher, Tarek},
  booktitle={2021 International Conference on Computer Communications and Networks (ICCCN)},
  pages={1--10},
  year={2021},
  organization={IEEE}
}

@inproceedings{kara2024freqmae,
  title={FreqMAE: Frequency-Aware Masked Autoencoder for Multi-Modal IoT Sensing},
  author={Kara, Denizhan and Kimura, Tomoyoshi and Liu, Shengzhong and Li, Jinyang and Liu, Dongxin and Wang, Tianshi and Wang, Ruijie and Chen, Yizhuo and Hu, Yigong and Abdelzaher, Tarek},
  booktitle={Proceedings of the ACM on Web Conference 2024},
  pages={2795--2806},
  year={2024}
}

@article{brown2020language,
  title={Language models are few-shot learners},
  author={Brown, Tom and Mann, Benjamin and Ryder, Nick and Subbiah, Melanie and Kaplan, Jared D and Dhariwal, Prafulla and Neelakantan, Arvind and Shyam, Pranav and Sastry, Girish and Askell, Amanda and others},
  journal={Advances in neural information processing systems},
  volume={33},
  pages={1877--1901},
  year={2020}
}

@article{yao2018sensegan,
  title={Sensegan: Enabling deep learning for internet of things with a semi-supervised framework},
  author={Yao, Shuochao and Zhao, Yiran and Shao, Huajie and Zhang, Chao and Zhang, Aston and Hu, Shaohan and Liu, Dongxin and Liu, Shengzhong and Su, Lu and Abdelzaher, Tarek},
  journal={Proceedings of the ACM on interactive, mobile, wearable and ubiquitous technologies},
  volume={2},
  number={3},
  pages={1--21},
  year={2018},
  publisher={ACM New York, NY, USA}
}

@article{liu2024focal,
  title={FOCAL: Contrastive learning for multimodal time-series sensing signals in factorized orthogonal latent space},
  author={Liu, Shengzhong and Kimura, Tomoyoshi and Liu, Dongxin and Wang, Ruijie and Li, Jinyang and Diggavi, Suhas and Srivastava, Mani and Abdelzaher, Tarek},
  journal={Advances in Neural Information Processing Systems},
  volume={36},
  year={2024}
}

@inproceedings{caron2021emerging,
  title={Emerging properties in self-supervised vision transformers},
  author={Caron, Mathilde and Touvron, Hugo and Misra, Ishan and J{\'e}gou, Herv{\'e} and Mairal, Julien and Bojanowski, Piotr and Joulin, Armand},
  booktitle={Proceedings of the IEEE/CVF international conference on computer vision},
  pages={9650--9660},
  year={2021}
}


%% file: refs/ref-wenhao.bib
@article{prorok2021beyond,
  title={Beyond robustness: A taxonomy of approaches towards resilient multi-robot systems},
  author={Prorok, Amanda and Malencia, Matthew and Carlone, Luca and Sukhatme, Gaurav S and Sadler, Brian M and Kumar, Vijay},
  journal={arXiv preprint arXiv:2109.12343},
  year={2021}
}

@inproceedings{luo2020behavior,
  title={Behavior mixing with minimum global and subgroup connectivity maintenance for large-scale multi-robot systems},
  author={Luo, Wenhao and Yi, Sha and Sycara, Katia},
  booktitle={2020 IEEE International Conference on Robotics and Automation (ICRA)},
  pages={9845--9851},
  year={2020},
  organization={IEEE}
}

@INPROCEEDINGS{yang2024decentralized, 
    AUTHOR    = {Yupeng Yang AND Yiwei Lyu AND Yanze Zhang AND Sha Yi AND Wenhao Luo}, 
    TITLE     = {{Decentralized Multi-Robot Line-of-Sight Connectivity Maintenance under Uncertainty}}, 
    BOOKTITLE = {Proceedings of Robotics: Science and Systems}, 
    YEAR      = {2024}, 
    ADDRESS   = {Delft, Netherlands}, 
    MONTH     = {July}, 
    DOI       = {10.15607/RSS.2024.XX.005} 
}

@article{cavorsi2023multirobot,
  title={Multirobot adversarial resilience using control barrier functions},
  author={Cavorsi, Matthew and Sabattini, Lorenzo and Gil, Stephanie},
  journal={IEEE Transactions on Robotics},
  volume={40},
  pages={797--815},
  year={2023},
  publisher={IEEE}
}

@inproceedings{luo2019minimum,
  title={Minimum k-connectivity maintenance for robust multi-robot systems},
  author={Luo, Wenhao and Sycara, Katia},
  booktitle={IEEE/RSJ International Conference on Intelligent Robots and Systems (IROS)},
  pages={7370--7377},
  year={2019},
  organization={IEEE}
}

@inproceedings{luo2020minimally,
  title={Minimally disruptive connectivity enhancement for resilient multi-robot teams},
  author={Luo, Wenhao and Chakraborty, Nilanjan and Sycara, Katia},
  booktitle={IEEE/RSJ International Conference on Intelligent Robots and Systems (IROS)},
  pages={11809--11816},
  year={2020},
  organization={IEEE}
}

@inproceedings{yang2024integrating,
  title={Integrating Online Learning and Connectivity Maintenance for Communication-Aware Multi-Robot Coordination},
  author={Yang, Yupeng and Lyu, Yiwei and Zhang, Yanze and Gao, Ian and Luo, Wenhao},
  booktitle={IEEE/RSJ International Conference on Intelligent Robots and Systems (IROS)},
  pages={5770--5776},
  year={2024},
  organization={IEEE}
}

@inproceedings{yang2023minimally,
  title={Minimally constrained multi-robot coordination with line-of-sight connectivity maintenance},
  author={Yang, Yupeng and Lyu, Yiwei and Luo, Wenhao},
  booktitle={IEEE International Conference on Robotics and Automation (ICRA)},
  pages={7684--7690},
  year={2023},
  organization={IEEE}
}

@article{ong2023nonsmooth,
  title={Nonsmooth control barrier function design of continuous constraints for network connectivity maintenance},
  author={Ong, Pio and Capelli, Beatrice and Sabattini, Lorenzo and Cort{\'e}s, Jorge},
  journal={Automatica},
  volume={156},
  pages={111209},
  year={2023},
  publisher={Elsevier}
}


%% file: refs/ref-xugui.bib
@article{rai2020driven,
  title={Driven by data or derived through physics? a review of hybrid physics guided machine learning techniques with cyber-physical system (cps) focus},
  author={Rai, Rahul and Sahu, Chandan K},
  journal={IEEe Access},
  volume={8},
  pages={71050--71073},
  year={2020},
  publisher={IEEE}
}

@inproceedings{zhou2022robustness,
  title={Robustness testing of data and knowledge driven anomaly detection in cyber-physical systems},
  author={Zhou, Xugui and Kouzel, Maxfield and Alemzadeh, Homa},
  booktitle={2022 52nd Annual IEEE/IFIP International Conference on Dependable Systems and Networks Workshops (DSN-W)},
  pages={44--51},
  year={2022},
  organization={IEEE}
}

@misc{xu2018semantic,
      title={A Semantic Loss Function for Deep Learning with Symbolic Knowledge}, 
      author={Jingyi Xu and Zilu Zhang and Tal Friedman and Yitao Liang and Guy Van den Broeck},
      year={2018},
      eprint={1711.11157},
      archivePrefix={arXiv},
      primaryClass={cs.AI}
}

@inproceedings{liang2019learning,
  title={Learning logistic circuits},
  author={Liang, Yitao and Van den Broeck, Guy},
  booktitle={Proceedings of the AAAI Conference on Artificial Intelligence},
  volume={33},
  number={01},
  pages={4277-4286},
  year={2019}
}

@inproceedings{guo2016jointly,
  title={Jointly embedding knowledge graphs and logical rules},
  author={Guo, Shu and Wang, Quan and Wang, Lihong and Wang, Bin and Guo, Li},
  booktitle={Proceedings of the 2016 conference on empirical methods in natural language processing},
  pages={192-202},
  year={2016}
}

@article{gou2021knowledge,
  title={Knowledge distillation: A survey},
  author={Gou, Jianping and Yu, Baosheng and Maybank, Stephen J and Tao, Dacheng},
  journal={International Journal of Computer Vision},
  volume={129},
  number={6},
  pages={1789-1819},
  year={2021},
  publisher={Springer}
}

@InProceedings{bartocci_2018,
author="Bartocci, Ezio",
editor="Colombo, Christian
and Leucker, Martin",
title="Monitoring, Learning and Control of Cyber-Physical Systems with STL (Tutorial)",
booktitle="Runtime Verification",
year="2018",
publisher="Springer International Publishing",
address="Cham",
pages="35-42",
isbn="978-3-030-03769-7"
}

@inproceedings{telexpaper,
author = {Jha, Susmit and Tiwari, Ashish and Seshia, Sanjit and Sahai, Tuhin and Shankar, Natarajan},
year = {2017},
month = {09},
pages = {208-224},
title = {TeLEx: Passive STL Learning Using Only Positive Examples},
isbn = {978-3-319-67530-5},
}

@INPROCEEDINGS{JonesAnomaly,  
author={A. {Jones} and Z. {Kong} and C. {Belta}},  booktitle={53rd IEEE Conference on Decision and Control},   
title={Anomaly detection in cyber-physical systems: A formal methods approach},   
year={2014},  
volume={},  
number={},  
pages={848-853}
}

@ARTICLE{embedlogic2019chen,
  author={Chen, Bingfeng and Hao, Zhifeng and Cai, Xiaofeng and Cai, Ruichu and Wen, Wen and Zhu, Jian and Xie, Guangqiang},
  journal={IEEE Access}, 
  title={Embedding Logic Rules Into Recurrent Neural Networks}, 
  year={2019},
  volume={7},
  number={},
  pages={},
  doi={10.1109/ACCESS.2019.2892140}}

@misc{rusu2016policy,
  doi = {10.48550/ARXIV.1511.06295}, 
  url = {https://arxiv.org/abs/1511.06295},
  author = {Rusu, Andrei A. and Colmenarejo, Sergio Gomez and Gulcehre, Caglar and Desjardins, Guillaume and Kirkpatrick, James and Pascanu, Razvan and Mnih, Volodymyr and Kavukcuoglu, Koray and Hadsell, Raia},
  title = {Policy Distillation},
  publisher = {arXiv},
  year = {2015},
  copyright = {arXiv.org perpetual, non-exclusive license}
}

@INPROCEEDINGS{dsn2021zhou,
  author={Zhou, Xugui and Ahmed, Bulbul and Aylor, James H. and Asare, Philip and Alemzadeh, Homa},
  booktitle={51st Annual IEEE/IFIP International Conference on Dependable Systems and Networks (DSN)}, 
  title={Data-driven Design of Context-aware Monitors for Hazard Prediction in Artificial Pancreas Systems}, 
  year={2021},
  volume={},
  number={},
  pages={484-496},
  doi={10.1109/DSN48987.2021.00058}
}

@book{leveson2011engineering,
  title={Engineering a safer world: Systems thinking applied to safety},
  author={Leveson, Nancy G.},
  year={2011},
  publisher={MIT Press},
  address={Cambridge, MA}
}

@article{lee2004trust,
  title={Trust in automation: Designing for appropriate reliance},
  author={Lee, John D. and See, Katrina A.},
  journal={Human Factors},
  volume={46},
  number={1},
  pages={50--80},
  year={2004},
  publisher={SAGE Publications}
}

@inproceedings{wischnewski2023measuring,
  title={Measuring and understanding trust calibrations for automated systems: A survey of the state-of-the-art and future directions},
  author={Wischnewski, Magdalena and Kr{\"a}mer, Nicole and M{\"u}ller, Emmanuel},
  booktitle={Proceedings of the 2023 CHI conference on human factors in computing systems},
  pages={1--16},
  year={2023}
}

@article{parasuraman2000model,
  title={A model for types and levels of human interaction with automation},
  author={Parasuraman, Raja and Sheridan, Thomas B. and Wickens, Christopher D.},
  journal={IEEE Transactions on Systems, Man, and Cybernetics - Part A: Systems and Humans},
  volume={30},
  number={3},
  pages={286--297},
  year={2000},
  publisher={IEEE}
}

@inproceedings{zhou2022strategic,
  title={Strategic safety-critical attacks against an advanced driver assistance system},
  author={Zhou, Xugui and Schmedding, Anna and Ren, Haotian and Yang, Lishan and Schowitz, Philip and Smirni, Evgenia and Alemzadeh, Homa},
  booktitle={2022 52nd Annual IEEE/IFIP International Conference on Dependable Systems and Networks (DSN)},
  pages={79--87},
  year={2022},
  organization={IEEE}
}

@article{wang2024enhancing,
  title={Enhancing Human--Machine Collaboration: A Trust-Aware Trajectory Planning Framework for Assistive Aerial Teleoperation},
  author={Wang, Z. and others},
  journal={Aerospace},
  volume={13},
  number={9},
  pages={876},
  year={2024},
  publisher={MDPI}
}

@article{li2025trusttriggered,
  title={Trust-Triggered Cyber-Physical-Human System for Human-Robot Collaboration in Flexible Manufacturing},
  author={Li, X. and others},
  journal={IEEE Transactions on Industrial Informatics},
  note={Early Access},
  year={2025},
  publisher={IEEE}
}

@article{wildman2024trust,
  title={Trust in Human-Agent Teams: A Multilevel Perspective and Future Research Agenda},
  author={Wildman, Jessica L. and Nguyen, T. and others},
  journal={Organizational Psychology Review},
  volume={14},
  number={3},
  year={2024},
  publisher={SAGE Publications}
}

@article{wang2024trustreflective,
  title={Trust-Aware Reflective Control for Fault-Resilient Dynamic Task Response in Human–Swarm Cooperation},
  author={Wang, Y. and others},
  journal={Robotics},
  volume={5},
  number={1},
  pages={22},
  year={2024},
  publisher={MDPI}
}

@article{stray2025explainability,
  title={Preliminary Quantitative Study on Explainability and Trust in AI Systems},
  author={Stray, J. and others},
  journal={arXiv preprint arXiv:2510.15769},
  year={2025}
}

@article{lai2025exploring,
  title={Exploring automation bias in human--AI collaboration: a review and implications for explainable AI},
  author={Lai, Y. and others},
  journal={AI \& Society},
  year={2025},
  publisher={Springer}
}

@article{mhapsekar2024building,
  author={Mhapsekar, Rahul Umesh et al.},
  title={Building Trust in AI-Driven Decision Making for Cyber-Physical Systems (CPS): A Comprehensive Review},
  journal={arXiv preprint arXiv:2405.06347},
  year={2024}
}

@article{tenhundfeld2022assessment,
  author={Tenhundfeld, Nathan et al.},
  title={Assessment of Trust in Automation in the “Real World”: Requirements for New Trust in Automation Measurement Techniques for Use by Practitioners},
  journal={Human Factors},
  volume={64},
  number={4},
  pages={534--555},
  year={2022}
}

@book{sheridan2014humans,
  title={Humans and Automation: System Design and Research Issues},
  author={Sheridan, Thomas B.},
  publisher={Wiley},
  year={2014}
}

@article{endsley2017here,
  title={From here to autonomy: lessons learned from human--automation research},
  author={Endsley, Mica R},
  journal={Human factors},
  volume={59},
  number={1},
  pages={5--27},
  year={2017},
  publisher={Sage Publications Sage CA: Los Angeles, CA}
}

@article{faroni2022safety,
  title={Safety-aware time-optimal motion planning with uncertain human state estimation},
  author={Faroni, Marco and Beschi, Manuel and Pedrocchi, Nicola},
  journal={IEEE Robotics and Automation Letters},
  volume={7},
  number={4},
  pages={12219--12226},
  year={2022},
  publisher={IEEE}
}

@inproceedings{fan2008influence,
  title={The influence of agent reliability on trust in human-agent collaboration},
  author={Fan, Xiaocong and Oh, Sooyoung and McNeese, Michael and Yen, John and Cuevas, Haydee and Strater, Laura and Endsley, Mica R},
  booktitle={Proceedings of the 15th European conference on Cognitive ergonomics: the ergonomics of cool interaction},
  pages={1--8},
  year={2008}
}

@article{zhou2023hybrid,
  title={Hybrid knowledge and data driven synthesis of runtime monitors for cyber-physical systems},
  author={Zhou, Xugui and Ahmed, Bulbul and Aylor, James H and Asare, Philip and Alemzadeh, Homa},
  journal={IEEE Transactions on Dependable and Secure Computing},
  volume={21},
  number={1},
  pages={12--30},
  year={2023},
  publisher={IEEE}
}

@article{DriveSim,
  author={NVIDIA},
  title={NVIDIA Drive Simulation},
  year={2018},
  howpublished = {\url{https://www.nvidia.com/en-us/self-driving-cars/drive-constellation/}}
}

@inproceedings{zhou2024runtime,
  title={{Runtime Stealthy Perception Attacks against DNN-Based Adaptive Cruise Control Systems}},
  author={Zhou, Xugui and Chen, Anqi and Kouzel, Maxfield and Ren, Haotian and McCarty, Morgan and Nita-Rotaru, Cristina and Alemzadeh, Homa},
  booktitle={ACM
Asia Conference on Computer and Communications Security (ASIA CCS)},
  year={2025}
}

@inproceedings{shao2024lmdrive,
  title={LMDrive: Closed-Loop End-to-End Driving with Large Language Models},
  author={Shao, Hao and Wang, Yuxuan and Chen, Ruo-Ping and others},
  booktitle={Proceedings of the IEEE/CVF Conference on Computer Vision and Pattern Recognition (CVPR)},
  year={2024}
}

@inproceedings{wei2024chatsim,
  title={ChatSim: Editable Scene Simulation for Autonomous Driving via Collaborative LLM-Agents},
  author={Wei, Yuxi and Wang, Zi and Lu, Yifan and others},
  booktitle={Proceedings of the IEEE/CVF Conference on Computer Vision and Pattern Recognition (CVPR)},
  year={2024}
}

@article{shen2023simonwheels,
  title={Sim-on-wheels: physical world in the loop simulation for self-driving},
  author={Shen, Yuan and Chandaka, Bhargav and Lin, Zhi-Hao and Zhai, Albert and Cui, Hang and Forsyth, David and Wang, Shenlong},
  journal={IEEE Robotics and Automation Letters},
  volume={8},
  number={12},
  pages={8192--8199},
  year={2023},
  publisher={IEEE}
}


%% file: refs/ref-yin.bib
@article{yao2024minicpm,
  title={{MiniCPM-v}: A {GPT-4v} level {MLLM} on your phone},
  author={Yao, Yuan and Yu, Tianyu and Zhang, Ao and Wang, Chongyi and Cui, Junbo and Zhu, Hongji and Cai, Tianchi and Li, Haoyu and Zhao, Weilin and He, Zhihui and others},
  journal={Nature Communication (in press)},
  year={2025}
}

@inproceedings{wu2023multimodal,
  title={Multimodal large language models: A survey},
  author={Wu, Jiayang and Gan, Wensheng and Chen, Zefeng and Wan, Shicheng and Philip, S Yu},
  booktitle={2023 IEEE International Conference on Big Data (BigData)},
  pages={2247--2256},
  year={2023},
  organization={IEEE}
}

@InProceedings{zitkovich2023rt2,
  title = 	 {RT-2: Vision-Language-Action Models Transfer Web Knowledge to Robotic Control},
  author =       {Zitkovich, Brianna and Yu, Tianhe and Xu, Sichun and Xu, Peng and Xiao, Ted and Xia, Fei and Wu, Jialin and Wohlhart, Paul and Welker, Stefan and Wahid, Ayzaan and Vuong, Quan and Vanhoucke, Vincent and Tran, Huong and Soricut, Radu and Singh, Anikait and Singh, Jaspiar and Sermanet, Pierre and Sanketi, Pannag R. and Salazar, Grecia and Ryoo, Michael S. and Reymann, Krista and Rao, Kanishka and Pertsch, Karl and Mordatch, Igor and Michalewski, Henryk and Lu, Yao and Levine, Sergey and Lee, Lisa and Lee, Tsang-Wei Edward and Leal, Isabel and Kuang, Yuheng and Kalashnikov, Dmitry and Julian, Ryan and Joshi, Nikhil J. and Irpan, Alex and Ichter, Brian and Hsu, Jasmine and Herzog, Alexander and Hausman, Karol and Gopalakrishnan, Keerthana and Fu, Chuyuan and Florence, Pete and Finn, Chelsea and Dubey, Kumar Avinava and Driess, Danny and Ding, Tianli and Choromanski, Krzysztof Marcin and Chen, Xi and Chebotar, Yevgen and Carbajal, Justice and Brown, Noah and Brohan, Anthony and Arenas, Montserrat Gonzalez and Han, Kehang},
  booktitle = 	 {Proceedings of The 7th Conference on Robot Learning},
  pages = 	 {2165--2183},
  year = 	 {2023},
  editor = 	 {Tan, Jie and Toussaint, Marc and Darvish, Kourosh},
  volume = 	 {229},
  series = 	 {Proceedings of Machine Learning Research},
  month = 	 {06--09 Nov},
  publisher =    {PMLR},
}

@inproceedings{xu2024towards,
  title={Towards Few-Shot Adaptation of Foundation Models via Multitask Finetuning},
  author={Xu, Zhuoyan and Shi, Zhenmei and Wei, Junyi and Mu, Fangzhou and Li, Yin and Liang, Yingyu},
  year={2024},
  booktitle={International Conference on Learning Representations}
}

@inproceedings{liu2025pave,
  title={PAVE: Patching and Adapting Video Large Language Models},
  author={Liu, Zhuoming and Li, Yiquan and Nguyen, Khoi Duc and Zhong, Yiwu and Li, Yin},
  booktitle={Proceedings of the Computer Vision and Pattern Recognition Conference},
  pages={3306--3317},
  year={2025}
}

@inproceedings{hu2021lora,
title={Lo{RA}: Low-Rank Adaptation of Large Language Models},
author={Edward J Hu and Yelong Shen and Phillip Wallis and Zeyuan Allen-Zhu and Yuanzhi Li and Shean Wang and Lu Wang and Weizhu Chen},
booktitle={International Conference on Learning Representations},
year={2022},
url={https://openreview.net/forum?id=nZeVKeeFYf9}
}

@inproceedings{wei2022finetuned,
  title={Finetuned Language Models are Zero-Shot Learners},
  author={Wei, Jason and Bosma, Maarten and Zhao, Vincent and Guu, Kelvin and Yu, Adams Wei and Lester, Brian and Du, Nan and Dai, Andrew M and Le, Quoc V},
  booktitle={International Conference on Learning Representations},
  year={2022},
}

@article{wang2020generalizing,
  title={Generalizing from a few examples: A survey on few-shot learning},
  author={Wang, Yaqing and Yao, Quanming and Kwok, James T and Ni, Lionel M},
  journal={ACM computing surveys},
  year={2020},
  publisher={ACM New York, NY, USA}
}

@inproceedings{jia2022visual,
  title={Visual prompt tuning},
  author={Jia, Menglin and Tang, Luming and Chen, Bor-Chun and Cardie, Claire and Belongie, Serge and Hariharan, Bharath and Lim, Ser-Nam},
  booktitle={European conference on computer vision},
  pages={709--727},
  year={2022},
  organization={Springer}
}

@article{hospedales2021meta,
  title={Meta-learning in neural networks: A survey},
  author={Hospedales, Timothy and Antoniou, Antreas and Micaelli, Paul and Storkey, Amos},
  journal={IEEE Transactions on Pattern Analysis and Machine Intelligence},
  year={2021},
  publisher={IEEE}
}

@inproceedings{sanh2022multitask,
  title={Multitask Prompted Training Enables Zero-Shot Task Generalization},
  author={Sanh, Victor and Webson, Albert and Raffel, Colin and Bach, Stephen H and Sutawika, Lintang and Alyafeai, Zaid and Chaffin, Antoine and Stiegler, Arnaud and Le Scao, Teven and Raja, Arun and others},
  booktitle={International Conference on Learning Representations},
  year={2022}
}

@inproceedings{muennighoff2023crosslingual,
  title={Crosslingual Generalization through Multitask Finetuning},
  author={Muennighoff, Niklas and Wang, Thomas and Sutawika, Lintang and Roberts, Adam and Biderman, Stella and Le Scao, Teven and Bari, M Saiful and Shen, Sheng and Yong, Zheng Xin and Schoelkopf, Hailey and others},
  booktitle={Proceedings of the 61st Annual Meeting of the Association for Computational Linguistics (Volume 1: Long Papers)},
  pages={15991--16111},
  year={2023}
}

@inproceedings{kumari2023multi,
  title={Multi-concept customization of text-to-image diffusion},
  author={Kumari, Nupur and Zhang, Bingliang and Zhang, Richard and Shechtman, Eli and Zhu, Jun-Yan},
  booktitle={Proceedings of the IEEE/CVF conference on computer vision and pattern recognition},
  pages={1931--1941},
  year={2023}
}


%% file: refs/ref-yong.bib
@book{tabuada2009verification,
  title={Verification and control of hybrid systems: a symbolic approach},
  author={Tabuada, Paulo},
  year={2009},
  publisher={Springer Science \& Business Media}
}

@book{belta2017formal,
  title={Formal methods for discrete-time dynamical systems},
  author={Belta, Calin and Yordanov, Boyan and Gol, Ebru Aydin},
  volume={89},
  year={2017},
  publisher={Springer}
}

@incollection{khajenejad2022resilient,
  title={Resilient state estimation and attack mitigation in cyber-physical systems},
  author={Khajenejad, Mohammad and Yong, Sze Zheng},
  booktitle={Security and Resilience in Cyber-Physical Systems: Detection, Estimation and Control},
  pages={149--185},
  year={2022},
  publisher={Springer}
}

@article{pajic2016attack,
  title={Attack-resilient state estimation for noisy dynamical systems},
  author={Pajic, Miroslav and Lee, Insup and Pappas, George J},
  journal={IEEE Transactions on Control of Network Systems},
  volume={4},
  number={1},
  pages={82--92},
  year={2016},
  publisher={IEEE}
}

@inproceedings{khajenejad2023resilient,
  title={Resilient state estimation for nonlinear discrete-time systems via input and state interval observer synthesis},
  author={Khajenejad, Mohammad and Jin, Zeyuan and Dinh, Thach Ngoc and Yong, Sze Zheng},
  booktitle={2023 62nd IEEE Conference on Decision and Control (CDC)},
  pages={1826--1832},
  year={2023},
  organization={IEEE}
}

@article{yong2018switching,
  title={Switching and data injection attacks on stochastic cyber-physical systems: Modeling, resilient estimation, and attack mitigation},
  author={Yong, Sze Zheng and Zhu, Minghui and Frazzoli, Emilio},
  journal={ACM Transactions on Cyber-Physical Systems},
  volume={2},
  number={2},
  pages={1--2},
  year={2018},
  publisher={ACM New York, NY, USA}
}

@inproceedings{murguia2016cusum,
  title={Cusum and chi-squared attack detection of compromised sensors},
  author={Murguia, Carlos and Ruths, Justin},
  booktitle={2016 IEEE Conference on Control Applications (CCA)},
  pages={474--480},
  year={2016},
  organization={IEEE}
}

@inproceedings{mo2010false,
  title={False data injection attacks in control systems},
  author={Mo, Yilin and Sinopoli, Bruno},
  booktitle={Preprints of the 1st workshop on Secure Control Systems},
  volume={1},
  year={2010}
}

@article{jin2017adaptive,
  title={An adaptive control architecture for mitigating sensor and actuator attacks in cyber-physical systems},
  author={Jin, Xu and Haddad, Wassim M and Yucelen, Tansel},
  journal={IEEE Transactions on Automatic Control},
  volume={62},
  number={11},
  pages={6058--6064},
  year={2017},
  publisher={IEEE}
}

@inproceedings{dan2010stealth,
  title={Stealth attacks and protection schemes for state estimators in power systems},
  author={D{\'a}n, Gy{\"o}rgy and Sandberg, Henrik},
  booktitle={2010 first IEEE international conference on smart grid communications},
  pages={214--219},
  year={2010},
  organization={IEEE}
}

@article{mishra2016secure,
  title={Secure state estimation against sensor attacks in the presence of noise},
  author={Mishra, Shaunak and Shoukry, Yasser and Karamchandani, Nikhil and Diggavi, Suhas N and Tabuada, Paulo},
  journal={IEEE Transactions on Control of Network Systems},
  volume={4},
  number={1},
  pages={49--59},
  year={2016},
  publisher={IEEE}
}

@inproceedings{yang2019sensor,
  title={Sensor redundancy for robustness in nonlinear state estimation},
  author={Yang, Guitao and Rezaee, Hamed and Parisini, Thomas},
  booktitle={2019 IEEE 58th Conference on Decision and Control (CDC)},
  pages={3865--3870},
  year={2019},
  organization={IEEE}
}

@article{yang2020multi,
  title={A multi-observer based estimation framework for nonlinear systems under sensor attacks},
  author={Yang, Tianci and Murguia, Carlos and Kuijper, Margreta and Ne{\v{s}}i{\'c}, Dragan},
  journal={Automatica},
  volume={119},
  pages={109043},
  year={2020},
  publisher={Elsevier}
}

@inproceedings{yeh1996triple,
  title={Triple-triple redundant 777 primary flight computer},
  author={Yeh, Ying C},
  booktitle={1996 IEEE Aerospace Applications Conference. Proceedings},
  volume={1},
  pages={293--307},
  year={1996},
  organization={IEEE}
}

@inproceedings{berg2016verification,
  title={Verification of triple modular redundancy (TMR) insertion for reliable and trusted systems},
  author={Berg, Melanie and LaBel, Kenneth A},
  booktitle={2016 MRQW Microelectronics Reliability and Qualification Working Meeting},
  number={GSFC-E-DAA-TN29375},
  year={2016}
}


%% file: refs/refs2.bib
@inproceedings{NawOrn20,
author={Farhad Nawaz and Melkior Ornik},
title={Explorative probabilistic planning with unknown target locations},
booktitle={59th IEEE Conference on Decision and Control},
pages={2732--2737},
year={2020}
}

@inproceedings{Woletal12, 
author={Eric M. {Wolff} and Ufuk {Topcu} and Richard M. {Murray}}, 
booktitle={51st IEEE Conference on Decision and Control}, 
title={Robust control of uncertain {Markov Decision Processes} with temporal logic specifications}, 
year={2012},
pages={3372--3379}}

@inproceedings{Ahmetal13,
  title={Regret based robust solutions for uncertain {Markov} decision processes},
  author={Ahmed, Asrar and Varakantham, Pradeep and Adulyasak, Yossiri and Jaillet, Patrick},
  booktitle={27th International Conference on Neural Information Processing Systems},
  year={2013}
}

@article{OrnTop21,
  title={Learning and planning for time-varying {MDPs} using maximum likelihood estimation},
  author={Ornik, Melkior and Topcu, Ufuk},
  journal={Journal of Machine Learning Research},
  volume={22},
  year={2021}
}

@article{Thaetal22,
  title={Expedited Online Learning with Spatial Side Information},
  author={Thangeda, Pranay and Ornik, Melkior and Topcu, Ufuk},
  journal={IEEE Transactions on Automatic Control},
  year={2022}
}

@inproceedings{Ornetal18,
  title={Expedited learning in {MDPs} with side information},
  author={Ornik, Melkior and Fu, Jie and Lauffer, Niklas T. and Perera, W. K. and Alshiekh, Mohammed and Ono, Masahiro and Topcu, Ufuk},
  booktitle={57th IEEE Conference on Decision and Control},
  pages={1941--1948},
  year={2018}
}

@article{ShaOrn22,
  title={Reachability of Nonlinear Systems with Unknown Dynamics},
  author={Shafa, Taha and Ornik, Melkior},
  journal={IEEE Transactions on Automatic Control},
  year={2022}
}

@inproceedings{Putetal24,
  title={Weathering ongoing uncertainty: Learning and planning in a time-varying partially observable environment},
  author={G. Puthumanaillam and X. Liu and N. Mehr and M. Ornik},
  booktitle={2024 IEEE International Conference on Robotics and Automation},
  year={2024}
}

@article{Cheetal08,
  title={Stochastic maximum-likelihood {DOA} estimation in the presence of unknown nonuniform noise},
  author={Chen, Chiao En and Lorenzelli, Flavio and Hudson, Ralph E. and Yao, Kung},
  journal={IEEE Transactions on Signal Processing},
  volume={56},
  number={7},
  pages={3038--3044},
  year={2008}
}

@article{Joh88,
  title={Maximum likelihood estimation of discrete control processes},
  author={John, Rust},
  journal={SIAM Journal on Control and Optimization},
  volume={26},
  number={5},
  pages={1006--1024},
  year={1988}
}

@inproceedings{FosSim20,
  title={Logarithmic regret for adversarial online control},
  author={Foster, Dylan and Simchowitz, Max},
  booktitle={International Conference on Machine Learning},
  pages={3211--3221},
  year={2020}
}

@inproceedings{Shaetal23,
  title={Lm-nav: Robotic navigation with large pre-trained models of language, vision, and action},
  author={Shah, Dhruv and Osi{\'n}ski, B{\l}a{\.z}ej and Levine, Sergey and others},
  booktitle={Conference on Robot Learning},
  pages={492--504},
  year={2023}
}

@inproceedings{Zhaetal24,
  title={Exploring generative {AI} for sim2real in driving data synthesis},
  author={Zhao, Haonan and Wang, Yiting and Bashford-Rogers, Thomas and Donzella, Valentina and Debattista, Kurt},
  booktitle={2024 IEEE Intelligent Vehicles Symposium (IV)},
  pages={3071--3077},
  year={2024},
  organization={IEEE}
}

@inproceedings{Pazetal16,
  title={Efficient PAC-optimal exploration in concurrent, continuous state MDPs with delayed updates},
  author={Pazis, Jason and Parr, Ronald},
  booktitle={Proceedings of the AAAI Conference on Artificial Intelligence},
  volume={30},
  number={1},
  year={2016}
}

@inproceedings{ThaOrn22,
  title={Adaptive sampling site selection for robotic exploration in unknown environments},
  author={Thangeda, Pranay and Ornik, Melkior},
  booktitle={2022 IEEE/RSJ International Conference on Intelligent Robots and Systems (IROS)},
  pages={4120--4125},
  year={2022},
  organization={IEEE}
}

@inproceedings{Fujetal18,
  title={Addressing function approximation error in actor-critic methods},
  author={Fujimoto, Scott and Hoof, Herke and Meger, David},
  booktitle={International conference on machine learning},
  pages={1587--1596},
  year={2018}
}

@article{Adcetal21,
  title={The gap between theory and practice in function approximation with deep neural networks},
  author={Adcock, Ben and Dexter, Nick},
  journal={SIAM Journal on Mathematics of Data Science},
  volume={3},
  number={2},
  pages={624--655},
  year={2021},
  publisher={SIAM}
}

@inproceedings{Dixetal20,
  title={Artificial intelligence and machine learning in sparse/inaccurate data situations},
  author={Dixit, Rahul and Chinnam, Ratne Babu and Singh, Harpreet},
  booktitle={2020 IEEE Aerospace Conference},
  year={2020},
  organization={IEEE}
}

@inproceedings{Putetal24d,
  title={Online Learning and Planning in Time-Varying Environments: An Aircraft Case Study},
  author={Puthumanaillam, Gokul and Mamik, Yuvraj and Ornik, Melkior},
  booktitle={AIAA SCITECH 2024 Forum},
  year={2024}
}

@inproceedings{Putetal25,
  title={Enhancing Robot Navigation Policies with Task-Specific Uncertainty Management},
  author={Puthumanaillam, Gokul and Padrao, Paolo and Fuentes, Jose and Bobadilla, Leonardo and Ornik, Melkior},
  booktitle={24th International Conference on Autonomous Agents and Multiagent Systems},
  year={2025}
}

@inproceedings{Putetal25b,
  title={TAB-Fields: A maximum entropy framework for mission-aware adversarial planning},
  author={Puthumanaillam, Gokul and Song, Jae Hyuk and Yesmagambet, Nurzhan and Park, Shinkyu and Ornik, Melkior},
  booktitle={7th Annual Learning for Dynamics \& Control Conference},
  year={2025}
}

@inproceedings{PadOrn25,
  title={Energetic resilience of linear driftless systems},
  author={Padmanabhan, Ram and Ornik, Melkior},
  booktitle={11th IFAC Symposium on Robust Control Design},
  year={2025}
}

@article{Aueetal08,
  title={Near-optimal regret bounds for reinforcement learning},
  author={Auer, Peter and Jaksch, Thomas and Ortner, Ronald},
  journal={Advances in neural information processing systems},
  volume={21},
  year={2008}
}

@article{Foretal22,
  title={Intrinsically motivated goal exploration processes with automatic curriculum learning},
  author={Forestier, S{\'e}bastien and Portelas, R{\'e}my and Mollard, Yoan and Oudeyer, Pierre-Yves},
  journal={Journal of Machine Learning Research},
  volume={23},
  number={152},
  pages={1--41},
  year={2022}
}

@inproceedings{Benetal20,
  title={Multi-robot coordination for estimation and coverage of unknown spatial fields},
  author={Benevento, Alessia and Santos, Mar{\'\i}a and Notarstefano, Giuseppe and Paynabar, Kamran and Bloch, Matthieu and Egerstedt, Magnus},
  booktitle={2020 IEEE International Conference on Robotics and Automation},
  pages={7740--7746},
  year={2020}
}

@article{Yuetal19,
  author={Yue Yu and Honglun Wang and Na Li},
  title={Fault-tolerant control for over-actuated hypersonic reentry vehicle subject to multiple disturbances and actuator faults},
  journal={Aerospace Science and Technology},
  volume={87},
  pages={230--243},
  year={2019}
}

@inproceedings{Ornetal19,
author = {Melkior Ornik and Steven Carr and Arie Israel and Ufuk Topcu},
title = {Myopic control of systems with unknown dynamics},
booktitle = {American Control Conference},
pages = {1064--1071},
year = {2019}
}

@inproceedings{ThaOrn20,
  title={{PROTRIP}: Probabilistic risk-aware optimal transit planner},
  author={Thangeda, Pranay and Ornik, Melkior},
  booktitle={23rd IEEE International Conference on Intelligent Transportation Systems},
  year={2020}
}
